\documentclass[twocolumn,floatfix]{revtex4-2}
\pdfoutput=1
\usepackage{graphicx}
\usepackage{dcolumn}
\usepackage{longtable}
\usepackage{amsmath}
\usepackage{amssymb}
\usepackage{color}
\usepackage{isotope}

\begin{document}
\title{Triaxial shapes in even-even nuclei: A theoretical overview}

\author
{Dennis Bonatsos$^1$, Andriana Martinou$^1$,  S.K. Peroulis$^1$, D.~Petrellis$^2$, P. Vasileiou$^3$,  T.J. Mertzimekis$^3$, and N. Minkov$^4$ }

\affiliation
{$^1$ Institute of Nuclear and Particle Physics, National Centre for Scientific Research ``Demokritos'', GR-15310 Aghia Paraskevi, Attiki, Greece}

\affiliation
{$^2$ Nuclear Physics Institute, Czech Academy of Sciences, CZ-250 68 \v{R}e\v{z} near Prague, Czech Republic}

\affiliation
{$^3$  Department of Physics, National and Kapodistrian University of Athens, Zografou Campus, GR-15784 Athens, Greece}

\affiliation
{$^4$ Institute of Nuclear Research and Nuclear Energy, Bulgarian Academy of Sciences, 72 Tzarigrad Road, 1784 Sofia, Bulgaria}

\begin{abstract}

Triaxial shapes in even-even nuclei have been considered since the early days of the nuclear collective model. Although many theoretical approaches have been used over the years for their description, no effort appears to have been made for grouping them together and identifying regions on the nuclear chart where the appearance of triaxiality might be favored. In addition, over the last few years, discussion has started on the appearance of small triaxiality in nuclei considered so far as purely axial rotors. In the present work we collect 
the predictions made by various theoretical approaches and show that pronounced triaxiality appears to be favored within specific stripes on the nuclear chart, with low triaxiality being present in the regions between these stripes, in agreement with
 parameter-free predictions made by the proxy-SU(3) approximation to the shell model, based on the Pauli principle and the short-range nature of the nucleon-nucleon interaction. The robustness of triaxiality within these stripes is supported by global calculations made in the framework of the Finite-Range Droplet Model (FRDM), which is based on completely different assumptions and possesses parameters fitted in order to reproduce fundamental nuclear properties.   

 \end{abstract}

\maketitle

%%%%%%%%%%%%%%%%%%%%%%%%%%%%%%%%%%%%%%%%%%
\section{Introduction}   % Sec. 1 

Atomic nuclei composed by even number of protons and even number of neutrons, called even-even nuclei, can exhibit a variety of properties and effects, like quadrupole, octupole (pear-like), and hexadecapole axially symmetric deformations, triaxial deformations, existence of isomeric states, shape coexistence, and many others. 
Global calculations over large parts of the nuclear chart have been performed using various theoretical approaches, with their results compared to compilations of experimental data, and the regions of the nuclear chart in which these effects are favored having been pointed out. For example, quadrupole and hexadecapole deformations have been calculated in Relativistic Mean Field (RMF) theory \cite{Lalazissis1999}, while relevant experimental data for the quadrupole deformation have been tabulated in Refs. \cite{Raman2001,Pritychenko2016}. Quadrupole deformations taking into account deviations from axial symmetry have been obtained by mean-field calculations involving the D1S Gogny interaction and taking into account departure from triaxiality \cite{Delaroche2010}, while parameter-independent predictions for the deformation variables have been obtained microscopically on a symmetry basis in Refs. \cite{Bonatsos2017b,Bonatsos2023}. Quadrupole, hexadecapole, and in addition octupole deformations have been calculated in the Finite-Range Droplet Model (FRDM) \cite{Moeller2016}, while relevant experimental data for the octupole deformation have been tabulated in Refs. \cite{Spear1989,Kibedi2002}. Regions in which octupole deformation appears have been reviewed in Refs. \cite{Butler1996,Butler2016}, while a microscopic derivation of octupole magic numbers has been recently given \cite{Martinou2024}.  
Shape isomers have been calculated within the Finite-Range Liquid-Drop Model (FRLDM) \cite{Moeller2009,Moeller2012} and reviewed in Ref. \cite{Dracoulis2016,Garg2023,Walker2024}. 
Shape coexistence has been reviewed in Refs. \cite{Heyde1983,Wood1992,Heyde2011,Garrett2022,Bonatsos2023b,Leoni2024}, with islands on the nuclear chart on which shape coexistence can occur determined microscopically on a symmetry basis in a parameter-free way in Refs. \cite{Martinou2021,Martinou2023}.
     
Global calculations for triaxial nuclei, i.e., nuclei violating axial symmetry, have been performed within the Finite-Range Liquid-Drop Model \cite{Moeller2006,Moeller2008}, while stripes within which triaxiality is favored have been microscopically predicted recently in a parameter-free way in Ref. \cite{Bonatsos2025}. In parallel, it has been suggested by computationally demanding Monte Carlo Shell Model (MCSM) calculations \cite{Otsuka2019,Tsunoda2021,Otsuka2023},
as well as by more accessible Triaxial Projected Shell Model (TPSM) calculations \cite{Rouoof2024}, that some amount of triaxiality is present in almost all nuclei across the nuclear chart.  

It is the purpose of the present work to review existing theoretical predictions for triaxiality in even-even nuclei and detect regions of the nuclear chart in which triaxiality is favored. In Sections \ref{BohrMottelson}-\ref{sig} the theoretical approaches used in the study of triaxiality are exposed, while in {Appendices}  
\ref{Z2426}-\ref{Z8898} a detailed account of existing theoretical work for each series of isotopes in the region with $Z=24$-98 is separately given. Finally, in Sections 
\ref{global}-\ref{conclusion} the findings of the present compilation are summarized and an outlook for further work is given.   

\section{Collective Model of Bohr and Mottelson} \label{BohrMottelson}

\subsection{The  Rigid Triaxial Rotor Model}  \label{RTRM}

In the collective model of Bohr and Mottelson \cite{Bohr1952,Bohr1953,Bohr1998a,Bohr1998b}, introduced in 1952 \cite{Bohr1952}, nuclear properties are described in terms of the collective variables $\beta$ and $\gamma$, describing the departure from sphericity and the departure from axial symmetry, respectively. For the latter, the values 
$\gamma=0$ and $\gamma=60^{\rm o}$ correspond to ellipsoids with prolate (rugby-ball like) and oblate (pancake like) axially symmetric shapes, while the intermediate values $0< \gamma < 60^{\rm o}$ correspond to triaxial shapes with three unequal semi-axes, with maximum triaxiality occurring at $\gamma=30^{\rm o}$.   

Soon thereafter two radically different special cases have been considered. On one hand, Wilets and Jean in  1956 studied the $\gamma$-unstable case \cite{Wilets1956}, in which the nuclear potential within the Bohr Hamiltonian depends only on $\beta$ and not on $\gamma$, allowing exact separation of variables in the relevant wave function as a by-product.  This limit has been found useful in the description of the spectra of several nuclei known at that time  \cite{ScharffGoldhaber1955,Rohozinski1974}, later called $\gamma$-soft or $\gamma$-unstable nuclei, in which the value of $\gamma$ can vary without any energy expense. 

On the other hand, Davydov and Filippov \cite{Davydov1958} in 1958 considered the case of rigid triaxiality, with the nuclear shape corresponding to an asymmetric top with fixed values of $\beta$ and $\gamma$, and studied spectra and transition probabilities within it \cite{Davydov1958,Davydov1959}. This model has been called the Rigid Triaxial Rotor Model (RTRM) \cite{Esser1997}. Soon thereafter the cases of a constant value of $\gamma$ within the Bohr Hamiltonian \cite{Davydov1960}, as well as of a $\gamma$ potential corresponding to small oscillations around a non-vanishing value $\gamma_0$ \cite{Davydov1961} have been considered. The analytical solution for the triaxial rotor with $\gamma=30^{\rm o}$ has been given by Meyer-ter-Vehn \cite{MeyerterVehn1975} in 1975.  

Several extensions of the RTRM have been introduced over the years, including the two-rotor model for the description of the scissors mode, formed by the separate  collective motion of protons and neutrons \cite{Filippov1998,Pan1999}, the presentation of the RTRM as multiple $Q$-excitations built on the ground state \cite{Jolos1996}, as well as the semiclassical versions of a cranked triaxial rotor \cite{Gheorghe1998} and a triaxial rotor model treated by a time-dependent variational principle \cite{Raduta2007}, while for odd-$A$ nuclei the particle-rotor model has been introduced \cite{MeyerterVehn1974,MeyerterVehn1975}. 

An interesting application of the RTRM regards the application of the analytical expressions of the RTRM for spectra \cite{Varshni1970,Yan1993, Esser1997,Bindra2018} and/or transition rates \cite{Yan1993,Esser1997} in order to extract values of $\gamma$ from the relevant data for several series of isotopes. The values extracted from the spectra and separately from the transition rates have been found to be consistent with each other and also different from zero, usually above $10^{\rm o}$, serving as an early sign for the existence of some degree of triaxiality over extended regions of the nuclear chart (see Tables I and II of \cite{Yan1993}, Table II of \cite{Esser1997}, Tables 1-3 of \cite{Bindra2018}) .

In the RTRM the three moments of inertia along the three axes are determined by the single parameter $\gamma$, representing an irrotational flow \cite{Rowe1970} within the nucleus. In 2004 this restriction has been lifted by Wood \textit{et al.} \cite{Wood2004}, which allowed for independent inertia and electric quadrupole tensors within the triaxial rotor model, thus obtaining the generalized triaxial rotor model (GTRM). The GTRM has been applied for the study of many nuclei \cite{Allmond2008,Allmond2009,Allmond2010,Allmond2017,Pittel2017,Sugawara2018,Sugawara2019} extracting the $\gamma$ values characterizing them (see, for example, Table 1 of \cite{Allmond2017}).    

An early alternative description of triaxial nuclei in the Xe-Ba and Th regions has been provided within the framework of the Triaxial Rotation Vibration Model (TRVM) \cite{Meyer1997,Meyer1998}, which is an extension of the Rotation Vibration Model (RVM), in which the deviations of the shape coordinates around their static values are considered \cite{Faessler1962a,Faessler1962b,Faessler1964a,Faessler1964b,Faessler1965}. Triaxiality is obtained by taking into account the interaction between rotations and vibrations \cite{Meyer1997,Meyer1998}.   

In recent years the interpretation of $\beta$ and $\gamma$ collective bands as quadrupole vibrations has been questioned
 \cite{Sharpey2008,Sharpey2010,Sharpey2011a,Sharpey2011b,Sharpey2019,Sharpey2023a,Sharpey2023b}, with $\beta$ bands attributed to two-particle--two hole 
(2p2h) excitations \cite{Sharpey2023a}, and $\gamma$ bands considered as a consequence of breaking of axial symmetry leading to triaxial shapes \cite{Sharpey2019,Sharpey2023b}. 

\subsection{The Algebraic Collective Model}\label{ACM}

Extensive numerical calculations over several regions of the nuclear chart \cite{Hess1980,Hess1981,Petkov1995} have been performed in the framework of the Generalized Collective Model (GCM), introduced by the Frankfurt group in 1971 \cite{Gneuss1971}, in the potential of which all independent terms up to sixth order in $\beta$ are taken into account. A comprehensive review of work done in the framework of the Bohr collective Hamiltonian has been given in Ref. \cite{Prochniak2009}.

A major step forward in numerical calculations related to the Bohr collective model has been taken through the introduction of the Algebraic Collective Model 
\cite{Rowe2004,Rowe2005,Rowe2009}, triggered by the realization \cite{Rowe1998} that the Davidson potential \cite{Davidson1932}, initially used for the description of rotation-vibration spectra of diatomic molecules, has a SU(1,1)$\times$SO(5) structure \cite{Szpikowski1980,Rowe2005b}, which can be applied to atomic nuclei, resulting in much faster convergence of the relevant numerical calculations, especially in the case of deformed nuclei, due to the optimal choice of the basis 
\cite{DeBaerdemacker2007,DeBaerdemacker2009} in which the diagonalization of the Hamiltonian is performed. 

The availability of a computer code for the ACM \cite{Welsh2016} greatly facilitated its application in the case of triaxial nuclei \cite{Caprio2009,Caprio2011,Thiamova2014,Thiamova2015}, in which the presence of a $\cos^2 3\gamma$ term has been found necessary.

\subsection{Shape/Phase Transitions and Critical Point Symmetries}\label{SPT}

The introduction by Iachello of the critical point symmetries (CPSs)  E(5) \cite{Iachello2000} in 2000  and X(5) \cite{Iachello2001} in 2001, related to shape-phase transitions (SPTs) \cite{Casten2006,Casten2007,Cejnar2010} (also called quantum phase transitions (QPTs) \cite{Iachello2006b}) from vibrational (near-spherical) nuclei to $\gamma$-unstable and to prolate axially deformed nuclei respectively, have raised much interest on special solutions of the Bohr Hamiltonian \cite{Fortunato2005,Buganu2016}, several of them related to triaxial shapes. 

A shape/phase transition in the $\gamma$ variable, leading from axial to triaxial nuclei, has been described by the critical point symmetry Y(5) \cite{Iachello2003,Caprio2009}, introduced by Iachello in 2003 \cite{Iachello2003}, with $^{164}$Dy and $^{166,168}$Er suggested as possible candidates for it, while the shape/phase transition from prolate to oblate shapes \cite{Jolie2003} has been described in terms of the Z(5) \cite{Bonatsos2004} critical point symmetry, in which the $\gamma$ variable is confined in a steep harmonic oscillator potential around the value of maximal triaxiality, $\gamma =30^{\rm o}$, with the transition taking place in $^{192-196}$Pt, suggested as a possible experimental manifestation. A $\gamma$-rigid version of Z(5), called Z(4), has also been introduced \cite{Bonatsos2005}, in which $\gamma$ is fixed at $30^{\rm o}$, with the transition taking place in $^{128-132}$Xe, suggested as a possible experimental manifestation. In both Z(5) and Z(4) an infinite square well potential is used in the $\beta$ variable, in analogy to E(5) and X(5). Exact separation of the $\beta$ and $\gamma$ variables is obtained in Z(4), while in Z(5) an approximate separation of variables is achieved by assuming potentials of the form $u(\beta,\gamma)=u_1 (\beta) + u_2(\gamma)$, in analogy to the X(5) case \cite{Iachello2001,Caprio2005b}. A critical point symmetry called T(5) has also been introduced \cite{Zhang2015a}, containing X(5) and Z(5) as special cases. In the T(5) CPS approximate separation of variables is achieved, with an infinite square well potential used in $\beta$ and a steep harmonic oscillator potential centered around a specific value $\gamma_e$ being employed in $\gamma$. Furthermore, a special solution of the Bohr Hamiltonian called T(4) has been developed \cite{Zhang2017}, which is a $\gamma$-rigid solution intermediate between Z(4) and X(4) \cite{Budaca2016}, the latter being a solution intermediate between X(5) and X(3) \cite{Bonatsos2006}, where X(3) is a $\gamma$-rigid version of X(5).  

\subsection{Special Solutions of the Bohr Hamiltonian} \label{specialB}

Several special solutions similar to Z(5) have been suggested, using in $\gamma$ the same steep harmonic oscillator centered around $30^{\rm o}$ used in Z(5), but replacing in $\beta$ the infinite square well potential by  the Davidson potential \cite{Yigitoglu2011}, a hyperbolic P\"oschl-Teller potential \cite{Naderi2017}, 
an inverse square potential \cite{Ajulo2022}, or by a finite well potential \cite{Inci2012}. 
In addition, similar solutions achieving exact separation of variables by using potentials of the form $u(\beta,\gamma)=u_1 (\beta) + u_2(\gamma)/\beta^2$ have been considered, by using  in $\gamma$ a steep harmonic oscillator potential around $\gamma =30^{\rm o}$ and in $\beta$ the Coulomb potential \cite{Fortunato2004,Fortunato2006}, the  Kratzer potential \cite{Fortunato2004,Fortunato2005b,Fortunato2006}, the Davidson potential \cite{Fortunato2006,Lee2013}, the infinite square well potential \cite{Fortunato2006},  the Morse potential \cite{Inci2014}, the Hulth\'{e}n potential \cite{Chabab2015}, or the Killingbeck potential \cite{Neyazi2016}. 
Furthermore, exact separation of variables has been used with a sextic potential in $\beta$ and a Mathieu potential in $\gamma$  \cite{Raduta2011,Buganu2012,Buganu2012b,Raduta2013,Raduta2014}, while a generalized Gneuss-Greiner potential has been used in Ref. \cite{Alimohammadi2019}.   

In addition, several exactly separable special solutions similar to Z(4) \cite{Bonatsos2005} have been suggested, in which the infinite square well potential in the variable $\beta$ is replaced by the Davidson potential (Z(4)-D, \cite{Yigitoglu2018}), or by a sextic potential (Z(4)-sextic, \cite{Buganu2015,Buganu2015b,Buganu2015c,Budaca2016b}). A special solution of the Bohr Hamiltonian providing a bridge between Z(4) \cite{Bonatsos2005} and X(4) \cite{Budaca2016} has also been developed \cite{Ajulo2024}. 

In most of the above mentioned special solutions, a harmonic oscillator potential centered around $\gamma =30^{\rm o}$ is used. A much sharper potential well, centered
around  $\gamma =30^{\rm o}$, is the potential $1/\sin^2(3\gamma)$, the former potential corresponding to the second order Taylor expansion of the latter \cite{DeBaerdemacker2006}.   

\subsection{Modifications of the Bohr Hamiltonian} \label{modifiedB}

A well known problem of the Bohr collective model \cite{Bohr1952,Bohr1953} is that the nuclear moments of inertia are predicted to increase proportionally to $\beta^2$, while experimental evidence indicates that the increase should be much slower, especially for deformed nuclei \cite{Ring1980}. One way to moderate the increase of the moment of inertia with increasing deformation $\beta$ is to allow the mass to depend on the deformation. The general formalism for quantum systems with mass depending on the coordinates has already been developed \cite{Quesne2004,Bagchi2005,Quesne2007}, with analytical solutions obtainable for several physical systems through the techniques of Supersymmetric Quantum Mechanics (SUSYQM) \cite{Cooper1995,Cooper2001}. Deformation-dependent masses (DDM) have been introduced in the Bohr Hamiltonian with a Davidson potential \cite{Bonatsos2010,Bonatsos2011}, and a Kratzer potential \cite{Bonatsos2013}. Special solutions regarding triaxial nuclei using the DDM concept have been developed for the analog of the Z(4) solution using the infinite square well potential (Z(4)-DDM, \cite{AitElKorchi2020}), the Davidson potential (Z(4)-DDM-D, \cite{Buganu2017}) and the Kratzer potential (Z(4)-DDM-K, \cite{AitElKorchi2021,AitElKorchi2022}). A special solution using the DDM concept has also been developed for the X(3) solution (X(3)-DDM, \cite{AitElKorchi2020}). 

Another attempt of modifying the moments of inertia in the Bohr Hamiltonian has been made by using the concept of minimal length \cite{Chabab2016,Chabab2018}, which comes from non-commutative geometry \cite{Witten1986,Seiberg1999} leading to a generalized uncertainty principle \cite{Kempf1994,Kempf1997} involving deformed canonical commutation relations. This concept has been applied in many fields, including string theory \cite{Gross1988} and quantum gravity \cite{Mead1964}. In relation to triaxial nuclei, the Z(4) special solution of the Bohr Hamiltonian has been modified by including the minimal length (Z(4)-ML, \cite{AitElKorchi2020}). A similar modification of the X(3) solution has also been worked out (X(3)-ML, \cite{AitElKorchi2020}).  

It should be remarked that the connections between three different forms of unconventional Schr\"{o}dinger equations involving position-dependent mass, or deformed canonical commutation relations, or a curved space, have been clarified in Ref. \cite{Quesne2004}. Therefore the DDM and ML methods described in the last two paragraphs 
are not unrelated \cite{AitElKorchi2023}. 

Another family of solutions of the Bohr Hamiltonian can be obtained by considering energy-dependent potentials, already used in other fields of physics \cite{Formanek2004,Lombard2004}. Solutions of the Bohr Hamiltonian involving a coupling constant linearly dependent on the energy have been provided for the harmonic oscillator \cite{Budaca2015,Budaca2017,Budaca2019} and Coulomb \cite{Budaca2016c} potentials. In relation to triaxiality, a solution for the Kratzer potential possessing a coupling constant linearly dependent on the energy has been provided \cite{Budaca2020}.   

A novel approach to SPTs has been introduced by Hammad in 2021 \cite{Hammad2021a,Hammad2021b}, making use of the fractional calculus \cite{Miller1993,Podlubny1999,Herrmann2011,Khalil2014}. Using fractional derivatives \cite{Miller1993,Podlubny1999,Herrmann2011} in the Bohr equation, the fractional order of the derivative can be used  in order to approach the critical point closer than what is allowed by derivatives of integer order, something useful since the neutron number, which is used as  the control parameter in a series of even-even isotopes, is increasing by steps of 2. In addition, the use of conformable fractional derivatives \cite{Khalil2014} guarantees that the usual rules valid for derivatives in the usual calculus, continue to be valid in the case of the fractional calculus as well. Analogues of the Z(5) CPS using conformable fractional derivatives have been considered for the Kratzer potential \cite{Hammad2021c}, as well as for the Morse \cite{Hammad2023}, Tietz-Hua \cite{Hammad2023}, and various multi-parameter exponential-type potentials \cite{Hammad2023} and a four inverse power terms potential \cite{Ahmadou2022}.     
 
 \section{The Nuclear Shell Model} \label{SM}

The nuclear spherical shell model \cite{Mayer1955,Heyde1990,Talmi1993} has been introduced in 1949 \cite{Mayer1949,Haxel1949,Mayer1950} in order to explain the extreme stability observed at the nuclear magic numbers 2, 8, 20, 28, 50, 82, 126, \dots \cite{Mayer1948}. It is based on a three-dimensional isotropic harmonic oscillator (3D-HO), to which a spin-orbit interaction \cite{Mayer1950} is added, resulting in the modification of the 3D-HO magic numbers 2, 8, 20, 40, 70, 112, 168, \dots \cite{Wybourne1974,Moshinsky1996,Iachello2006} beyond the $sd$ shell. 

While giving a satisfactory description of nuclear properties in the vicinity of the magic numbers, where nuclei of nearly-spherical shape are expected to occur, the spherical shell model fails to account for the large quadrupole moments observed in nuclei away from closed shells. This has been remedied through the introduction of the deformed shell model by Nilsson \cite{Nilsson1955,Nilsson1995}, in which nuclei are allowed to obtain axially symmetric shapes. The fact that spheroidal shapes offer greater stability than spherical shapes, being energetically preferable, has been pointed out by Rainwater in 1950 \cite{Rainwater1950}. 

Based on the spheroidal shapes introduced by Nilsson, the macroscopic-microscopic Finite-Range Droplet Model (FRDM) and Finite-Range Liquid-Drop Model (FRLDM) have been developed \cite{Moeller1981a,Moeller1981b,Moeller1995,Moeller2006,Moeller2008,Moeller2016}, which allow for the determination of the ground-state masses and deformation parameters of a large number of nuclei.    

The concept of cranking of a spheroidal collective field about a fixed axis has been introduced by Inglis in 1954 \cite{Inglis1954,Inglis1956} and used in several early shell model calculations (see, for example, \cite{Shen2011}), in some of which the 3D-HO potential has been  replaced by a Woods-Saxon potential \cite{Woods1954}. An authoritative review of the development of the nuclear shell model can be found in Ref. \cite{Caurier2005}. The main difficulty associated with nuclear shell model calculations is the large size of the model space, which makes the computational needs extremely large.   

A computationally tractable approach which has been widely used for studying triaxiality in several nuclei is the Triaxial Projected Shell Model (TPSM) \cite{Sheikh1999,Sheikh2016}, in which a triaxial Nilsson + BCS basis is used. The TPSM is a generalization of the axially symmetric Projected Shell Model (PSM) approach \cite{Hara1995}. In PSM (TPSM) a Nilsson potential with axial (triaxial) deformation is used in order to generate the deformed single-particle states, while pairing correlations are treated by the usual BCS approximation. A three-dimensional angular momentum projection \cite{Ring1980} is then carried out on the Nilsson+BCS quasiparticle states in order to obtain the many-body basis with states having good angular momentum, and subsequently the Hamiltonian is diagonalized in the projected basis. TPSM calculations have been carried out for Ge \cite{Bhat2014}, Se \cite{Bhat2014}, Mo \cite{Zhang2015b}, Ru \cite{Zhang2015b,Nazir2023}, Ce \cite{Sheikh2009}, Nd \cite{Sheikh2009}, Gd \cite{Rajput2022}, Dy \cite{Sheikh2001,Sun2002} and Er \cite{Sun2000,Sheikh2001,Sheikh2008} series of isotopes, while detailed results for 30 triaxial nuclei ranging from Ge to U have been summarized in Refs. \cite{Jehangir2021,Rouoof2024} (see Table 3 of \cite{Rouoof2024}). 

Large-scale shell model calculations have become feasible by the introduction  of the Quantum Monte Carlo Diagonalization (QMCD) method \cite{Honma1995,Mizusaki1996,Honma1996,Otsuka1998}, reviewed in \cite{Otsuka2001,Shimizu2012}. Recent Monte Carlo Shell Model (MCSM) calculations have suggested rigid triaxiality in $^{166}$Er \cite{Tsunoda2021}, taking advantage of self-organization in quantum systems \cite{Otsuka2019}. Prevalence of triaxial shapes in heavy nuclei has been recently suggested \cite{Otsuka2023}.  

\subsection{The SU(3) Symmetry} \label{SU3}

The shells of the 3D-HO used in the original spherical shell model are known to possess the symmetry U($(N+1)(N+2)/2$) \cite{Wybourne1974,Moshinsky1996,Iachello2006}, where $N$ is the number of oscillator quanta. In each shell an SU(3) subalgebra exists \cite{Bonatsos1986}. 

Elliott in 1958, working in the $sd$ ($N=2$) shell with U(6) symmetry, managed  \cite{Elliott1958a,Elliott1958b,Elliott1963,Elliott1968,Harvey1968} to build a bridge between the spherical shell model and the deformation described by the collective model. Initially \cite{Elliott1958a} he classified the shell model states using the 
U(6)$\supset$SU(3)$\supset$SO(3) chain of subalgebras, in which nuclear bands occur within the irreducible representations (irreps) of SU(3), which are characterized by the Elliott quantum numbers $(\lambda,\mu)$. Nuclear states are labeled by the angular momentum $L$ (which characterizes the irreps of the angular momentum algebra SO(3)) and $K$, the ``missing quantum number'' \cite{Iachello1987,Iachello2006} in the decomposition from SU(3) to SO(3). The quantum number $K$ divides the states into rotational bands with increasing values of $L$ appearing in each of them.  Subsequently \cite{Elliott1958b} he proved that all states belonging to a given band come from the same intrinsic state, obtaining simple expressions for the quadrupole moments, resembling those of a rotational model with permanent deformation. In different words, Elliott proved the existence of deformed states corresponding to quadrupole deformation within the spherical shell model. SU(3) has played since then a major role in nuclear structure, as summarized in the recent book by Kota \cite{Kota2020}.      

It should be noticed at this point that a mapping of the invariants of SU(3) onto the invariants of the collective model provides a connection between the Elliott labels $\lambda$, $\mu$ and the collective variables $\beta$, $\gamma$, which read \cite{Castanos1988,Draayer1989}
\begin{equation} \label{mu}
\gamma = \arctan \left( {\sqrt{3} (\mu+1) \over 2\lambda+\mu+3}  \right),
\end{equation}
 \begin{equation} 
	\beta^2  \propto (\lambda^2+\lambda \mu + \mu^2+ 3\lambda +3 \mu +3) \propto  (C_2(\lambda,\mu)+2), 
	\end{equation}
where $C_2$ is the second order Casimir operator of SU(3) \cite{Iachello1987,Iachello2006}
 \begin{equation}\label{C2} 
  C_2(\lambda,\mu) = {2\over 3} (\lambda^2+\lambda \mu + \mu^2+ 3\lambda +3 \mu).
	\end{equation}

The SU(3) symmetry is broken beyond the $sd$ shell by the spin-orbit interaction, which lowers in each shell the orbital with the highest eigenvalue of the total angular momentum $j$ into the shell below. As a result, each shell, beyond the $sd$ one, ends up consisting of its remaining original orbitals, minus the defecting orbital which went to the shell below, plus the invading orbital (called the intruder orbital) which came from the shell above.

\subsection{The Pseudo-SU(3) Symmetry}\label{pseudo}

Approximate restoration of the SU(3) symmetry can be achieved in different ways. In the pseudo-SU(3) scheme \cite{Arima1969,Hecht1969,RatnaRaju1973,Draayer1982,Draayer1983,Draayer1984,Bahri1992,Ginocchio1997}, introduced in 1973 \cite{RatnaRaju1973}, a unitary transformation \cite{Castanos1992a,Castanos1992b,Castanos1994} is used in order to map the remaining original orbitals of each shell onto the full set of the  orbitals of the shell below it, thus restoring the SU(3) symmetry for the remaining orbitals, which now live in the same shell with the intruder orbitals next to them. The intruder orbitals are not included in the recovered SU(3) symmetry and have to be taken into account separately, through shell model techniques \cite{Draayer1983,Draayer1984}.     
  
\subsection{The Proxy-SU(3) Symmetry}   \label{proxy}
  
A different approximate restoration of the SU(3) symmetry is the proxy-SU(3) symmetry \cite{Bonatsos2017a,Bonatsos2017b,Bonatsos2023},  introduced in 2017 \cite{Bonatsos2017a,Bonatsos2017b}. In the proxy-SU(3) scheme, a unitary transformation \cite{Martinou2020} is used for the intruder levels, except the sublevel (accommodating two nucleons) which possesses the highest eigenvalue of the projection $j_m$ of the total angular momentum $j$. This unitary transformation  \cite{Martinou2020} maps the rest of the intruder levels onto the defecting orbitals, which had gone into the shell below. As a result, SU(3) is restored in the shell under study, in which the orbitals obeying the SU(3) symmetry now live with the isolated orbital with the highest $j_m$, which, however, can in most cases be neglected, since from the standard Nilsson diagrams \cite{Lederer1978} one can see that it lies at the top of the shell, thus it will be empty for most nuclei to be considered within this shell. 

An important finding realized in the proxy-SU(3) framework is that the most important irreps, which lie lowest in energy and therefore accommodate the ground state band, also called most leading irreps in earlier literature \cite{Draayer1983,Draayer1984}, are \textit{not} the ones possessing the highest eigenvalue of the second order Casimir of SU(3), given by Eq. (\ref{C2}), but the highest weight (hw) irreps \cite{Martinou2021b}, which are the most symmetric irreps allowed by the Pauli principle and the short-range nature of the nucleon-nucleon interaction. Although the two sets of irreps are identical in the lower half of the shell, they become different beyond midshell (see, for example, Table I of Ref. \cite{Bonatsos2017b}). Although this fact has been known since the early days of the use of SU(3) in nuclear physics,  as seen for example in Table 5 of Ref. \cite{RatnaRaju1973} for the $pf$ shell, its consequences have been rather ignored. The dominance of the highest weight irreps offers an explanation for the dominance of prolate over oblate shapes in the ground states of even-even nuclei, a long-standing problem considered unresolved until recently \cite{Hamamoto2009,Hamamoto2012}, it predicts a prolate to oblate transition in the heavy rare earths around $N=114$ \cite{Bonatsos2017b}, and paves the way for the determination of certain regions on the nuclear chart in which islands of shape coexistence can occur \cite{Martinou2021,Martinou2023,Bonatsos2023b}, as corroborated through covariant density functional theory calculations \cite{Bonatsos2022a,Bonatsos2022b} and experimental evidence \cite{Bonatsos2023a,Bonatsos2023b}.  

In addition, the dominance of the highest weight irreps predicts non-negligible values of the shape variable $\gamma$ over extended regions of the nuclear chart. 
This can be seen, for example, in Tables II and III of Ref. \cite{Bonatsos2017b}, in which it is clear that most nuclei possess $\mu$ values different from zero, resulting in non-zero values of $\gamma$ through Eq. (\ref{mu}). These predictions are in good agreement with $\gamma$ values obtained from the data, as seen in Figs. 5, 6 of Ref. \cite{Bonatsos2017b}. A detailed comparison of the proxy-SU(3) predictions for the collective variable $\gamma$ and its empirical values extracted from the experimental data has been given recently in Ref. \cite{Bonatsos2025}. 

It should be emphasized that these results are obtained in a completely parameter-free way, based only on the consequences of the Pauli principle and the short-range nature of the nucleon-nucleon interaction. The basic property leading to these results is the fact that the net result of the restrictions imposed by the Pauli principle on 5 (or more) nucleons (protons or neutrons) in a given shell, which can accommodate $N$ nucleons, is not the same as the net result of the restrictions imposed by the Pauli principle on $N-5$  (or fewer) nucleons in the same shell, as seen, for example, in Table I of Ref. \cite{Bonatsos2017b}. 

It should be noticed that the use of the highest weight irreps within the pseudo-SU(3) scheme leads to predictions \cite{Bonatsos2020a} compatible with these of the proxy-SU(3) approach. It is remarkable that two considerably different approximation schemes, involving different fractions of the valence nucleons, lead to very similar conclusions. This might be understood on the basis that both schemes are based on unitary transformations of some of the orbitals onto their pseudo- or proxy- counterparts.  
 
It has been recently realized that the next highest weight (nhw) irreps within the proxy-SU(3) symmetry might have important contributions in some special cases \cite{Bonatsos2025}. Extensive results for the nhw irreps in several regions of the nuclear chart have been given in Ref. \cite{Bonatsos2024}, in order to be used in future irrep-mixing calculations.  
 
\subsection{Regions of High Triaxiality Predicted by the Proxy-SU(3) Symmetry} \label{regi}

 The proxy-SU(3) symmetry predicts non-vanishing values of the collective variable $\gamma$ almost everywhere across the nuclear chart \cite{Bonatsos2017b,Bonatsos2025}. 
 However, if one is restricted to $\gamma$ values between $15^{\rm o}$ and  $45^{\rm o}$, i.e., values close to the value $\gamma=30^{\rm o}$ which corresponds to maximum triaxiality, one sees that these occur within certain horizontal and vertical stripes on the nuclear chart, lying within the proton or neutron numbers 22-26, 34-48, 74-80, 116-124, 172-182 \cite{Bonatsos2025}. Empirical information \cite{Bonatsos2025} supports the interpretation that strong triaxiality should appear preferably within these stripes.
In sections \ref{Z2426}-\ref{Z8898} it will be tested to what extent these predictions are in agreement with theoretical work on triaxiality accumulated over the years.  
 
\section{Algebraic Models Using Bosons} \label{IBM}

\subsection{Interacting Boson Model-1} \label{IBM1}

In the Interacting Boson Model (IBM), introduced by Arima and Iachello in 1975 \cite{Arima1975,Iachello1987,Iachello1991,Casten1993,Frank2005}, collective nuclear properties are described in terms of bosons, which correspond to correlated proton pairs and neutron pairs, counted from the nearest closed shells and bearing angular momentum 0 ($s$-bosons) and 2 ($d$-bosons). The model has an overall U(6) symmetry, possessing three different dynamical symmetries: U(5) \cite{Arima1976}, corresponding to vibrational (near-spherical) nuclei, SU(3) \cite{Arima1978}, describing axially  symmetric deformed nuclei, and O(6) \cite{Arima1979}, appropriate for $\gamma$-unstable nuclei, which are soft towards triaxial deformation of their shape. Schematically the three dynamical symmetries are placed at the vertices of the symmetry triangle of IBM-1 \cite{Casten1990}, also called the Casten triangle.  

In the simplest version of the model, called IBM-1, no distinction is made between bosons coming from proton pairs or neutron pairs. When this distinction is made, IBM-2 is obtained. In the standard form of IBM-1 and IBM-2, only one-body and two-body terms are taken into account in the Hamiltonian. However, higher-order terms exist and can be included if necessary. 

The classical limit of IBM can be obtained through the use of coherent states \cite{Ginocchio1980a,Dieperink1980,Ginocchio1980b}. In the classical limit, a potential energy surface (PES) corresponding to the IBM Hamiltonian is obtained, with its shape depending on the values of the free parameters appearing in the IBM Hamiltonian. 
Studying the classical limit of IBM-1, Van Isacker and Chen in 1981 \cite{VanIsacker1981} proved that no triaxial shapes can be obtained within the standard IBM-1, in which only one-body and two-body terms appear in the Hamiltonian, while the addition of three-body (cubic) terms do give rise to triaxial shapes, with $^{104}$Ru suggested as an experimental example \cite{Heyde1984}. 

The relation between the O(6) dynamical symmetry and triaxiality has been clarified already in 1979 \cite{MeyerterVehn1979a,MeyerterVehn1979b}. It has been shown that the O(6) limit of IBM-1 corresponds to the $\gamma$-unstable model of Wilets and Jean \cite{Wilets1956}, if an infinite number of bosons is considered. While the RTRM with $\gamma=30^{\rm o}$ satisfies the same selection rules, the predictions of the two models for spectra and $B(E2)$ transition rates are considerably different \cite{MeyerterVehn1979a,MeyerterVehn1979b}. Experimental manifestations of the O(6) symmetry have been found in the Pt \cite{Cizewski1978,Casten1978} and Xe-Ba \cite{Casten1985a} regions. It has also been found that deviations from the O(6) symmetry in both regions can be accounted for easily by adding cubic terms to the IBM-1 Hamiltonian \cite{Casten1985b}. Looking the other way around, one concludes that a small degree of triaxiality is present in the best experimental manifestations of the O(6) dynamical symmetry. 

It has been further proved \cite{Otsuka1987,Sugita1989} within the O(6) limit of IBM-1 that the $\gamma$-unstable state can be generated from an intrinsic state with rigid triaxial deformation with $\gamma=30^{\rm o}$, thus demonstrating the formal equivalence between $\gamma$-instability and rigid triaxiality, especially in the ground state bands  \cite{Cohen1988}. Despite this formal equivalence, detailed investigations have shown that most nuclei possess  a $\gamma$-soft nature, as opposed to stable, triaxial ground-state shapes \cite{Gill1992,Sorgunlu2008}.   

The quadrupole operator in IBM-1 reads \cite{Iachello1987}
\begin{equation} \label{Qop}
Q_q = (d^\dagger s + s^\dagger \tilde d)^{(2)}_q + \chi (d^\dagger \tilde d)^{(2)}_q, 
\end{equation}
where $s^\dagger$ ($s$) and $d^\dagger $ ($\tilde d)$ are the creation (annihilation) operators for the $L=0$ and $L=2$ bosons respectively, while $\chi$ is a parameter which obtains the values $-\sqrt{7}/2$ and zero  in the SU(3) and O(6) limits respectively. It has been proved that the value of the collective variable $\gamma$ is connected to the choice of $\chi$ \cite{Elliott1986}. In more detail, the values of the collective variables $\beta$ and $\gamma$ can be estimated from the quadrupole shape invariants $QQ$, $QQQ$, $QQQQ$ introduced by Kumar in 1972 \cite{Kumar1972} and widely used since then \cite{Cline1986,Werner2001,Werner2005}. Recent studies \cite{Poves2020} of the fluctuations of $\beta$ and $\gamma$, estimated through calculations of the quadrupole shape invariants within  the configuration-interaction shell model, indicate that there is a certain degree of softness for $\beta$, while for $\gamma$ the fluctuations can become large, rendering the effective value of 
$\gamma$ not meaningful \cite{Poves2020}.     

A mapping between the triaxial rotor and the SU(3) limit of IBM-1 has been established \cite{Smirnov2000,Zhang2014,Zhang2022,Pan2024,Zhang2024}, in which the inclusion of higher-order terms of the type $LQL$ and $LQQL$ (where $L$ is the angular momentum operator) in the IBM Hamiltonian is required. A similar mapping regarding the O(6) limit of IBM-1 has been recently established \cite{Teng2024}. We conclude that in both the O(6) and SU(3) limits of IBM-1, higher order terms are required in order to produce rigid triaxiality. This has to be distinguished from the effective triaxiality stemming from the softness of the relevant classical potential seen in the framework of the standard IBM-1 \cite{Casten1984,Castanos1984}, for which the decrease of the effective $\gamma$ value with $\chi$ moving from the O(6) value $\chi=0$ to the SU(3) value $\chi=-\sqrt{7}/2$ has been established \cite{Casten1984}. The same dependence of $\gamma$ on $\chi$ has been corroborated by Hartree calculations with angular momentum projection performed before variation \cite{Dobes1985}.  

The need for cubic terms in IBM-1 for the production of triaxiality has also been demonstrated through studies within the Algebraic Collective Model (ACM) \cite{Rowe2004,Rowe2005,Rowe2009} (see subsec. \ref{ACM}), which is an algebraic version of the Bohr collective model allowing for rapidly converging calculations \cite{Caprio2009b,Welsh2016}. It has been seen that the occurrence of triaxiality within the ACM requires the inclusion of terms proportional to $\cos^2 (3\gamma)$, which in IBM-1 correspond to terms quadratic in  $QQQ$ \cite{Thiamova2010,Thiamova2012}.  

The study of shape/phase transitions from axial to triaxial shapes within IBM-1 has demonstrated the need for a three-dimensional parameter space \cite{Jolos2004a,Jolos2004b,Fortunato2011}, which can be obtained only with the inclusion of higher-order terms, as opposed to the two-dimensional parameter space sufficing in the case of IBM-1, called the Casten triangle \cite{Casten1990}. 

\subsection{Interacting Boson Model-2}\label{IBM2}

In the Interacting Boson Model-2 (IBM-2) distinction is made between bosons corresponding to valence proton pairs and to valence neutron pairs, counted from the nearest closed shells. Triaxial shapes occur in the SU(3) dynamical symmetry, when valence protons are holes and valence neutrons are particles, or vice versa \cite{Dieperink1982,Dieperink1983,Iachello1984,Dieperink1984,Dieperink1985,Walet1987}. 

In the SU(3) dynamical symmetry the parameter $\chi$ in the quadrupole operator of Eq. (\ref{Qop}) obtains the value $\chi=-\sqrt{7}/2$, which makes the $Q$ operator a generator of SU(3). However, the operator $Q$ with parameter value $\chi=+\sqrt{7}/2$ is also a generator of SU(3), called $\overline{\rm SU(3)}$ for distinction. SU(3) represents axially symmetric prolate (rugby-ball like) shapes corresponding in the classical limit to $\gamma=0$, while $\overline{\rm SU(3)}$ represents axially symmetric oblate (pancake like) shapes corresponding in the classical limit to $\gamma=60^{\rm o}$. Within IBM-2, SU(3) has been used for valence protons or neutrons outside closed shells and up to midshell, while $\overline{\rm SU(3)}$ has been used for valence protons or neutrons beyond midshell, which are counted from the closed shell above, i.e., they represent holes in their own shell. For $N$ bosons corresponding to valence particles the corresponding SU(3) irrep is $(2N,0)$, while for $N$ bosons corresponding to valence holes the corresponding SU(3) irrep is $(0,2N)$. Thus in a nucleus with $N_p$ bosons corresponding to proton holes and $N_n$ bosons corresponding to neutron particles the relevant irreps are $(0,2N_p)$ and $(2N_n,0)$ respectively, coupled into the most stretched irrep  $(2N_n,2N_p)$, which corresponds to a triaxial shape, as one can see from Eq. (\ref{mu}). The symmetry  which results from the coupling of SU(3) and $\overline{\rm SU(3)}$ has been called SU(3)$^*$ \cite{Dieperink1982,Dieperink1983,Dieperink1983b,Iachello1984,Dieperink1984,Dieperink1985,Walet1987}. $^{104}_{44}$Ru$_{60}$ \cite{Dieperink1982,Dieperink1983,Dieperink1984,Dieperink1985} and $^{192}_{76}$Os$_{116}$ \cite{Walet1987} have been suggested as experimental manifestations of the SU(3)$^*$ symmetry. 

The presence of SU(3)$^*$ in IBM-2 has as a consequence that the symmetry triangle of IBM-1 \cite{Casten1990} is becoming a tetrahedron, with U(5), O(6), SU(3) and SU(3)$^*$  at its vertices \cite{Arias2004a,Arias2004b,Caprio2004a,Caprio2004b}. Shape/phase transitions within the parameter space of IBM-2 have been considered in detail in Refs. \cite{Sevrin1987a,Sevrin1987b,Caprio2005}, while triaxial shapes have also been considered in Refs. \cite{Leviatan1990,Ginocchio1992}. Triaxiality in 
$^{108-112}$Ru has been considered within IBM-2 in Ref. \cite{Duarte1998}.   

The assumptions made in relation to the SU(3) irreps used in SU(3)$^*$ are supported microscopically by the findings of the proxy-SU(3) scheme, described in subsec. \ref{proxy}. In Table I of Ref. \cite{Bonatsos2017b} one sees that in proxy-SU(3) prolate ($\lambda > \mu$) irreps occur in the lower half of each shell, while oblate ($\lambda < \mu$) irreps occur in the upper halves of shells. The main difference is that in IBM the transition from prolate to oblate irreps occurs exactly at mid-shell, while in proxy-SU(3) it occurs further within the upper half of the shell.  This kind of asymmetry has also been seen recently within an IBM model including higher order interactions \cite{Wang2023,Wang2024}, called SU3-IBM for simplicity. The rich structure of the SU3-IBM model is expected to be the subject of intense investigations in the near future. 

\subsection{The Interacting Vector Boson Model}\label{IVBM}
 
Within the Interacting Vector Boson Model (IVBM) \cite{Georgieva1982,Georgieva1983,Georgieva2009} collective nuclear properties are described in terms of two vector bosons of angular momentum one, which can be considered as elementary excitations, using the Bargmann-Moshinsky basis \cite{Bargmann1960,Bargmann1961,Roussev1990}. Its overall symmetry is Sp(12,R), which is a noncompact algebra, thus in general the number of bosons is not conserved. However, it does contain a U(6) subalgebra, within which the boson number is conserved \cite{Georgieva2009}. Within the IVBM an SU(3)$^*$ symmetry has been found \cite{Ganev2011}, analogous to the one existing in IBM-2, and triaxial shapes, as well as shape/phase transitions involving triaxial shapes, have been considered within it \cite{Ganev2011}, with $^{192}$Os used as an example of experimental manifestation of this symmetry. 

Furthermore, a Vector Boson Model with the  SU(3) symmetry broken by the presence of a pairing interaction in the nuclear Hamiltonian has been introduced \cite{Minkov1997}, in which the odd-even staggering within the $\gamma$-bands, which is a hallmark of triaxial behavior, as discussed in Section \ref{stagg}, can be satisfactorily described in terms of mixing of the $\gamma$-band with the ground state band \cite{Minkov1999,Minkov2000}. In these considerations, the SU(3) irrep accommodating the ground state band and the $\gamma$ band (therefore having $\mu \geq 2$) has been treated as a free parameter. It would be interesting to examine what the consequences will be of using \cite{Minkov2024} a microscopically derived SU(3) irrep, as the one occurring in the proxy-SU(3) scheme \cite{Bonatsos2017b,Bonatsos2023} within the VBM scheme. 

\subsection{The Coherent State Model}\label{CSM}

The Coherent State Model \cite{Raduta1981,Raduta1982,Raduta2015} is a boson model providing a description of the ground, $\gamma$ and $\beta$ bands based on the coherent state formalism. It has been used for the description of triaxial nuclei in Refs. \cite{Raduta2011,Buganu2012,Buganu2012b,Raduta2013,Raduta2014}.  
 
\section{Self-Consistent Mean-Field Methods}\label{meanfield}

In the microscopic nuclear structure calculations based on the shell model, the nucleons are assumed to live in a specific single-particle potential, which represents the nuclear mean field. Then, configuration mixing is performed, involving all possible states in the realm of its application, called the shell model space. In the self-consistent mean-field methods, the opposite approach is used, namely the nuclear mean-field is determined though a self-consistent calculation, fitting it to the nuclear structure data \cite{Bender2003}. The simplest variational method for the determination of the wave functions of a many-body system is the Hartree--Fock (HF) method \cite{Ring1980,Greiner1996}, first used in atomic and molecular physics. In the case of nuclei, the pairing interaction has to be included, leading to the Hartree--Fock--Bogoliubov (HFB) approach \cite{Ring1980,Greiner1996}. The inclusion of pairing is facilitated by the BCS approximation, which allows the formation of pairs of degenerate states connected through time reversal, as in the BCS theory of superconductivity in solids \cite{Bardeen1957}.  Three different choices have been made for the effective nucleon-nucleon interaction: the Gogny interaction 
\cite{Gogny1973,Gogny1975,Delaroche2010} and  the Skyrme interaction \cite{Skyrme1956,Skyrme1959,Kluepfel2009,Erler2011,Dobaczewski2021,Chen2022}, used in a non-relativistic framework, and the relativistic mean field (RMF) models \cite{Reinhard1989,Ring1996,Lalazissis1997,Ring1997,Lalazissis2005,Vretenar2005,Niksic2006a,Niksic2006b,Tian2009a,Tian2009b,Niksic2009}, 
the latter having the advantage of providing a natural explanation of the spin-orbit force. A compilation of various parametrizations of the effective interactions in use can be found in Table I  of the review article \cite{Bender2003}. Extensive codes have been published, facilitating the use of the Skyrme forces \cite{Maruhn2014,Reinhard2021} and the RMF \cite{Ring1997,Niksic2014} all over the nuclear chart.

The self-consistent mean field approach used in nuclear structure \cite{Dobaczewski2011,Dobaczewski2012} strongly resembles the density functional theory used for the description of many-electron systems \cite{Hohenberg1964,Kohn1965}. The main difference is that in electronic systems one can derive very accurately the energy density functionals by {\it ab initio} methods from the theory of electron gas, while in nuclear structure the effective energy density functionals are motivated by {\it ab initio} theory \cite{Drut2010}, but they contain free parameters fitted to the nuclear structure data. 
                                                                                            
In early theoretical work on triaxial nuclei, modified harmonic oscillators \cite{Aberg1985,Ragnarsson1989} and generalized forms \cite{Beuschel1997} of the Nilsson model have been used. In addition, an early microscopic-macroscopic approach for the calculation of nuclear potential energy surfaces is the Nilsson-Strutinsky model.
In  phenomenological models, like the liquid drop model of vibrations and rotations (see Appendix 6A of \cite{Bohr1998b}), the distribution of nucleons in phase space is supposed to be homogeneous, which is not the case in the nuclear shell model, in which the distribution becomes inhomogeneous. Strutinsky's method 
(\cite{Strutinsky1967}, see also sec. 2.9 of \cite{Ring1980}) allows for calculations of the shell-model corrections to the liquid drop energy of the nucleus, starting from a modified Nilsson's level scheme \cite{Nilsson1955,Ragnarsson1978,Nilsson1995}. The corrections  depend on the occupation number and on the deformation.  The Nilsson-Strutinsky model has been used for the description of triaxial nuclei for example in \cite{Chasman1986,Luo1993,Juodagalvis2000,Nayak2022}.

The self-consistent mean field approaches determine the ground state properties of nuclei. A method allowing the calculation of excitation spectra and B(E2) transition rates has been introduced by Otsuka and Nomura \cite{Nomura2008,Nomura2010,Nomura2011b,Nomura2011c,Nomura2011aa,Nomura2011d,Nomura2012a}. The parameters of the Interacting Boson Model (IBM) \cite{Arima1975,Iachello1987,Iachello1991,Casten1993,Frank2005} are determined by mapping the total energy surface obtained from the relativistic mean field approach onto the total energy surface determined by IBM. One is then free to use the IBM Hamiltonian with these microscopically derived parameters for the calculation of spectra and electromagnetic transition rates. This method has already been applied to some regions of the nuclear chart exhibiting triaxiality  
\cite{Nomura2012a,Nomura2021a,Nomura2021b}. A similar method has been recently used in the non-relativistic mean field framework employing the Skyrme interaction
\cite{Vasileiou2024,Vasileiou2025}.  

Triaxiality in various nuclei has been considered within the HF method in Refs. \cite{Parikh1968,Giraud1969,Flerackers1977,Sahu1979,Gupta1982,Bonche1985,Redon1986,Dutta2000,vanDalen2014}, 
and within HFB in Refs. \cite{Girod1978,Hayashi1984,Bonche1991,Oi2003,Hinohara2011,Chen2017,Schuck2019}.
Studies of triaxility in various nuclei using the Gogny force can be found in Refs. 
\cite{Girod1983,Girod2009,Robledo2009,RodriguezGuzman2010,RodriguezGuzman2010b,Rodriguez2010,Tagami2016,Suzuki2021}, 
while the Skyrme force has been used in studying triaxiality in specific nuclei in Refs. 
\cite{Heenen1993,Shi2012,Fracasso2012,Zhang2015b,Benrabia2017,Scamps2021,Washiyama2023,Vasileiou2024}.
Studies of triaxiality in specific nuclei using the relativistic mean field approach can be found in Refs. 
\cite{Koepf1988,Hirata1996,Yao2010,Niksic2010,Yao2011,Nomura2012a,Xiang2016,Abusara2017a,Naz2018,Yang2021a,Nomura2021a,Nomura2021b,ElBassem2024}.  
These will be discussed in the relevant sections regarding the corresponding series of isotopes.

\section{Empirical Signatures of Triaxiality} \label{sig} 

\subsection{The Shape Parameter $\gamma$} \label{shapepar}

{
Within the Rigid Triaxial Rotor Model \cite{Davydov1958,Davydov1959}, described in subsec. \ref{RTRM}, the value of the shape parameter $\gamma$ can be obtained
 directly from the ratio of the energy of the second $2^+$ state, $2_2^+$,  over the energy of the first $2^+$ state, $2_1^+$  
\begin{equation} \label{R}
R= {E(2_2^+) \over E(2_1^+)} 
\end{equation}
through the expression \cite{Casten1990,Bonatsos2025}
\begin{equation} \label{g}
\gamma = {1\over 3} \sin^{-1} \left( {3\over R+1} \sqrt{R\over 2}  \right). 
\end{equation}
One can also obtain $\gamma$ from the ratio of transition rates (branching ratio) 
\begin{equation} \label{R2}
R_2= {B(E2; 2_2^+ \to 2_1^+)  \over B(E2; 2_2^+ \to 0_1^+)} 
\end{equation}
through the expression \cite{Casten1990,Bonatsos2025}
\begin{equation}
R_2 = {20\over 7} { {\sin^2 3\gamma  \over 9-8 \sin^2 3\gamma} \over   1-{3-2\sin^2 3\gamma \over \sqrt{9-8\sin^2 3\gamma} }  } . 
\end{equation} }

{
The ratios $R$ and $R_2$ are shown in Fig. 3 of Ref. \cite{Bonatsos2025}. The ratio $R$ starts from the value 2 at $\gamma=30^{\rm o}$ and raises towards infinity for $\gamma=0$, while the ratio $R_2$ shows the opposite behavior, starting from 1.43 at $\gamma=0$ and raising towards infinity at  $\gamma=30^{\rm o}$. At $\gamma=15^{\rm o}$ one has $R=6.85$ and $R_2=2.71$. Nuclei with $R<6.85$ and $R_2>2.71$ are expected to have $\gamma>15^{\rm o}$, while nuclei with $R>6.85$ and $R_2<2.71$ are expected to have $\gamma<15^{\rm o}$. }

{However,} atomic nuclei are not rigid bodies, therefore the collective variables $\beta$ and $\gamma$ do not obtain a fixed value for each nucleus. In addition, $\beta$ and $\gamma$ are not directly measurable quantities. It is therefore desirable to try to estimate the shape of the nucleus through model-independent quantities which would be direct observables. It turns out that the quadrupole shape invariants introduced by Kumar in 1972 \cite{Kumar1972,Cline1986} can serve this purpose. 

The quadrupole shape invariants \cite{Kumar1972,Cline1986,Werner2001,Werner2005} are the expectation values of higher powers of the quadrupole operator, calculated in a given state, which can be taken to be the ground state. The second order quadrupole shape invariant, $q_2$, which corresponds to the matrix elements of $Q\cdot Q$, turns out to be proportional to $\beta^2$, while the third order invariant, $q_3$, which corresponds to the matrix element of $(QQQ)^0$, turns out to be proportional to $\beta^3 \cos (3\gamma)$. Therefore $q_2$ can serve in estimating the value of $\beta$, while the ratio $q_3/q_2^{3/2}$ can serve for estimating the value of $\cos (3\gamma)$. $n$th order invariants of the form $q_n/q_2^{n/2}$ can also be used \cite{Werner2005}. 

In the framework of the IBM, one can easily see that the collective variable $\gamma$ is connected \cite{Elliott1986} to the value of the parameter $\chi$, appearing in the expression for the quadrupole operator given in Eq.~(\ref{Qop}). The question of equivalence between $\gamma$-instability and rigid triaxility within the O(6) dynamical symmetry of the IBM has been posed \cite{Otsuka1987}, concluding that the ground states of the two cases are equivalent, but the excited states are not \cite{Cohen1988}. 

Recent calculations \cite{Poves2020} in the framework of the configuration-interaction shell model, allowing for the calculation of higher-order quadrupole invariants and therefore of the fluctuations of the shape variables $\beta$ and $\gamma$,  show that $\beta$ has a non-negligible degree of softness, while $\gamma$ is usually characterized by large fluctuations, rendering its effective value not meaningful \cite{Poves2020}. A by-product of this method is that doubly magic nuclei cannot be considered as spherical, because the notion of a well-defined shape does not apply to them \cite{Poves2020}.       

\subsection{Rigid Triaxiality vs. $\gamma$-Softness} \label{stagg}

The relative displacement of the levels with odd angular momentum $L$ relative to their neighbors with even $L$ within the $\gamma$ band, called the odd-even staggering,
 has long been considered as an indicator of triaxiality \cite{Zamfir1991,McCutchan2007}. 

Odd-even staggering in $\gamma$ bands is studied using the quantity \cite{Zamfir1991,McCutchan2007}
\begin{equation} \label{SL}
S(L) =  { E(L^+_\gamma) + E[(L-2)^+_\gamma] - 2  E[(L-1)^+_\gamma] \over E(2_1^+)}, 
\end{equation}
which measures the displacement of the $(L-1)$ level in relation to the average of its neighbors, $L$ and $(L-2)$, normalized to the energy of the first excited state of the nucleus.

In Fig. 1(a) the odd-even staggering occurring in the $\gamma$-rigid Davydov model \cite{Davydov1958,Davydov1959} for various values of $\gamma$ is shown. It is observed that irrespectively of the value of $\gamma$, minima occur at odd values of the angular momentum $L$. 

%%%%%%  FIG. 1 %%%%%%%%%%%%%%%%%%%%%%%%%

\begin{figure} [htb]

 \includegraphics[width=75mm]{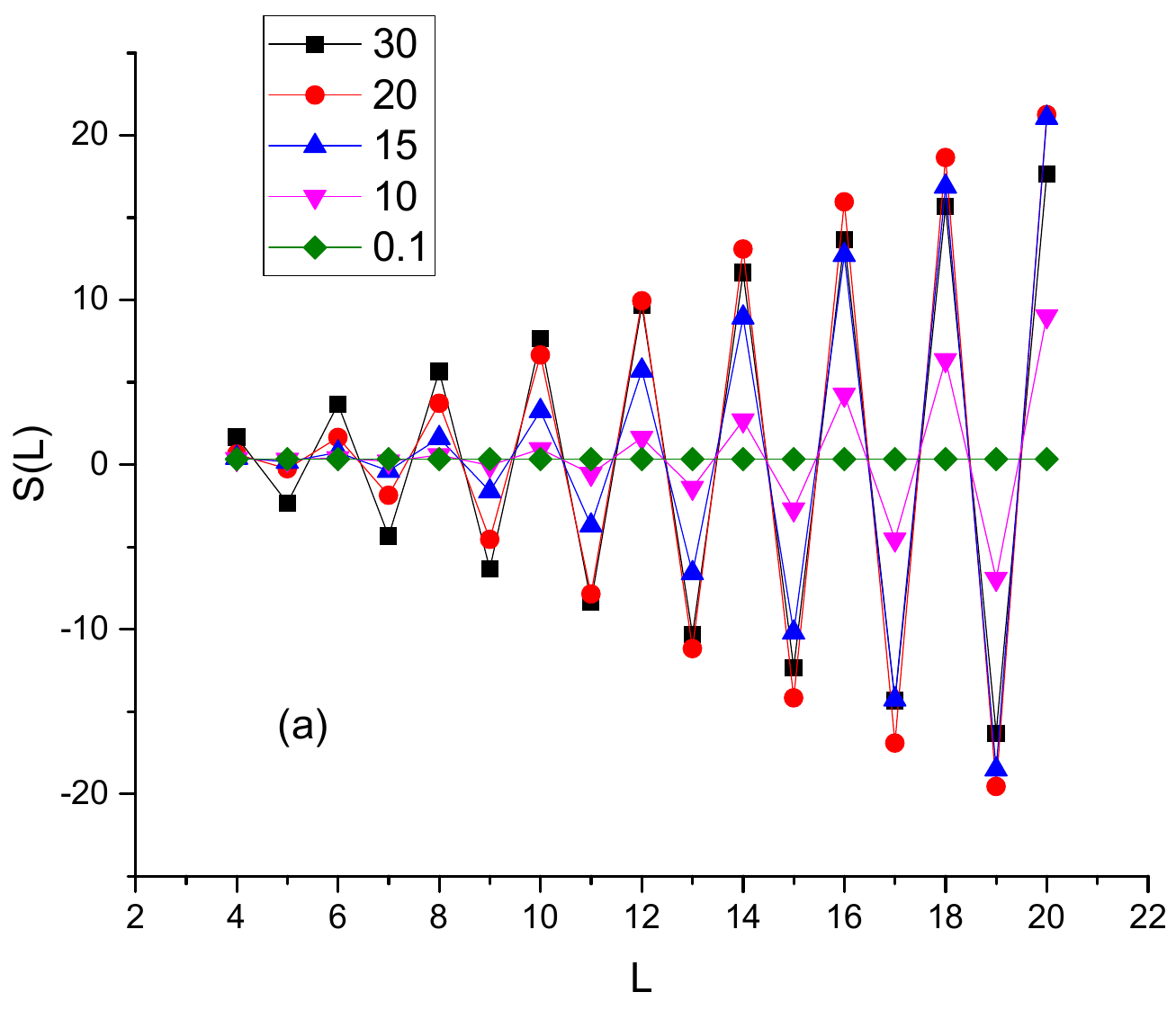}\hspace{5mm}
    \includegraphics[width=75mm]{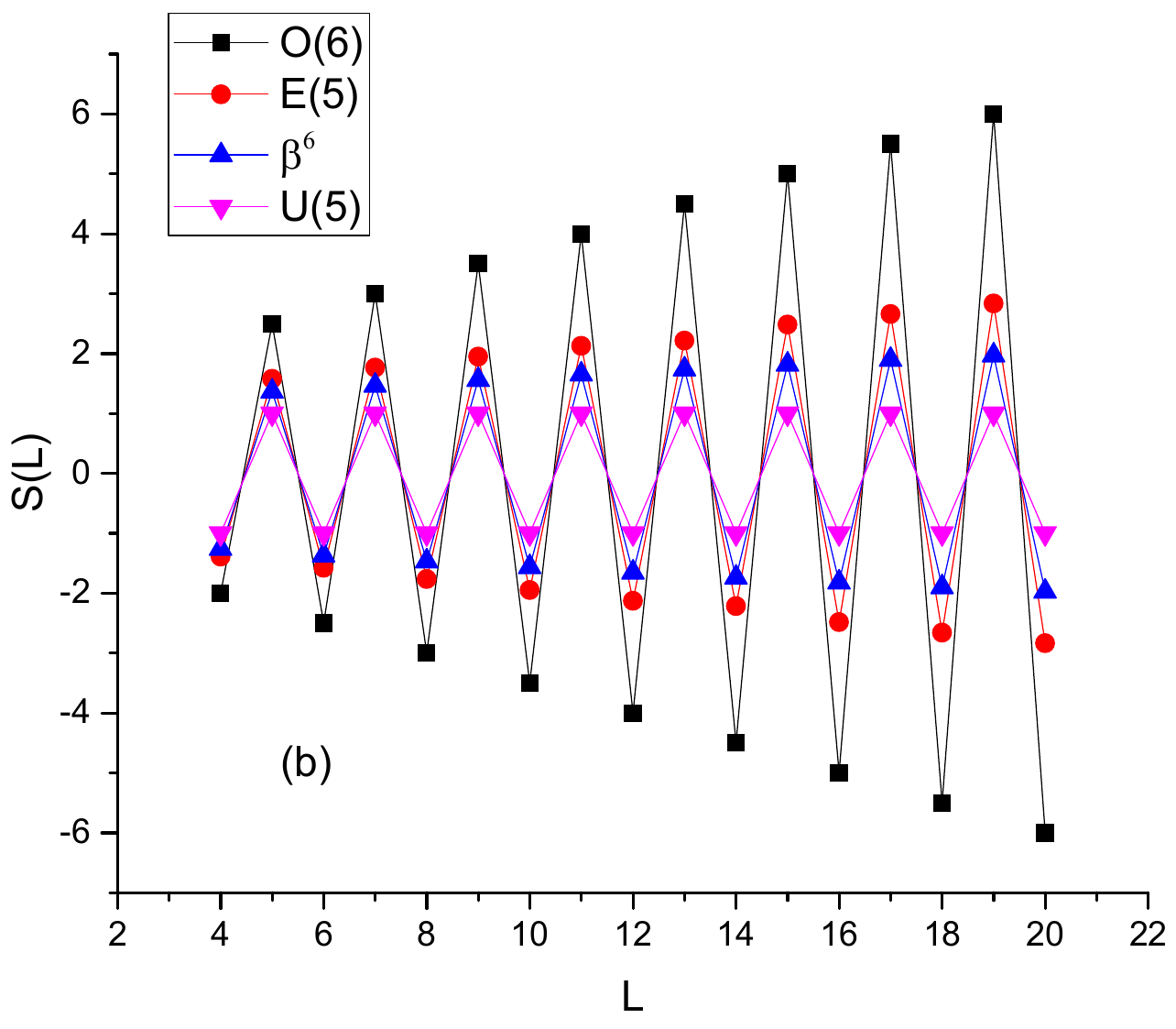}
  
    \caption{Odd-even staggering in $\gamma$-bands, given by Eq. (\ref{SL}), (a) for the Davydov model \cite{Davydov1958,Davydov1959} at various values of $\gamma$, and (b) for $\gamma$-soft models ranging from U(5) to the O(6) of the IBM \cite{Iachello1987}, including the critical point symmetry E(5) \cite{Iachello2000} and the E(5)-$\beta^6$ model \cite{Bonatsos2004b}. See sec. \ref{stagg} for further discussion.} 
    
\end{figure}

In Fig. 1(b) the odd-even staggering occurring in several $\gamma$-soft models, ranging from the vibrator with U(5) symmetry to the O(6) limit of the interacting boson model (IBM) \cite{Iachello1987}, and including the critical point symmetry E(5) \cite{Iachello2000} and the E(5)-$\beta^6$ model \cite{Bonatsos2004b}, in which the infinite square well potential in the deformation variable $\beta$, used in E(5), is replaced by a $\beta^6$ potential. It is observed that for all models minima occur at even values of $L$, which is the opposite behavior in comparison to what is seen in the case of the rigid triaxial models in Fig. 1(a). This behavior can be easily attributed to the O(5) symmetry underlying the U(5) and O(6) dynamical symmetries of IBM \cite{Iachello1987}, which implies that the levels of the $\gamma$-band are grouped by the degeneracies imposed by O(5) \cite{Rakavy1957,Bes1959}  as 2, (3,4), (5,6), \dots (see Table I and Fig. 3 of Ref. \cite{Bonatsos2004b} for details), while in the rigid triaxial rotor of Davydov \cite{Davydov1958,Davydov1959} the relevant grouping is (2,3), (4,5), (6,7), \dots. 

In addition to the opposite sign of staggering, in Fig. 1 it is seen that the size of the odd-even staggering in the rigid triaxial cases is an order of magnitude higher than the staggering in the $\gamma$-soft cases.  

It should be noticed that  although very few nuclei exhibit odd-even staggering corresponding to rigid triaxial shapes (see Fig. 4 of Ref. \cite{McCutchan2007}), many nuclei exhibit odd-even staggering corresponding to soft triaxiality \cite{Liao1995}. 

In corroboration of these observations, consistent modeling \cite{Grosse2022} of several observables of heavy nuclei reveals that for the triaxiality parameter $\gamma$ the zero value does not occur even in deformed nuclei, but rather the $\gamma$ values exhibit a variance of around 8$^{\rm o}$ \cite{Grosse2022}, while acquiring values of 10$^{\rm o}$ or higher (see, for example, Fig. 5 of Ref. \cite{Bonatsos2017b}).  

Another quantity exhibiting different behavior in the $\gamma$-rigid and $\gamma$-soft cases is the ratio \cite{Casten2020,Bonatsos2021}
\begin{equation} \label{RL}
R(L) = { E(L^+_\gamma)-E(L^+_g) \over E(2^+_\gamma)-E(2^+_g) }, 
\end{equation}
based on the energy differences between the $\gamma$-band and the ground state band (g) at the same angular momentum $L$. In Fig. 2(a) one sees that in the rigid triaxial case the ratios R(L) increase rapidly and gradually bending upwards, while in Fig. 2(b) one observes that in the $\gamma$-soft cases the ratios R(L) increase much more slowly and linearly. The qualitative differences between these two cases can be clearly seen in Fig. 3.  

%%%%%%%%%%%%%%%%%   FIG. 2 %%%%%%%%%%%%%%%%%%%%%%%%%%%%

\begin{figure} [htb]    

\centering %% If there is a figure in wide page, please release command \centering
\includegraphics[width=75mm]{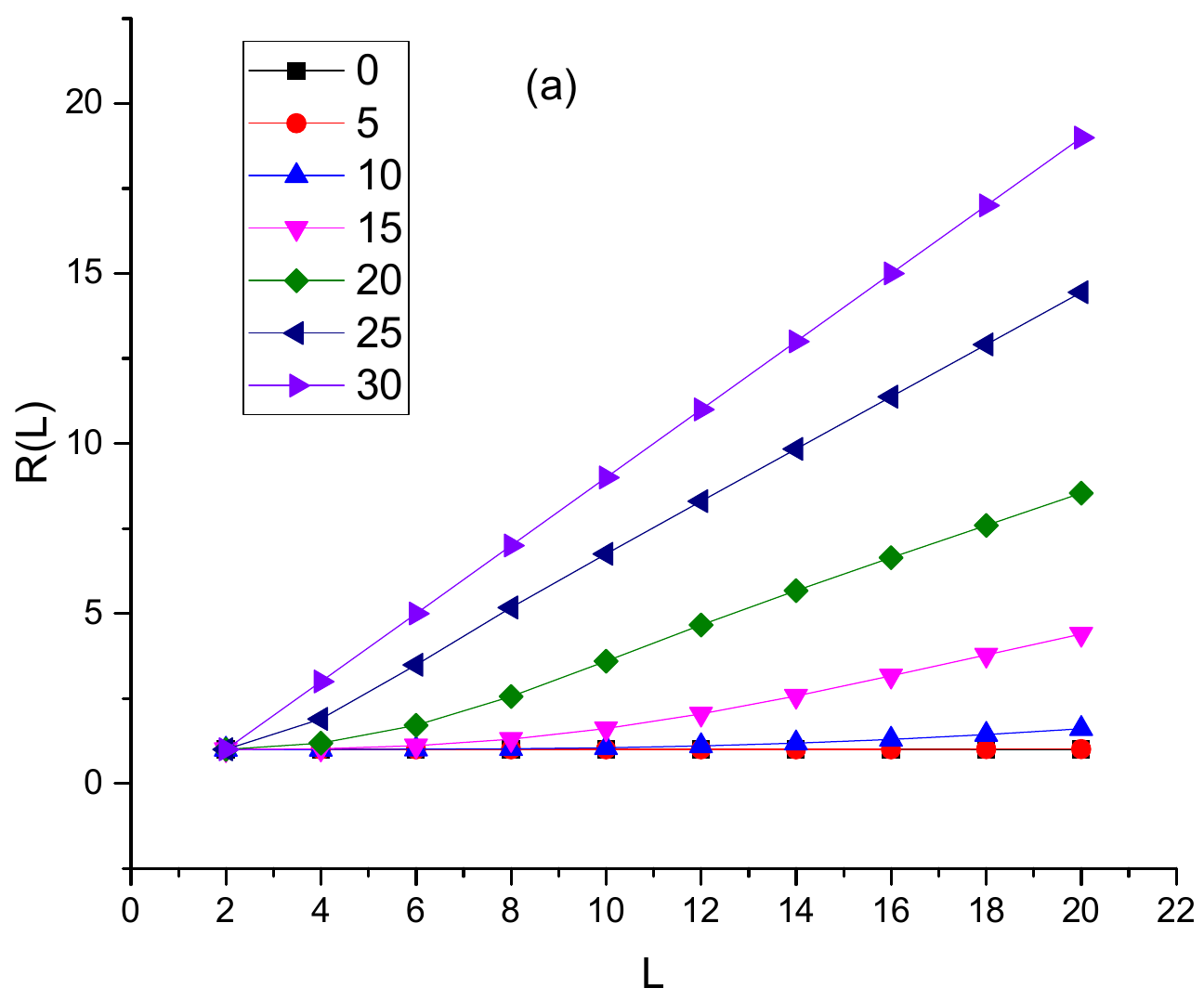}\hspace{5mm}
    \includegraphics[width=75mm]{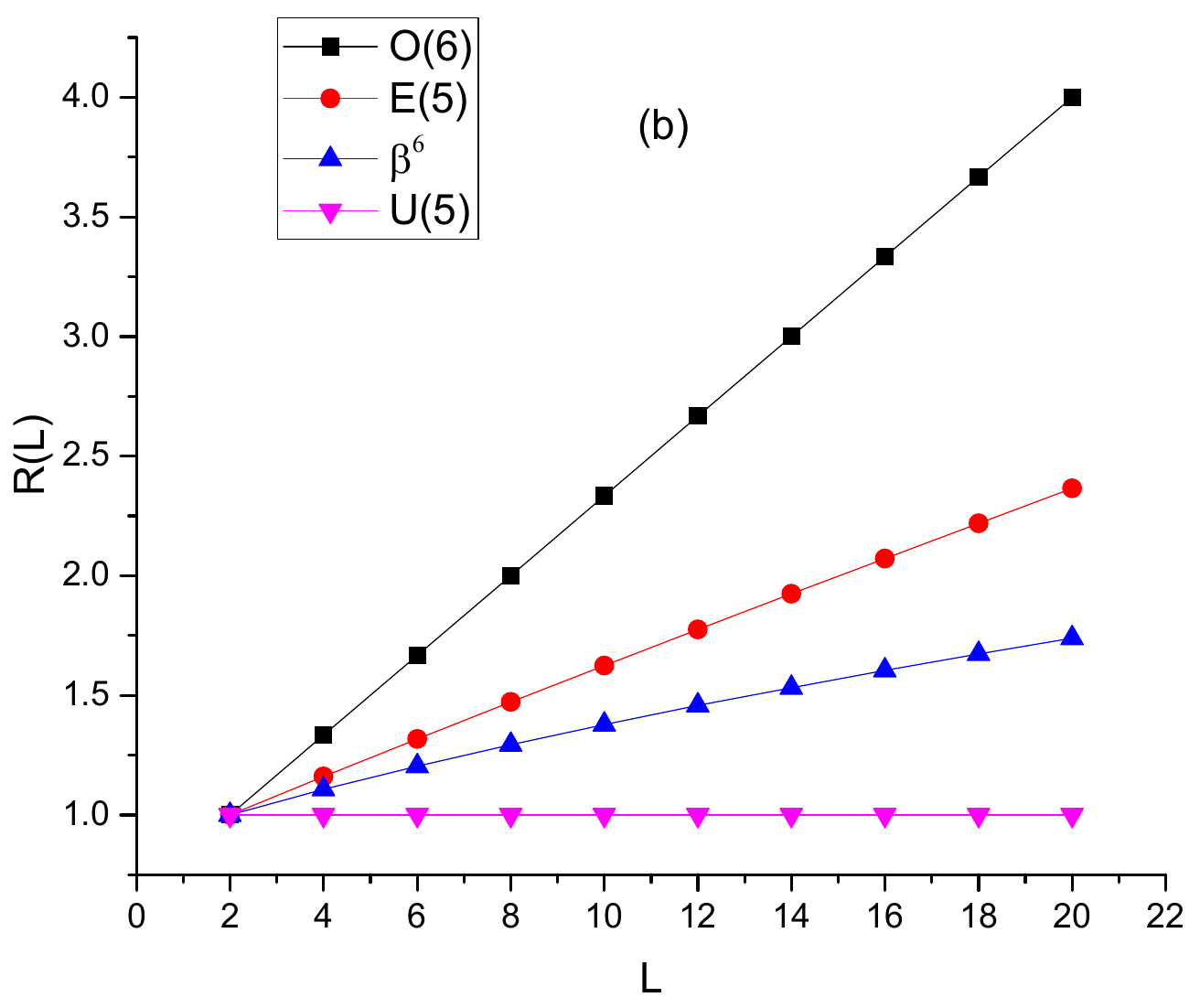}

    \caption{Ratios $R(L)$ given by Eq. (\ref{RL}), (a) for the Davydov model \cite{Davydov1958,Davydov1959} at various values of $\gamma$, and (b) for $\gamma$-soft models ranging from U(5) to the O(6) of the IBM \cite{Iachello1987}, including the critical point symmetry E(5) \cite{Iachello2000} and the E(5)-$\beta^6$ model \cite{Bonatsos2004b}. See sec. \ref{stagg} for further discussion.}  
    
\end{figure}

%%%%%%%%%%%%%%%%%%%%%%%%%%%%%%%%%%%%%%%%%%% FIG. 3 %%%%%%%%%%%%%%%%%%%%%%%%%%%%%%%%%%%%%%%%%%%%%

\begin{figure} [htb]

    \includegraphics[width=75mm]{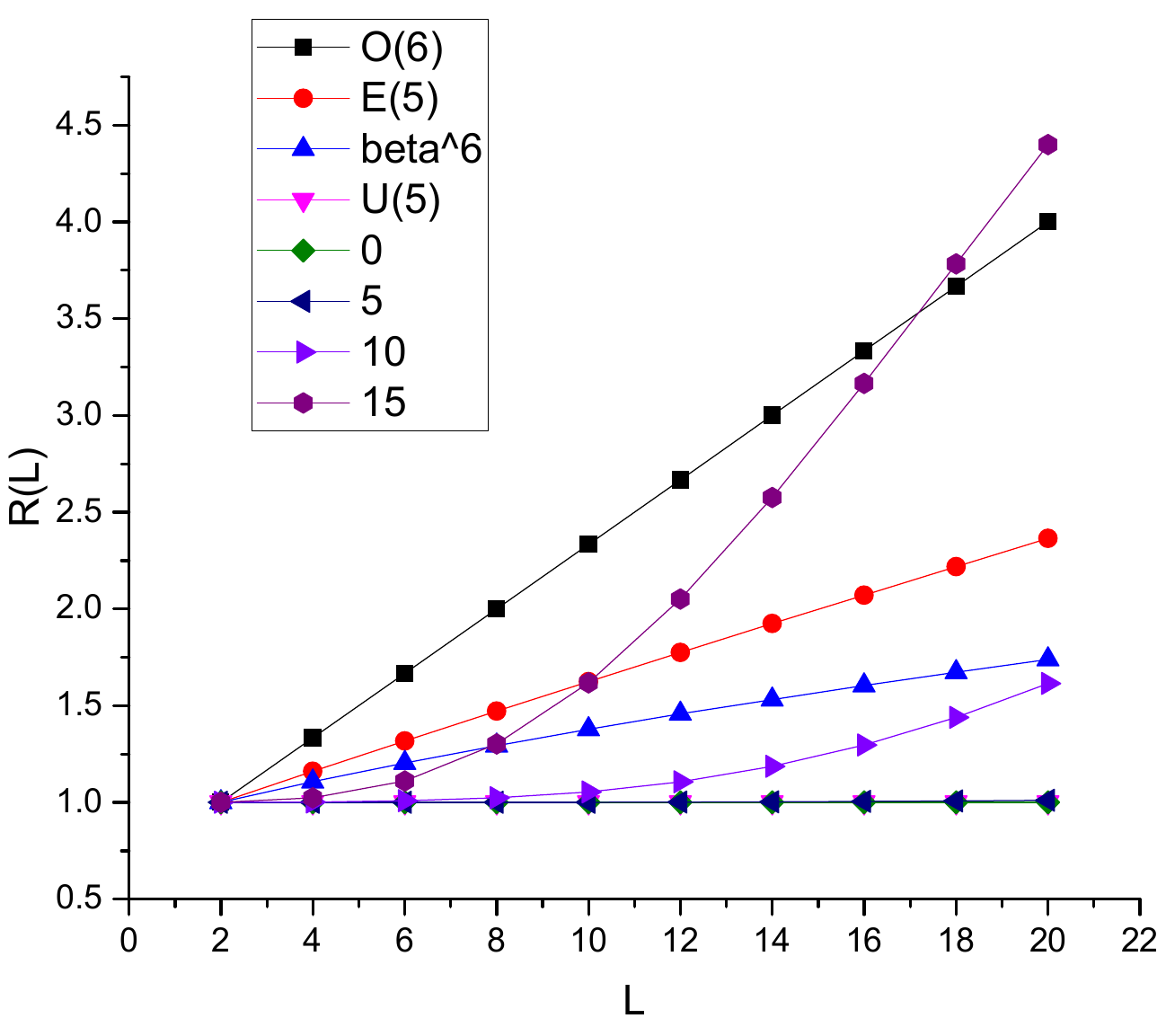}

    \caption{Comparison of ratios $R(L)$ given by Eq. (\ref{RL}), for the Davydov model \cite{Davydov1958,Davydov1959} at various values of $\gamma$, and for $\gamma$-soft models ranging from the U(5) to the O(6) of the IBM \cite{Iachello1987}, including the critical point symmetry E(5) \cite{Iachello2000} and the E(5)-$\beta^6$ model \cite{Bonatsos2004b}. See sec. \ref{stagg} for further discussion. } 
    
\end{figure}

It should be noticed that the rigid triaxial rotor of Davydov \cite{Davydov1958,Davydov1959} at $\gamma =30^{\rm o}$ has the same selection rules for electric quadrupole transitions as the quadrupole vibrator model and the $\gamma$-unstable model of Wilets and Jean \cite{Wilets1956}. Therefore the selection rules of electric quadrupole transitions cannot help in distinguishing rigid triaxial behavior from the $\gamma$-soft one (see p. 191 of Ref. \cite{Casten1990}). 

\section{Global Systematics of Triaxiality} \label{global}

After examining {in Appendices \ref{Z2426}-\ref{Z8898}} separately each isotopic chain in the region $Z=24$-98, and drawing partial conclusions in various specific regions, we are now going to try to build the general picture regarding areas on the nuclear chart in which triaxiality is favored. 

{
We are going to consider first the predictions made within the proxy-SU(3) approximation to the shell model. A review of the proxy-SU(3) symmetry scheme and its connection to the Nilsson model and to the spherical shell model has been given in Ref. \cite{Bonatsos2023}, while the group theoretical details regarding the highest weight irreducible representations of SU(3) and their dominance have been given in Refs. \cite{Martinou2018,Martinou2021b}.} 

In Fig. 4 the proxy-SU(3) predictions, taken from Eq. (\ref{mu}),  for the deformation variable $\gamma$ for all experimentally known \cite{ensdf} nuclei with $18\leq Z \leq 80$ and $18\leq N \leq 124$ are shown \cite{Bonatsos2025}.  The horizontal and vertical stripes covering the nucleon numbers 22-26, 34-48, 74-80, 116-124, 172-182, within which substantial triaxiality is expected to occur \cite{Bonatsos2025}, are also shown. We remark that non-zero values of $\gamma$ occur almost everywhere across the nuclear chart, in agreement with recent suggestions in the framework of the Monte Carlo Shell Model \cite{Otsuka2019,Tsunoda2021,Otsuka2023} and the Triaxial Projected Shell Model \cite{Rouoof2024} that a non-zero degree of triaxiality appears everywhere on the nuclear chart, including regions of prolate deformed nuclei considered as ideal examples of axial deformation \cite{Otsuka2019,Tsunoda2021,Otsuka2023}. In addition, recent studies of consistent modelling of several observables in heavy nuclei has also demonstrated the need for broken axial symmetry, with the variance of the triaxiality parameter $\gamma$ centered in the range of $\gamma \sim 8^{\rm o}$ \cite{Grosse2022}.

%%%%%   FIG. 4 %%%%%%%%%%%%%%%%%%%%%%%%%%%%%%%

\begin{figure*} [htb]    

\includegraphics[width=170mm]{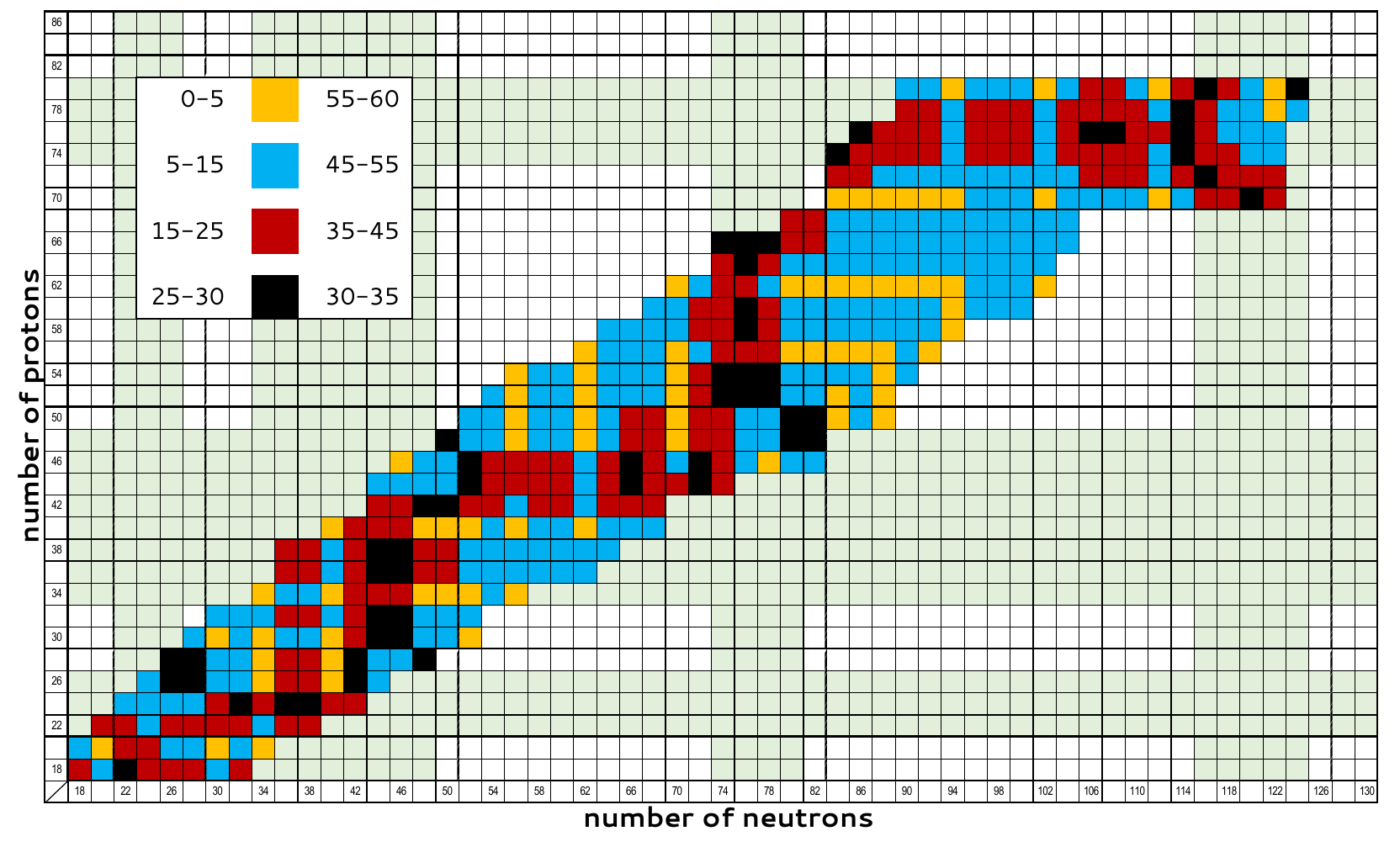}

 \caption{Proxy-SU(3) predictions, taken from Eq. (\ref{mu}),  for the deformation variable $\gamma$ (in degrees) for all experimentally known \cite{ensdf} nuclei with $18\leq Z \leq 80$ and $18\leq N \leq 124$.  The horizontal and vertical stripes covering the nucleon numbers 22-26, 34-48, 74-80, 116-124, 172-182, within which substantial triaxiality is expected to occur \cite{Bonatsos2025}, are also shown. Adapted from Ref. \cite{Bonatsos2025}.  See Sec. \ref{global} for further discussion.} 
    
\end{figure*}

Fig. 5 is derived from Fig. 4 by eliminating nuclei with low triaxiality, i.e. nuclei with $\gamma=0$-15$^{\rm o}$, being close to prolate ($\gamma=0$) shapes, and  nuclei with $\gamma=45^{\rm o}$-60$^{\rm o}$, being close to oblate ($\gamma=60^{\rm o}$) shapes, and showing only nuclei with high triaxiality, i.e.  nuclei with $\gamma=15^{\rm o}$-45$^{\rm o}$, being close to maximally triaxial ($\gamma=30^{\rm o}$) shapes. We see that a ''staircase'' pattern appears, falling within the stripes in which triaxiality is expected to be favored according to the proxy-SU(3) predictions \cite{Bonatsos2025}. 

%%%% FIG. 5 %%%%%%%%%%%%%%%%%%%%%%%%%%%

\begin{figure*} [htb]
    \includegraphics[width=170mm]{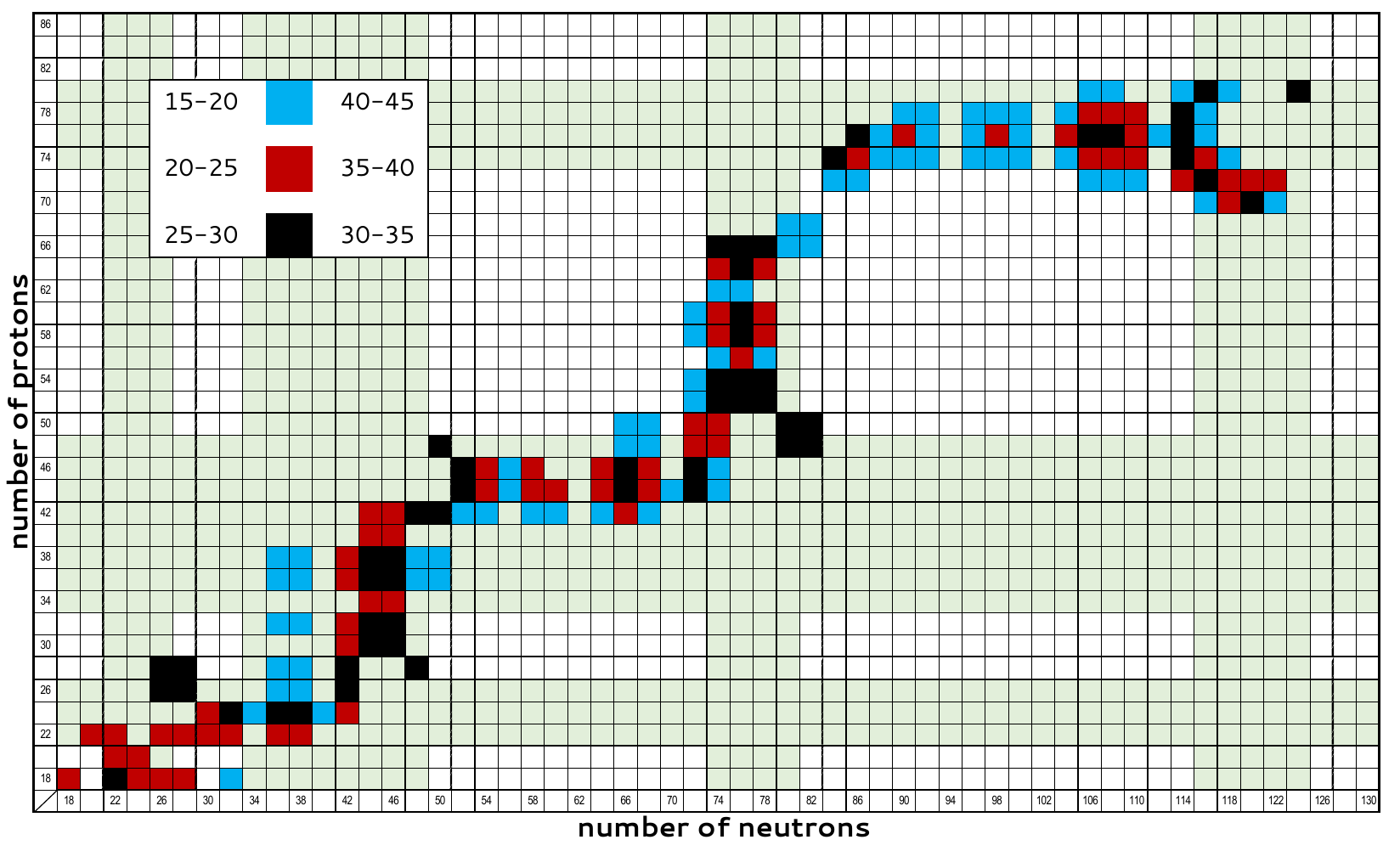}
  \caption{Same as Fig. 4, but only for nuclei having $15^{\rm o} \leq \gamma \leq 45^{\rm o} $ included. The horizontal and vertical stripes covering the nucleon numbers 22-26, 34-48, 74-80, 116-124, 172-182, within which substantial triaxiality is expected to occur \cite{Bonatsos2025}, are also shown. Adapted from Ref. \cite{Bonatsos2025}. 
    See Sec. \ref{global} for further discussion.} 
    
\end{figure*}

{
At this point it is worth examining to which extent these proxy-SU(3) predictions agree to available experimental data \cite{ensdf}. }

{
In Fig. 6 all nuclei with known \cite{ensdf} $2^+_1$ and $2_2^+$ levels are shown, divided into two groups, these with $R<6.85$ (expected to have $\gamma > 15^{\rm o}$, as discussed in subsec. \ref{shapepar}) and those with $R>6.85$ (expected to have $\gamma < 15^{\rm o}$). We see that many nuclei belong to the first group. However, not all of them are expected to exhibit strong triaxiality, since $R=2$ also corresponds to the simple vibrator with U(5) symmetry \cite{Iachello1987}.}

{
In order to confront this problem,  all nuclei up to $Z=80$ with experimentally known $2^+_1$ and $2_2^+$ levels, as well as $B(E2; 2^+_2\to 0^+_1)$ and $B(E2; 2^+_2\to 2^+_1)$ transition rates, have been collected in Table I of Ref. \cite{Bonatsos2025}.
The experimental $R_{4/2}=E(4_1^+)/E(2_1^+)$ ratio, a well-known \cite{Casten1990} indicator of collectivity, is also shown, in order to facilitate the recognition of nuclei being close to the simple vibrator value of $R_{4/2}=2$, mentioned above. }
 
%%%%%%%%%%%%%%%%%%%%%%%%%%%%%%%%%%%%%%%%%%% FIG. 6 %%%%%%%%%%%%%%%%%%%%%%%%%%%%%%%%%%%%%%%%%%%%%

\begin{figure*} [htb]

    \includegraphics[width=170mm]{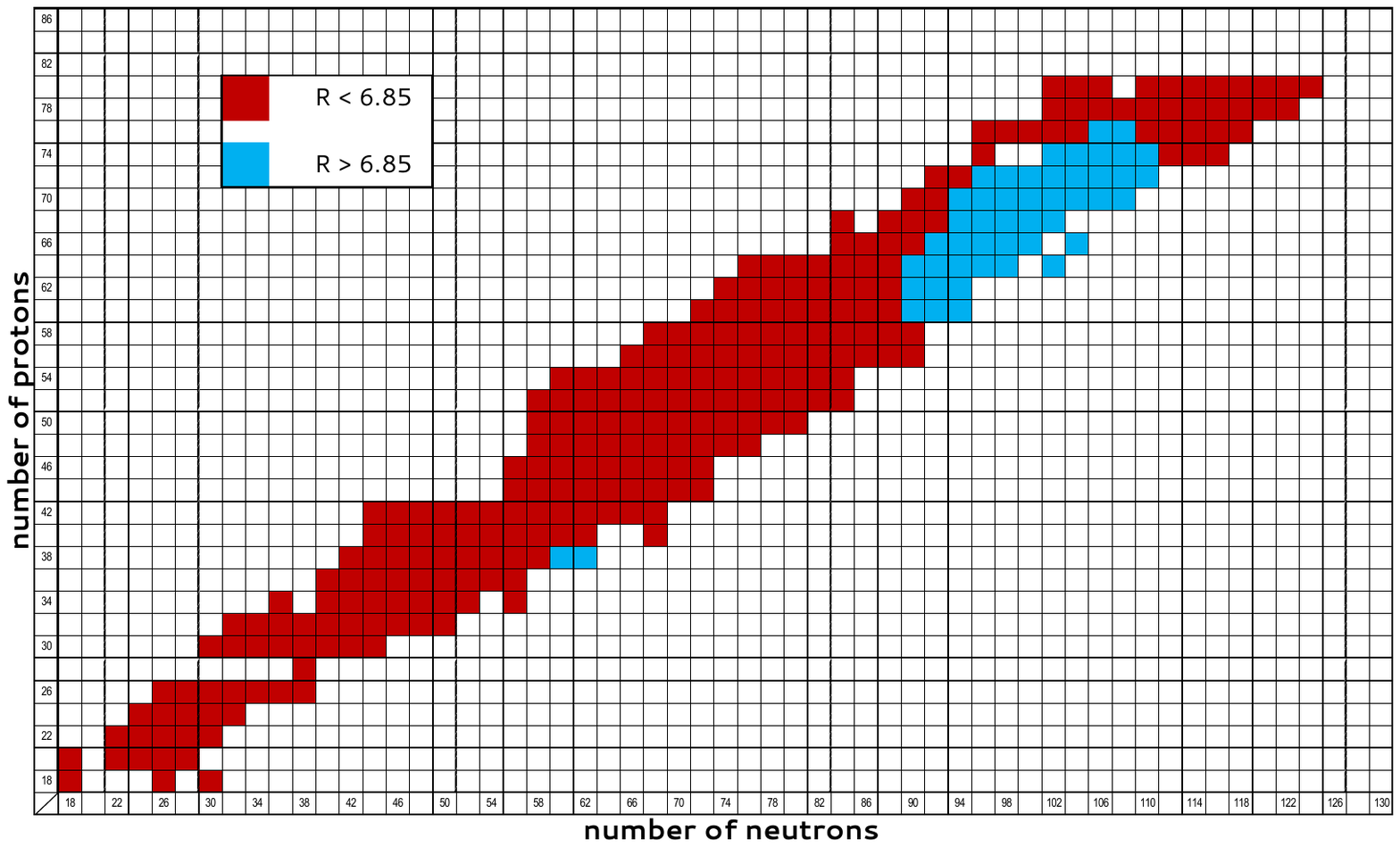}
   
    \caption{{Nuclei with experimentally known $2_1^+$ and $2_2^+$ levels, subdivided into these with $R<6.85$ (expected to have $\gamma > 15^{\rm o}$, as discussed in subsec. \ref{shapepar}) and those with $R>6.85$ (expected to have $\gamma < 15^{\rm o}$). Data have been taken from Ref. \cite{ensdf}. In nuclei in which $\beta$- and $\gamma$-bands are assigned in Ref. \cite{ensdf}, the $2^+$ state of the $\gamma$-band is chosen as the $2_2^+$.  Adapted from Ref. \cite{Bonatsos2025}. 
    See Sec. \ref{global} for further discussion.}} 
    
\end{figure*} 
 
{
In Fig. 7 the nuclei of Table I of Ref. \cite{Bonatsos2025} are subdivided into two groups, these with $R<6.85$ and $R_2>2.71$ (expected to have $\gamma > 15^{\rm o}$,as discussed in subsec. \ref{shapepar}) and those with $R>6.85$ and $R_2<2.71$ (expected to have $\gamma < 15^{\rm o}$). 
 We see that by taking into account the value of the branching ratio, the number of candidates for substantial triaxiality is drastically reduced, while most of the remaining candidates are aligned along the regions (indicated in Fig. 5 by green stripes) for which oblate SU(3) irreps are predicted by the proxy-SU(3) symmetry, in  agreement with the situation appearing in Fig. 2. }

%%%%%%%%%%%%%%%%%%%%%%%%%%%%%%%%%%%%%%%%%%% FIG. 7 %%%%%%%%%%%%%%%%%%%%%%%%%%%%%%%%%%%%%%%%%%%%%

\begin{figure*} [htb]

    \includegraphics[width=170mm]{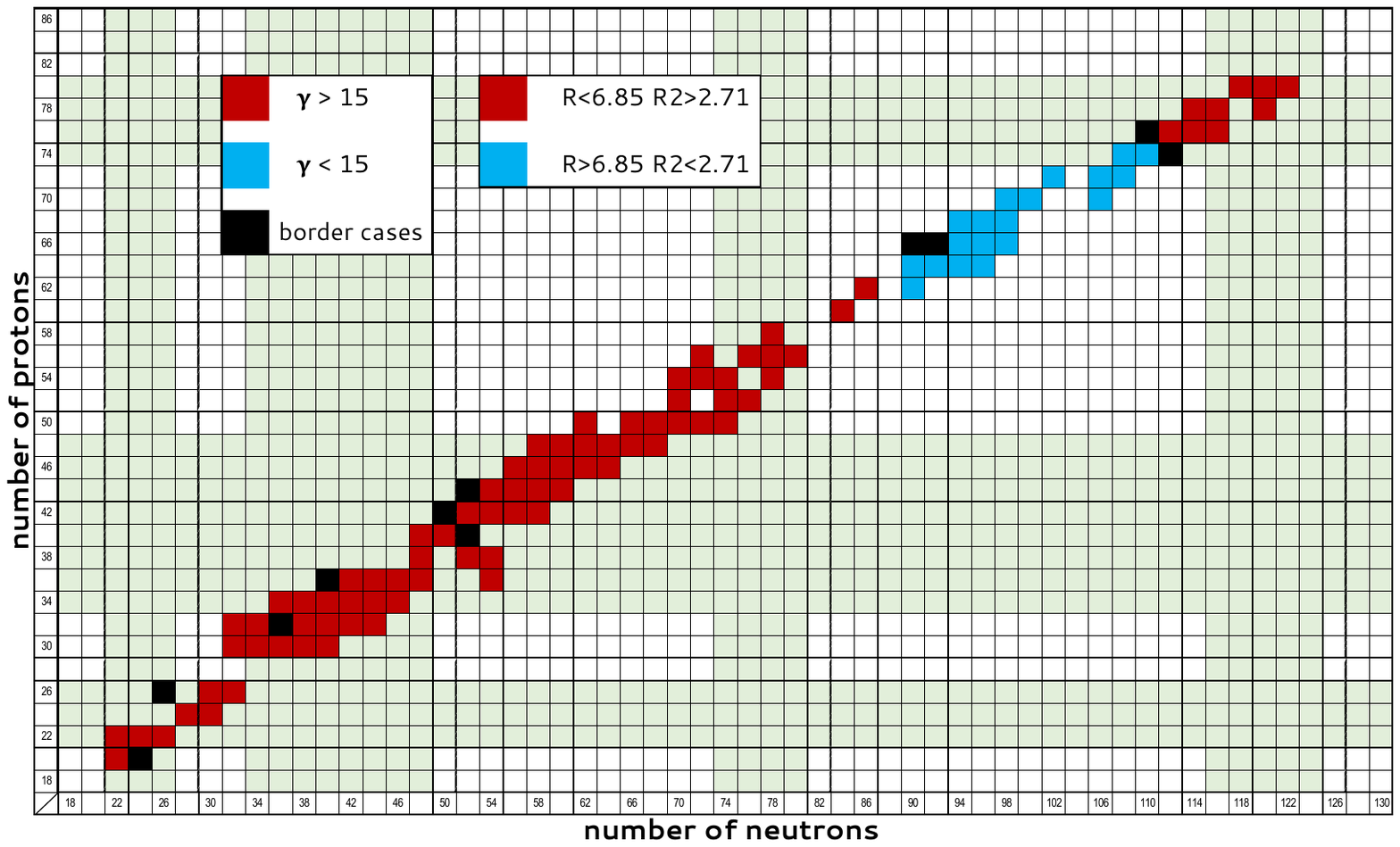}
    
    \caption{{Nuclei with experimentally known $2_1^+$ and $2_2^+$ levels, as well as known $B(E2; 2_2^+\to 0_1^+)$ and $B(E2; 2_2^+\to 2_1^+)$ transition rates,  subdivided into these with $R<6.85$ and $R_2>2.71$ (expected to have $\gamma > 15^{\rm o}$) and those with $R>6.85$ and $R_2<2.71$ (expected to have $\gamma < 15^{\rm o}$).  
 Data have been taken from Ref. \cite{ensdf} and collected in Table I of Ref. \cite{Bonatsos2025}. In nuclei in which $\beta$- and $\gamma$-bands are assigned in Ref. \cite{ensdf}, the $2^+$ state of the $\gamma$-band is chosen as the $2_2^+$. Ten border-line nuclei with $R<6.85$ and $R_2<2.71$, as well as one nucleus ($^{158}$Dy$_{92}$) with $R>6.85$ and $R_2>2.71$ are also shown. Most of the nuclei expected to have $\gamma > 15^{\rm o}$ lie within the horizontal and vertical stripes predicted by the proxy-SU(3) symmetry, covering the nucleon numbers  22-26, 34-48, 74-80, 116-124, 172-182, also shown in Fig. 5.  A substantial deviation is seen in the $Z=50$-56, $N=62$-72 region, in which nine nuclei with $\gamma   > 15^{\rm o}$ are lying outside the green stripes. Adapted from Ref. \cite{Bonatsos2025}. 
    See Sec. \ref{global} for further discussion.}} 
    
\end{figure*}

The reasoning behind {Figs. 5 and 7} assumes that large values of the collective variable $\gamma$ imply large triaxiality. Thinking in terms of potential energy surfaces (PES), 
this means a minimum in the PES close to  $30^{\rm o}$. But the question remains if this minimum is deep, implying robust triaxiality, or shallow, suggesting 
$\gamma$-softness. 

One way to address this question is to consider  the decrease in energy due to triaxiality occurring within the macroscopic-microscopic approach of the Finite-Range Droplet Model (FRDM) or Finite-Range Liquid-Drop Model (FRLDM) \cite{Moeller2006,Moeller2008}. It is plausible that larger reduction in energy due to triaxiality would mean more robust triaxiality. 

Nuclei for which the FRDM predicts decrease in energy due to triaxiality equal or larger than 0.01 MeV, taken from Ref. \cite{Moeller2008}, are shown in 
{Fig. 8}. 
We see that the ``staircase'' structure regarding experimentally known nuclei seen in Fig. 5 is reproduced, thus corroborating the assumption that large values of $\gamma$ imply robust triaxiality. It should be noticed that these two predictions come from completely different models of completely different nature. The FRDM predictions come from a macroscopic-microscopic model with parameters fitted in order to reproduce a great variety of nuclear observables over the whole nuclear chart, while the proxy-SU(3) predictions come from symmetry arguments, based on the Pauli principle and the short-range nature of the nucleon-nucleon interaction \cite{Martinou2020,Martinou2021b}, in a completely parameter-free way. In addition, in 
{Fig. 8} we see regions of nuclei not appearing in Fig. 5, since they correspond to yet experimentally unknown nuclei 
with $Z=42$-46, $N=96$-110 and $Z=52$-62, $N=120$-122. These additional regions also fall within the stripes of favored triaxiality predicted by proxy-SU(3).

%%%% FIG. 8 %%%%%%%%%%%%%%%%%%%%%%%%%

\begin{figure*} [htb ]
 \includegraphics[width=170mm]{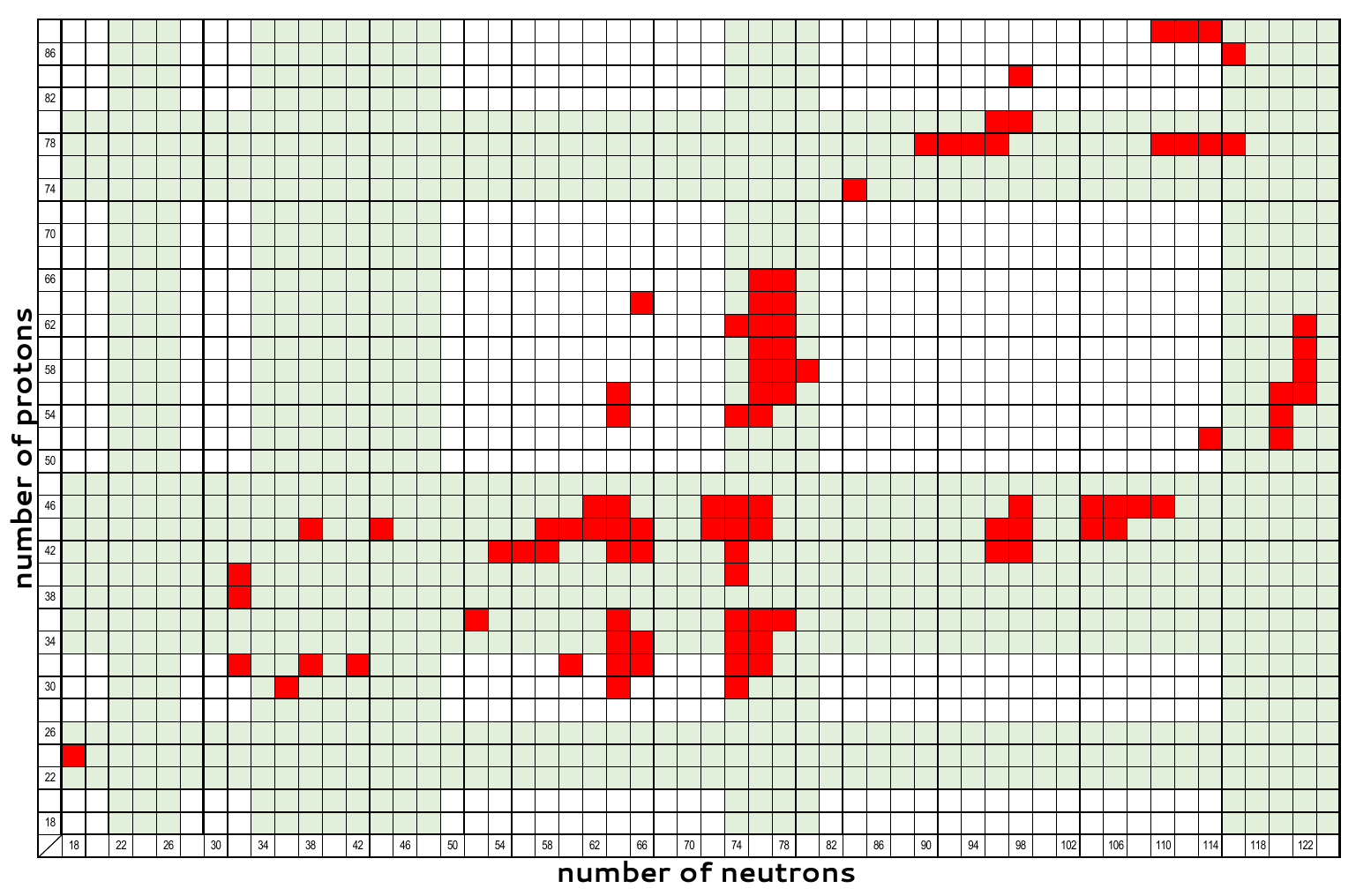}
\caption{Nuclei up to $Z=88$, $N=124$, for which the Finite Range Droplet Model (FRDM) predicts decrease in energy due to triaxiality equal or larger than 0.01 MeV, taken from Ref. \cite{Moeller2008}.     See Sec. \ref{global} for further discussion.} 
    
\end{figure*}

A different picture appears for the FRDM predictions beyond $Z=82$, depicted in 
{Fig. 9}, in which most of the nuclei with decrease in energy due to triaxiality equal or larger than 0.01 MeV are lying outside the stripes for favored triaxiality predicted by the proxy-SU(3) symmetry. In particular, this happens in the regions with $Z=82$-94, $N=154$-160, and $Z=88$-92, $N=108$-114, as well as in $Z=96$-108, $N=130$-142. 

%%%  FIG. 9 %%%%%%%%%%%%%%%%%%%%%

\begin{figure*} [htb]
    
 \includegraphics[width=170mm]{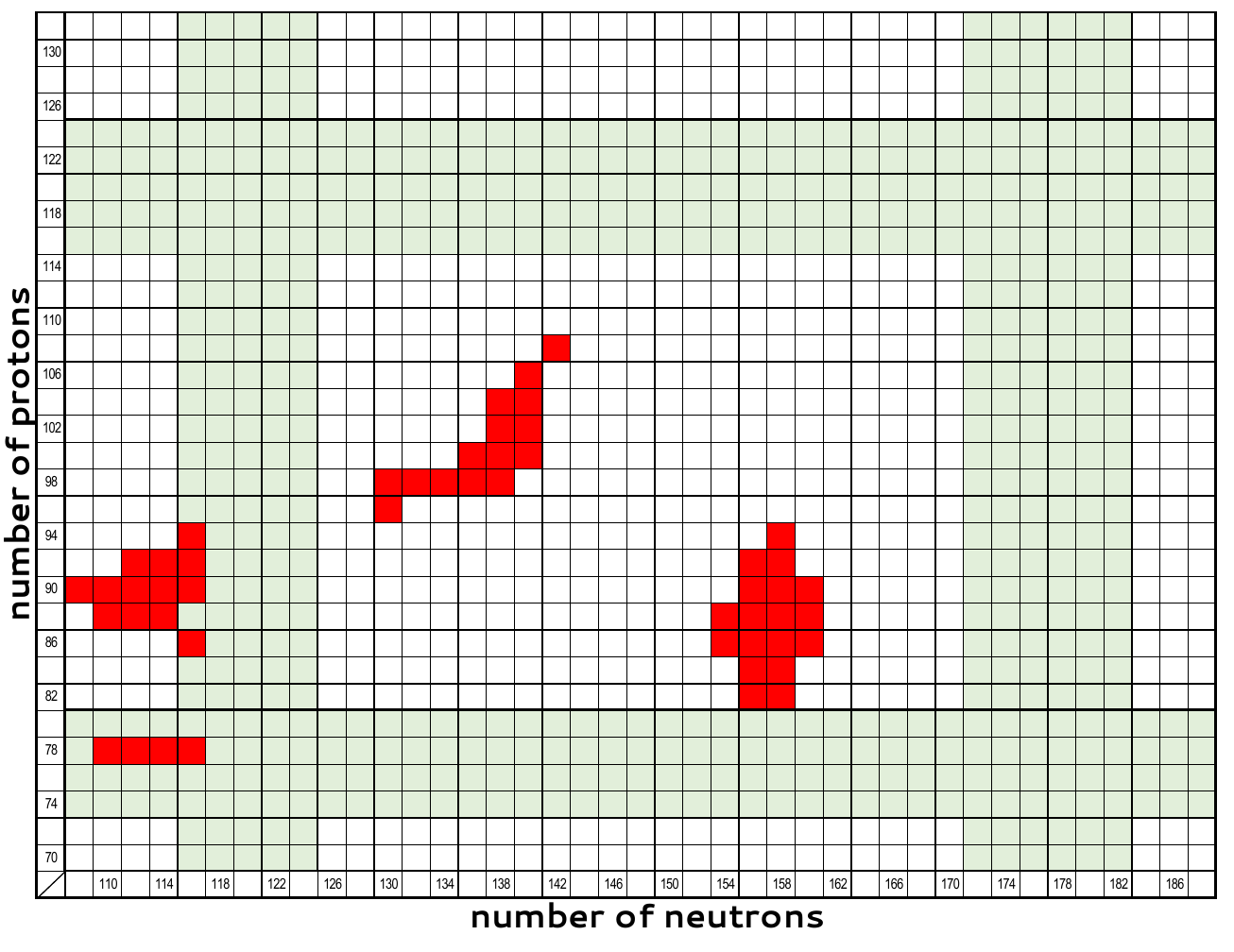}     
     
    \caption{Nuclei above $Z=70$, $N=108$, for which the Finite Range Droplet Model (FRDM) predicts decrease in energy due to triaxiality equal or larger than 0.01 MeV, taken from Ref. \cite{Moeller2008}.     See Sec. \ref{global} for further discussion.} 
    
\end{figure*}

In 
{Fig. 10} we have gathered the partial conclusions reached in {Appendices  \ref{Z2426}-\ref{Z7480}} of the present review up to $Z=86$, $N=124$. It is seen that nuclei for which substantial evidence for triaxiality exists fall within the stripes in which triaxiality is predicted to be favored by the proxy SU(3) symmetry, shown in Fig. 5. Agreement of these nuclei with {Fig. 8}, in which nuclei with FRDM predictions of the decrease in energy due to triaxiality equal or larger than 0.01 MeV are depicted, is also seen.  

%%%%%% FIG. 10 %%%%%%%%%%%%%%%%%%%%%%%%%%%%

\begin{figure*} [htb]
    
 \includegraphics[width=170mm]{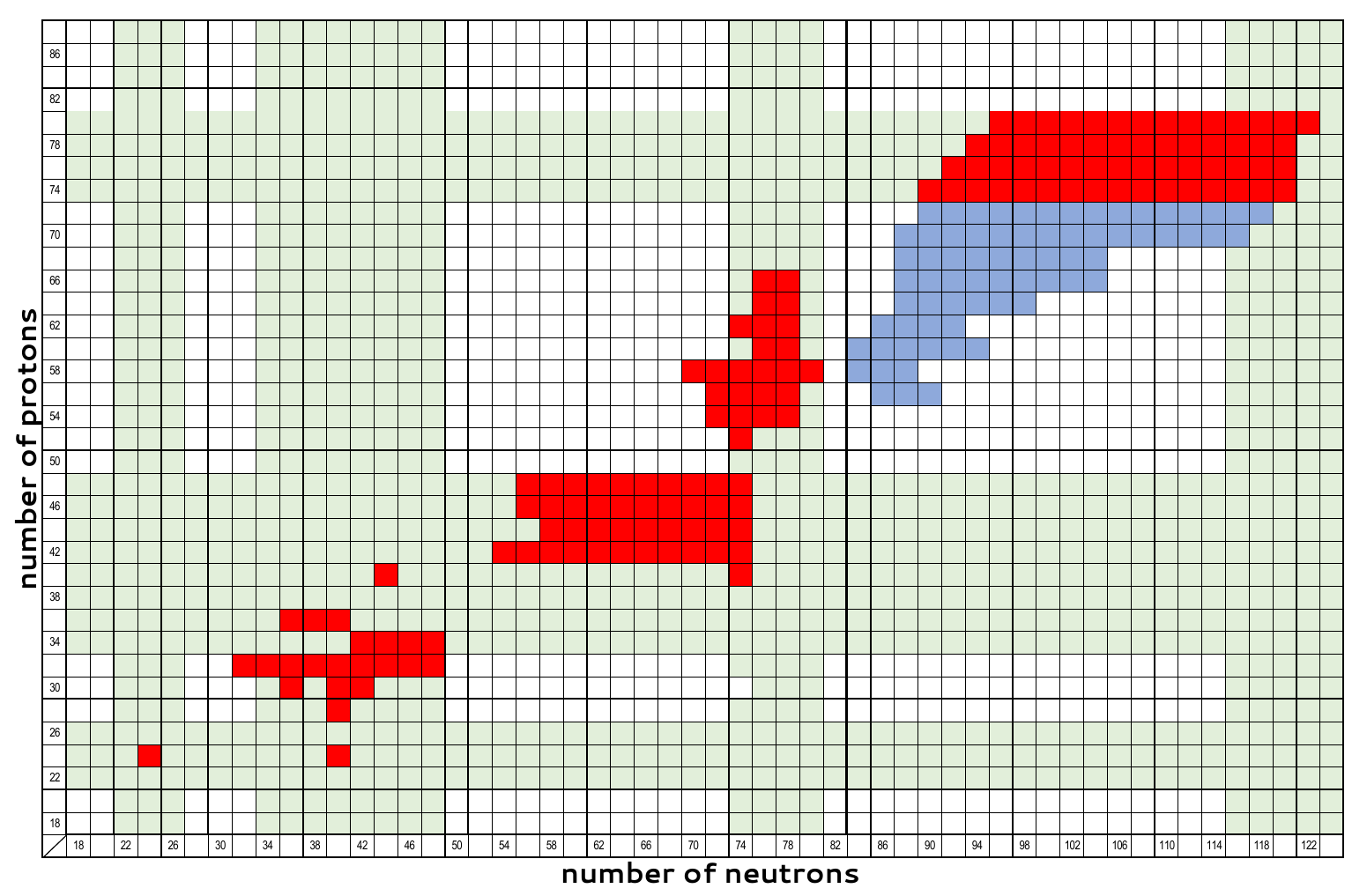}     
     
    \caption{Nuclei up to $Z=86$, $N=124$, for which evidence for substantial triaxiality (red boxes) or weak triaxiality (cyan boxes) has been found in {Appendices 
    \ref{Z2426}-\ref{Z7480}} of the present review.  See Sec. \ref{global} for further discussion.} 
    
\end{figure*}

Different conclusions are drawn from 
{Fig. 11}, in which the partial conclusions reached in {Appendices \ref{Z6872}-\ref{Z8898}} of the present review above $Z=70$, $N=108$ are depicted. In this case the partial conclusions of the present review indicate the presence of triaxiality within the region $Z=88$-98, $N=132$-154. There is neither overlap between these nuclei and the nuclei predicted  by FRDM to exhibit decrease in energy due to triaxiality equal or larger than 0.01 MeV, shown in 
{Fig. 9}, nor overlap with the stripes in which triaxiality is favored by the proxy-SU(3) symmetry. 

%%%%   FIG. 11 %%%%

\begin{figure*} [htb]
    
 \includegraphics[width=170mm]{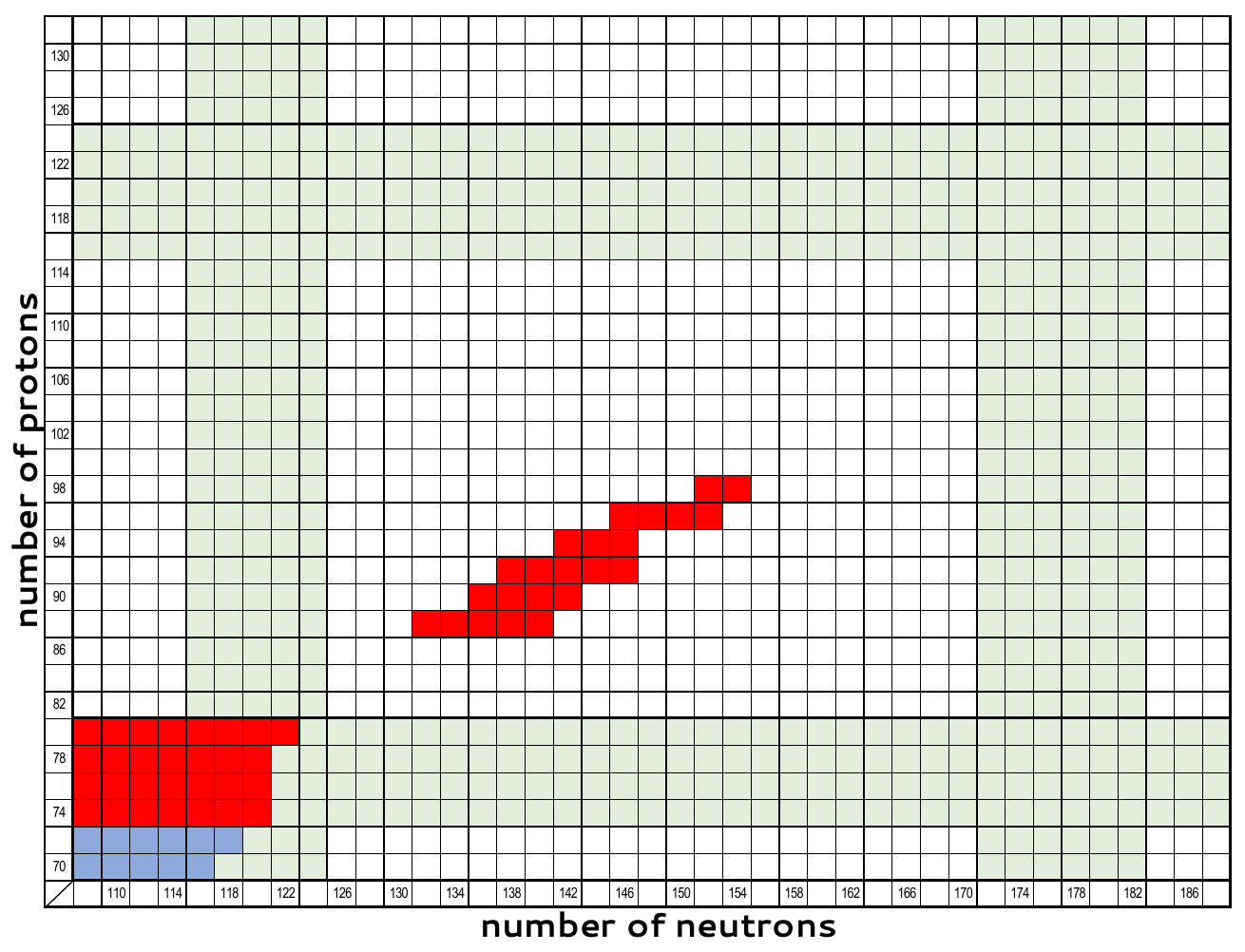}     
     
    \caption{Nuclei above $Z=70$, $N=108$, for which evidence for substantial triaxiality (red boxes) or weak triaxiality (cyan boxes) has been found in {Appendices 
    \ref{Z6872}-\ref{Z8898}} of the present review.  See Sec. \ref{global} for further discussion.} 
    
\end{figure*}

In summary, most of the theoretical work within various models considered in this review predicts substantial triaxiality for nuclei in the region up to $Z=82$, $N=126$ 
lying within the stripes favoring triaxiality according to the proxy-SU(3) symmetry, exhibiting in addition considerable (larger than 0.01 MeV) reduction in energy due to triaxiality according to the FRDM model. In contrast, for nuclei beyond $Z=82$, $N=126$, theoretical work within various models considered in this review predicts triaxiality in the region $Z=88$-98, $N=132$-154, in apparent disagreement with the predictions of  both the proxy-SU(3) symmetry and the FRDM model. 

{
In a first attempt to resolve this puzzle, we have extended Table 1 of Ref. \cite{Bonatsos2025} into the actinide region, including all nuclei for which the first two $2^+$ states, as well as the B(E2) transition rate among them is known \cite{ensdf}. As already discussed in section \ref{shapepar}, nuclei with $R>6.85$ and $R_2<2.71$ are expected to have 
$\gamma<15^{\rm o}$. This is indeed the case for all nuclei shown in Table 1, in agreement to  $\gamma$ values shown for actinides in the original Davydov papers \cite{Davydov1958,Davydov1959} (see Table 3 of \cite{Davydov1958} and Table 5 of \cite{Davydov1959}), as well as to $\gamma$ values provided by the extended Thomas-Fermi plus Strutinsky integral method \cite{Dutta2000} (see Tables II and III of \cite{Dutta2000}), discussed in Appendix \ref{Z8898}. In contrast, the Triaxial Projected Shell Model (TPSM)  \cite{Sheikh2016,Jehangir2021,Rouoof2024} appears to be providing in the actinide region $\gamma$ values systematically higher than these of the aforementioned approaches (see Table 2 of \cite{Sheikh2016}, Table 2 of \cite{Jehangir2021}, and Table 3 of \cite{Rouoof2024}). It should also be noticed that a method exists for extracting the value of $\gamma$ from the experimental $R_{4/2}$ ratio \cite{Varshni1970}. This is convenient, since the $R_{4/2}$ ratio is known experimentally in many more nuclei than the $R$ ratio of Eq. (\ref{R}), but has the handicap of providing systematically much higher values of $\gamma$ than these provided by the $R$ and $R_2$ (Eq. (\ref{R2})) ratios (see Table I of \cite{Varshni1970} for a long list of examples). The results provided by the TPSM and through the use of the $R_{4/2}$ ratio have also been taken into account in Appendix \ref{Z8898}, thus blurring the whole picture. It is clear that further microscopic work is needed in order to clarify this issue. 
}

%%% Table 1 %%%%%%%%%%%%%%%%%%%%%%%5

\begin{table*} [htb] 

\caption{{Nuclei with experimentally known $2^+_1$ and $2_2^+$ levels, as well as known $B(E2; 2^+_2\to 0^+_1)$ and $B(E2; 2^+_2\to 2^+_1)$ transition rates.  
Data have been taken from Ref. \cite{ensdf}.  In nuclei in which $\beta$- and $\gamma$-bands are assigned in Ref. \cite{ensdf}, the $2^+$ state of the $\gamma$-band is chosen as the $2_2^+$. Energies are given in MeV, while B(E2) transition rates are given in W.u. The ratios $R$ and $R_2$ are calculated from Eqs. (\ref{R}) and (\ref{R2}) respectively. The ratio $R_{4/2}$, a well-known \cite{Casten1990} indicator of collectivity, is also shown, along with the $\gamma$ value obtained from the ratio $R$ (called $\gamma_R$), from the ratio $R_2$ (called $\gamma_{R_2}$), and from the proxy-SU(3) symmetry using Eq. (\ref{mu}) (called $\gamma_{hw}$), using the SU(3) irreps given in Table 2. See Section \ref{global} for further discussion.} 
}

\begin{tabular}{ r   r  r r c c r r r r r }
\hline
  nucleus  & $2_1^+$ & $2_2^+$ & $R$ & $2_2^+\to 0_1^+$ & $2_2^+\to 2_1^+$ & $R_2$  & $R_{4/2}$ & $\gamma_R$& $\gamma_{R_2}$  & $\gamma_{hw}$\\ 
         & MeV     &  MeV    &     &   W.u.           &  W.u.            &        &           & & & \\ 
\hline

\isotope[230][90]{Th}$_{140}$ & 0.053 & 0.781 & 14.68 & 2.9 ${+6\choose -4}$ & 5.5 ${+10\choose -8}$ & 1.90 & 3.271 & 10.4 & 10.4 &  6.8 \\
\isotope[232][90]{Th}$_{142}$ & 0.049 & 0.785 & 15.91 & 2.9 (4)    & 7.2 (7)                         & 2.48 & 3.284 & 10.0 & 14.1 &  7.7 \\
 
\isotope[234][92]{U}$_{142}$  & 0.043 & 0.927 & 21.31 & 2.9 (5)    & 4.9 (8)                         & 1.69 & 3.296 &  8.7 &  8.1 &  7.3 \\
\isotope[238][92]{U}$_{146}$  & 0.045 & 1.060 & 23.61 & 3.04 (18)  & 5.3 (4)                         & 1.74 & 3.303 &  8.3 &  8.8 &  6.5 \\
 
\isotope[250][98]{Cf}$_{152}$ & 0.043 & 1.032 & 24.15 & 2.3 (3)    & 3.7 (4)                         & 1.61 & 3.321 &  8.2 &  6.9 & 10.4 \\

\hline
\end{tabular}
\end{table*}

%%% Table 2 %%%%%%%%%%%%%%

\begin{table} [htb] 

\caption{{Proxy-SU(3) hw irreps for protons [$(\lambda,\mu)_p$] and neutrons [$(\lambda,\mu)_n$], as well as total hw irreps  [$(\lambda,\mu)_{hw}$]
for the nuclei included in Table 1, obtained from Table I of Ref. \cite{Sarantopoulou2017}. As pointed out in Ref. \cite{Bonatsos2024}, when the hw irrep is fully symmetric
(i.e., has $\mu=0$), the next hw irrep (nhw) has to be taken into account. This happens in the case of neutrons of $^{238}$U, in which case the neutron nhw irrep is (50,14) and therefore the total nhw irrep is (80,18). The value of $\gamma_{hw}$ reported in Table 1 for  $^{238}$U is the average of the values 2.65 and 10.30, obtained from the total hw irrep (90,4) and the total nhw irrep (80,18), respectively, in accordance to the reasoning given in sec. 6 of  Ref. \cite{Bonatsos2024}. See Section \ref{global} for further discussion.} 
}

\begin{tabular}{ r   r  r r }
\hline
  nucleus  & $(\lambda,\mu)_p$ & $(\lambda,\mu)_n$ & $(\lambda,\mu)_{hw}$ \\
\hline

\isotope[230][90]{Th}$_{140}$ & 26,4 & 48,6 & 74,10 \\
\isotope[232][90]{Th}$_{142}$ & 26,4 & 50,8 & 76,12 \\
 
\isotope[234][92]{U}$_{142}$  & 30,4 & 50,8 & 80,12 \\
\isotope[238][92]{U}$_{146}$  & 30,4 & 60,0 &  90,4 \\
 
\isotope[250][98]{Cf}$_{152}$ & 34,8 & 54,12 & 88,20 \\

\hline
\end{tabular}
\end{table}

Finally, in 
{Figs. 12-15} we provide for quick qualitative reference the proxy-SU(3) predictions in several parts of the nuclear chart, obtained by using the numerical results 
for the highest weight (hw) and next highest weight (nhw) irreps given in tables 2-6 of Ref. \cite{Bonatsos2024}. The values shown are the ones corresponding to the hw irrep, 
except in the cases in which $\mu=0$ occurs for the proton hw irrep and/or the neutron hw irrep, in which case the values depicted are the average of the hw and nhw cases, 
in a simple schematic approximation, as described in Ref. \cite{Bonatsos2025}. The following observations can be made.   

a) In {Fig. 12}, the protons lie within the region 34-48, in which triaxiality is favored \cite{Bonatsos2025}. If the neutrons also lie in regions favoring triaxiality (34-48, 74-80),
the values of $\gamma$ raise above $30^{\rm o}$. 

b) A similar situation appears in {Fig. 15}. In W-Hg the protons lie  within the region 74-80, in which triaxiality is favored. If the neutrons also lie in the region 116-124, in which  triaxiality is favored, the values of $\gamma$ raise above $30^{\rm o}$. 

c) In {Fig. 13} the protons lie outside the regions in which triaxiality is favored (22-26, 34-48, 74-80, 116-124) \cite{Bonatsos2025}. If the neutrons lie in the region 74-80, in which  triaxiality is favored, the values of $\gamma$ raise above $15^{\rm o}$. 

d) A similar situation appears in {Fig. 14}. The protons lie outside the regions in which triaxiality is favored.   If the neutrons lie in the region 116-124, in which  triaxiality is favored, the values of $\gamma$ raise above $15^{\rm o}$. 

In short, we conclude that $\gamma$ values above $30^{\rm o}$ are reached when both the protons and the neutrons of a given nucleus lie within regions in which triaxiality is favored, while for  $\gamma$ values above $15^{\rm o}$ to be achieved, it is necessary for either the protons or the neutrons of a given nucleus to lie within regions in which triaxiality is favored. 

%%%%% FIG. 12 %%%%%%%%%%%%%%%%%%%%%%%%%%%5

\begin{figure} [htb]
    
 \includegraphics[width=75mm]{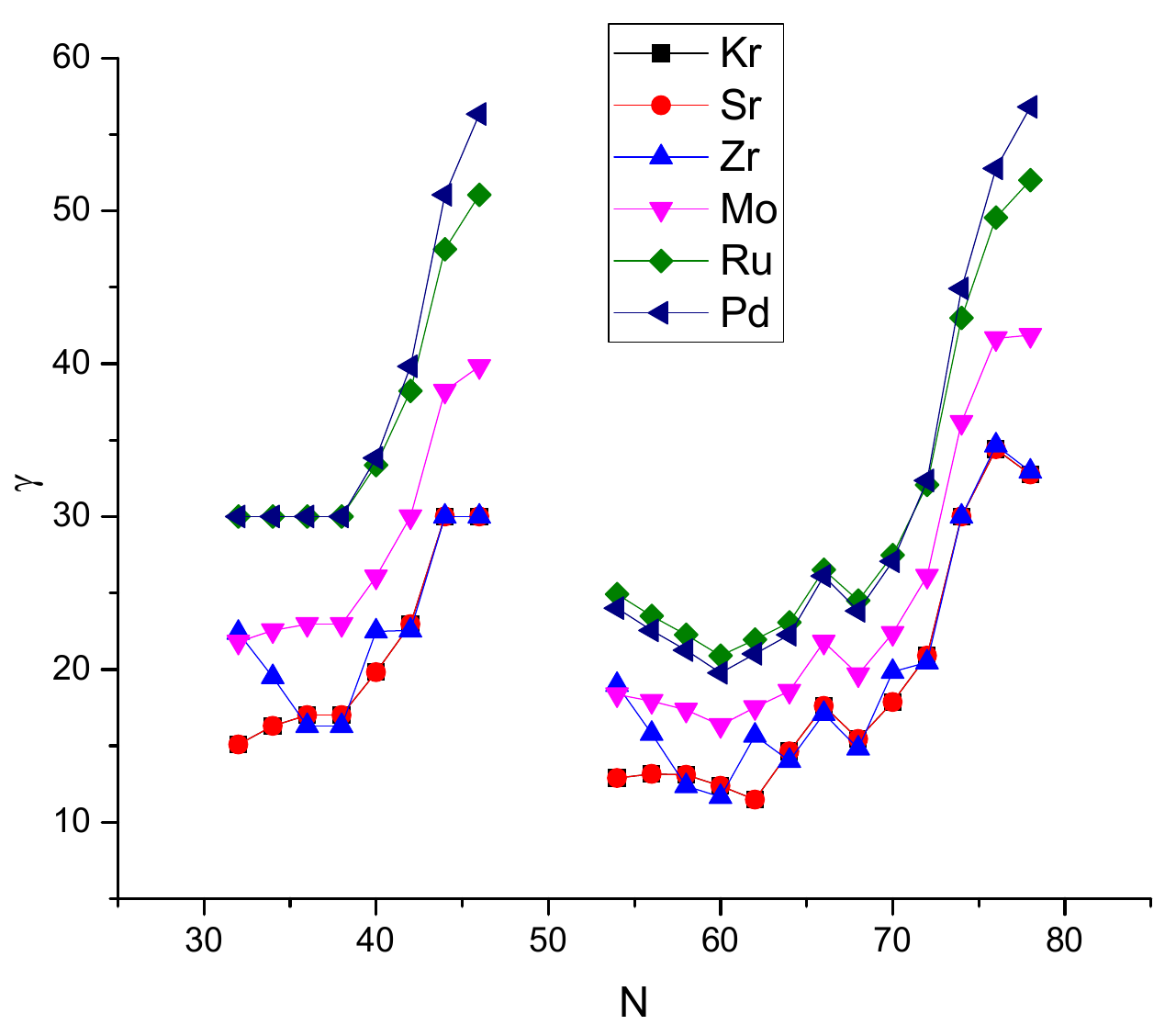}     
     
    \caption{Proxy-SU(3) predictions for the deformation variable $\gamma$ (in degrees) in the region with $Z=36$-46, $N=32$-46 and $N=54$-78, taken from Ref. \cite{Bonatsos2024}. 
      See Sec. \ref{global} for further discussion.} 
    
\end{figure}

%%%%%%%%%%%%%%%%%%%%%%%%%%%%%%%%%%%%%%%%%%% FIG. 13  %%%%%%%%%%%%%%%%%%%%%%%%%%%%%%%%%%%%%%%%%%%%%

\begin{figure} [htb]
    
 \includegraphics[width=75mm]{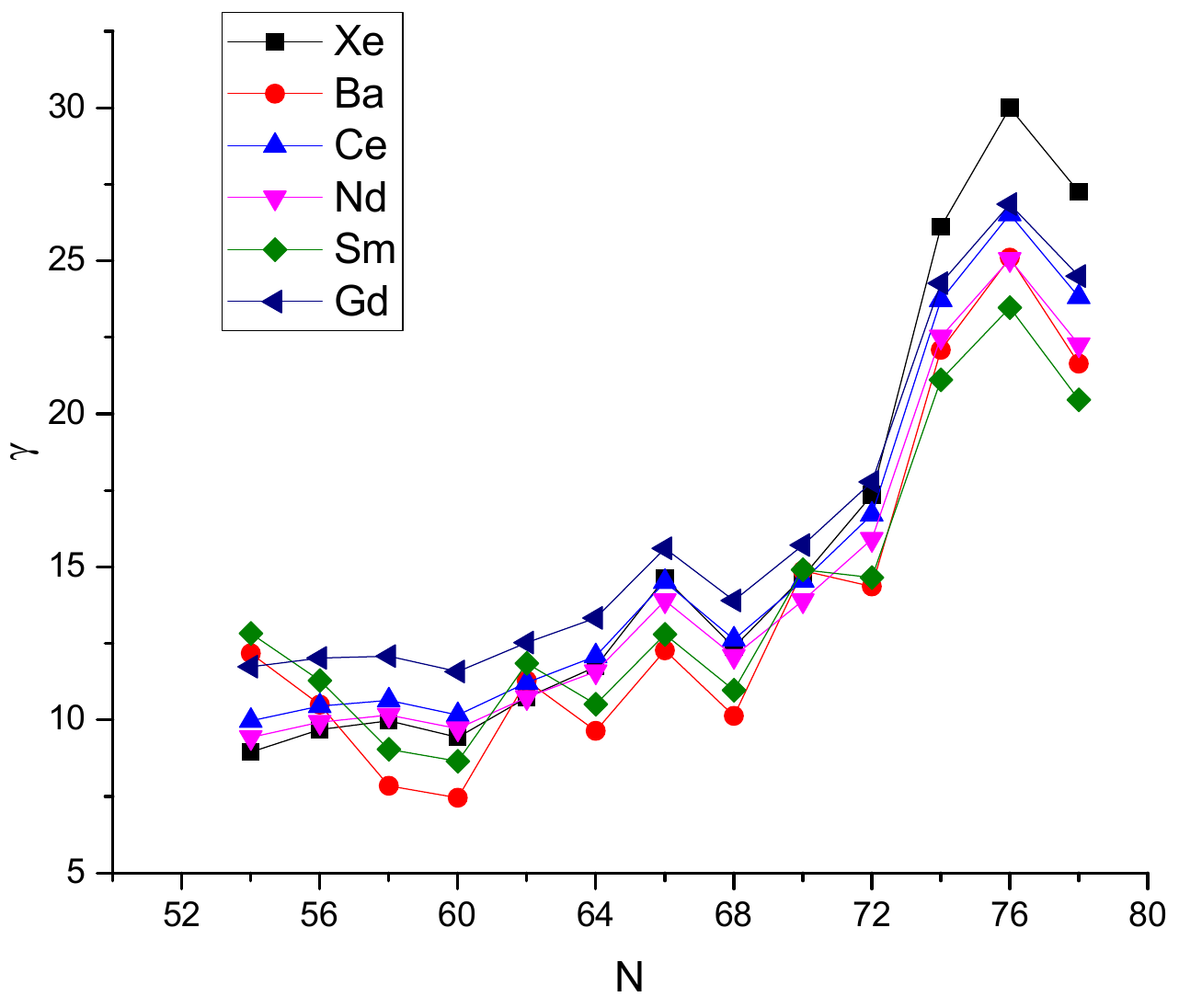}     
     
    \caption{Proxy-SU(3) predictions for the deformation variable $\gamma$ (in degrees) in the region with $Z=54$-64, $N=54$-78, taken from Ref. \cite{Bonatsos2024}.
      See Sec. \ref{global} for further discussion.} 
    
\end{figure}

%%%%%%%%%%%%%%%%%%%%%%%%%%%%%%%%%%%%%%%%%%% FIG. 14  %%%%%%%%%%%%%%%%%%%%%%%%%%%%%%%%%%%%%%%%%%%%%

\begin{figure} [htb]
    
 \includegraphics[width=75mm]{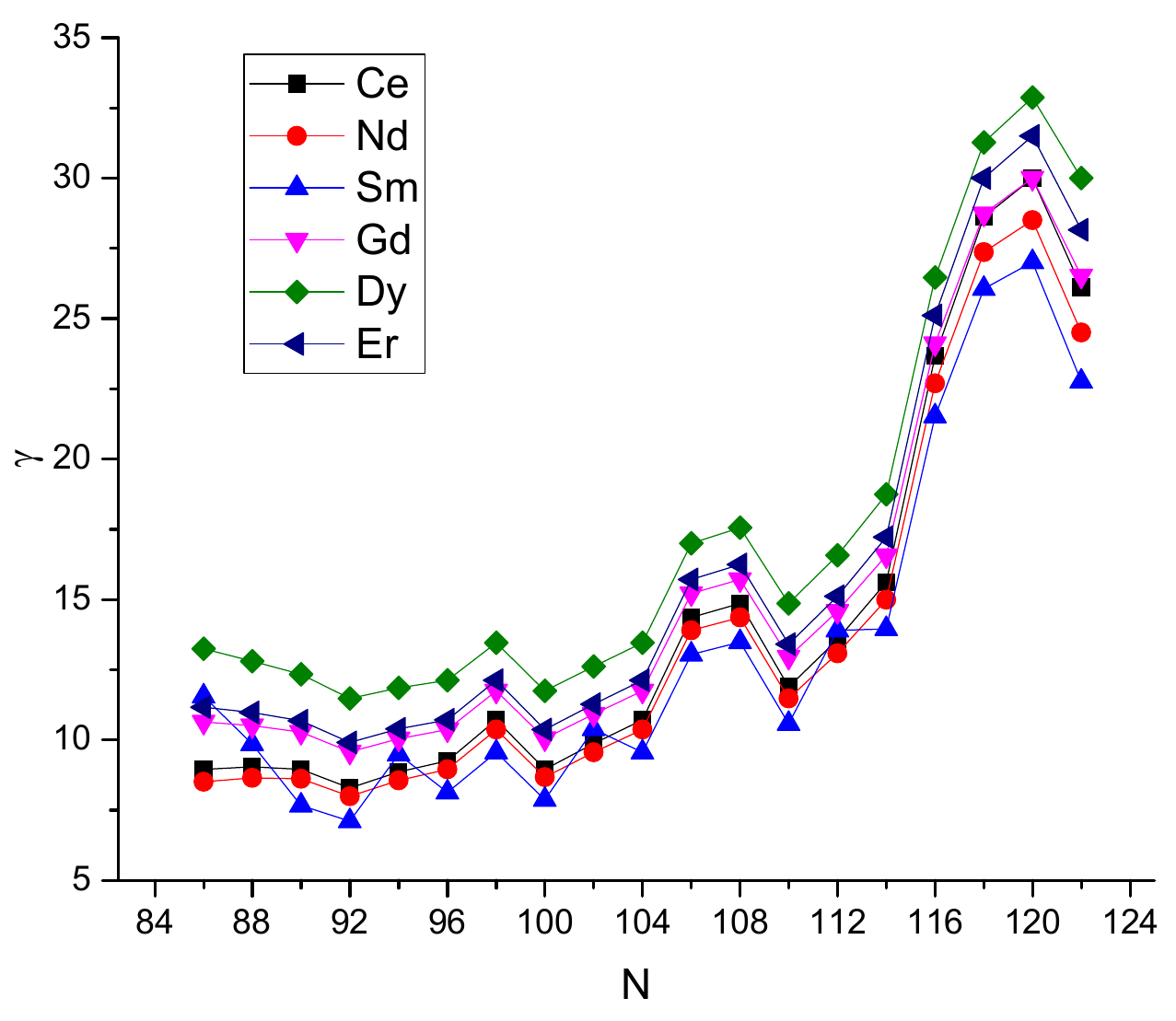}     
     
    \caption{Proxy-SU(3) predictions for the deformation variable $\gamma$ (in degrees) in the region with $Z=58$-68, $N=86$-122, taken from Ref. \cite{Bonatsos2024}.
      See Sec. \ref{global} for further discussion.} 
    
\end{figure}

%%%%%%%%%%%%%%%%%%%%%%%%%%%%%%%%%%%%%%%%%%% FIG. 15  %%%%%%%%%%%%%%%%%%%%%%%%%%%%%%%%%%%%%%%%%%%%%

\begin{figure} [htb]
    
 \includegraphics[width=75mm]{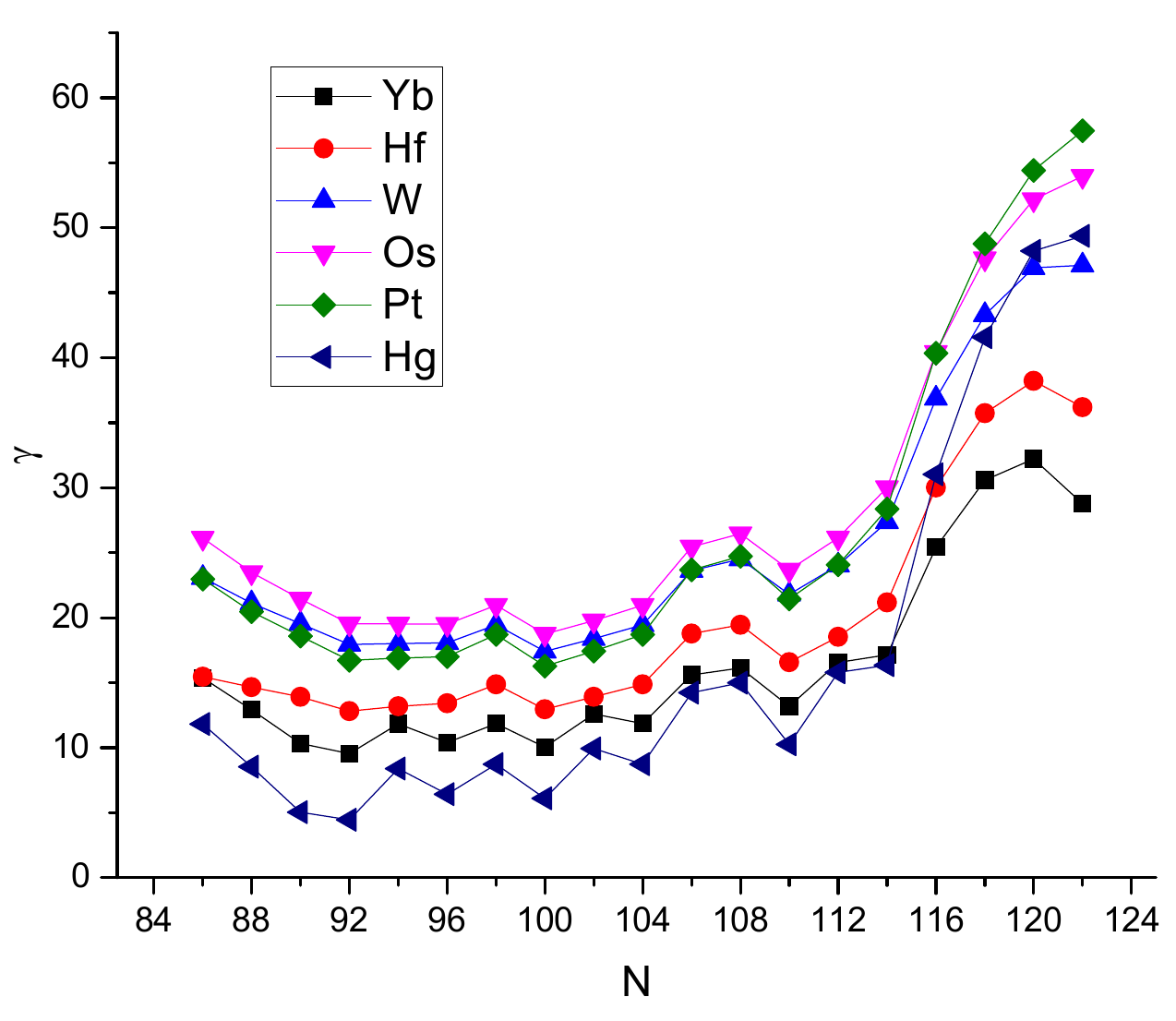}     
     
    \caption{Proxy-SU(3) predictions for the deformation variable $\gamma$ (in degrees) in the region with $Z=70$-80, $N=86$-122, taken from Ref. \cite{Bonatsos2024}. 
      See Sec. \ref{global} for further discussion.} 
    
\end{figure}

{
It should be mentioned that predictions for the collective variable $\gamma$ have also been derived \cite{Bonatsos2020a,Bonatsos2020b} using the pseudo-SU(3) symmetry \cite{RatnaRaju1973,Draayer1982,Draayer1983,Draayer1984,Bahri1992}. It is remarkable that although the pseudo-SU(3) and proxy-SU(3) schemes are based on different approximations to the shell model, discussed in sections \ref{pseudo} and \ref{proxy} respectively, their predictions for the $\beta$ and $\gamma$ collective variables are very similar, once the appropriate h.w. irreps are used in each of them (see Figs. 4 and 8 of \cite{Bonatsos2020a}, as well as Figs. 3 and 4 of \cite{Bonatsos2020b}, for comparisons between the 
pseudo-SU(3) and proxy-SU(3) predictions). In addition, detailed predictions for the values of the $\beta$ and $\gamma$ collective variables have been given using the D1S Gogny interaction \cite{Delaroche2010}. Comparisons between the D1S Gogny and proxy-SU(3) predictions have been carried out in Refs. \cite{Martinou2017a,Martinou2017b,Bonatsos2017c,Bonatsos2017d,Bonatsos2020a}. It is seen that in most cases the proxy-SU(3) predictions are lying close to the ones  provided by the D1S Gogny calculations, lying within the variance determined in the latter (see  Figs. 1 and 2 of  \cite{Martinou2017a}, Figs. 1 and 2 of \cite{Martinou2017b}, Figs. 2, 3, 8, and 9 of \cite{Bonatsos2017c}, Figs. 9 and 11 of \cite{Bonatsos2017d}, Figs. 3 and 7 of \cite{Bonatsos2020a}).  }

{
Since, as seen earlier in this section, the proxy-SU(3) symmetry seems to predict $\gamma$ values consistent with empirical data, it would have been interesting to be able to compare detailed proxy-SU(3) predictions to existing data for spectra and B(E2) transition rates, especially in nuclei in which the presence of triaxiality is experimentally confirmed. However, calculations of spectra \cite{Bonatsos2019} and B(E2)s \cite{Martinou2017a,Martinou2017b,Peroulis2021} within the proxy-SU(3) scheme are still under development. 
A by-pass is provided by detailed calculations of spectra and B(E2)s within the IBM-1 framework, in which the free parameters of IBM-1 are determined through self-consistent mean-field calculations using a Skyrme energy density functional, by forcing the potential energy surfaces of IBM-1 to agree with the microscopically derived ones. Detailed spectra 
and B(E2) transition rates have been provided for $^{162-184}$Hf \cite{Vasileiou2024,Vasileiou2025a}, $^{168-186}$W \cite{Vasileiou2024,Vasileiou2025a} and $^{160-180}$Er \cite{Vasileiou2025}, proving that the addition of triaxiality, as predicted by the proxy-SU(3) symmetry, largely improves the agreement of the calculated IBM-1 spectra and B(E2)s to the data. Extension of these calculations to neighboring series of isotopes, starting with the Yb ones, is in progress \cite{Vasileiou2025b}. It should be remembered, as discussed in Appendix \ref{Sm}, that recent experimental work \cite{Kleemann2025} on $^{154}$Sm, one of the deformed nuclei showing non-vanishing triaxiality according to Monte Carlo Shell Model calculations \cite{Otsuka2023}, determines the collective parameter values to be $\beta=0.2925(25)$ and $\gamma=5.0(15)^{\rm o}$, in close agreement to the proxy-SU(3) predictions \cite{Bonatsos2024} of $\beta=0.296$ and 
$\gamma=4.31^{\rm o}$ (see Table 3 of \cite{Bonatsos2024}), as well as to the Monte Carlo Shell Model predictions of $\beta=0.28$  and $\gamma=3.7^{\rm o}$ \cite{Otsuka2023} (see also Table III of \cite{Kleemann2025}). Detailed calculations of spectra and B(E2)s employing the proxy-SU(3) hw irreps in the framework of the Vector Boson Model have also been started \cite{Minkov2024}, providing an alternative by-pass. }

\section{Open Questions}

In the present review attention has been focused on nuclei exhibiting empirical and/or theoretical values of the collective variable $\gamma$ around $30^{\rm o}$
($15^{\rm o} < \gamma < 45^{\rm o}$), for which in addition the theoretical reduction \cite{Moeller2006,Moeller2008} in the potential energy of the ground state due to triaxiality is significant (larger than 0.01 MeV). As seen in Figs. 5,
{8, and 10}, there is indeed a strong correlation between values of $\gamma$ around $30^{\rm o}$ and substantial reduction in the potential energy of the ground state due to triaxiality. It is therefore safe to refer to these nuclei as triaxial. 

What has not been addressed in detail and should probably attract more attention in future work, is the distinction between rigid triaxiality and $\gamma$-softness. In the former case a potential exhibiting a deep minimum at a $\gamma$ value around  $30^{\rm o}$ is expected to occur, while in the latter case the potential in $\gamma$ is expected to have a shallow minimum, allowing $\gamma$ to obtain different values at minimal expense in energy.

The empirical feature most widely used for making a distinction between rigid triaxiality and $\gamma$-softness is the odd-even staggering within the $\gamma$-band 
\cite{Zamfir1991,McCutchan2007}, as described in subsec. \ref{stagg}. However, the very nature of the $\gamma$ band, as well as of the $\beta$ band, has been in recent years a point of dispute  \cite{Sharpey2008,Sharpey2010,Sharpey2011a,Sharpey2011b,Sharpey2019,Sharpey2023a,Sharpey2023b}. While in the framework of the collective model of Bohr and Mottelson these are considered as bands built on vibrational modes of the $\gamma$ and $\beta$ degrees of freedom respectively \cite{Bohr1998b}, recent work suggests a two-particle--two-hole (2p2h) nature for the $K=0$ ($\beta$) band \cite{Sharpey2023a} and breaking of the axial symmetry, i.e., triaxiality, for the $K=2$ ($\gamma$) band \cite{Sharpey2019,Sharpey2023b}.   

Many $K=0$ bands have been seen experimentally in several nuclei \cite{Aprahamian2002,Lesher2002,Lesher2007,Aprahamian2017,Wirth2004,Meyer2006a,Meyer2006b,Bucurescu2006},
for which various different theoretical interpretations have been suggested \cite{Zamfir2002c,LoIudice2004,LoIudice2005,Gerceklioglu2005,Gerceklioglu2010,Gerceklioglu2012,Hirsch2006,Popa2000,Popa2004,Popa2005,Popa2012}, making the identification of the microscopic mechanism underlying the first excited $K=0$ band in each nucleus a tricky point \cite{Garrett2001}. $K=2$ bands tend to have moments of inertia more similar to these of the ground state bands, but still corrections to this simple picture are required. The need for different mass coefficients in the ground state, $\gamma$, and 
$\beta$ bands has been demonstrated in Refs. \cite{Jolos2007,Jolos2008a,Jolos2008b,Jolos2009,Ermamatov2011,Ermamatov2012}.   

Recent calculations in the framework of the Monte Carlo Shell Model \cite{Otsuka2019,Tsunoda2021,Otsuka2023} and the Triaxial Projected Shell Model \cite{Rouoof2024} favor the interpretation of the $K=2$ bands as due to deviation from axial symmetry, which can be small in certain regions of the nuclear chart, but large in others, as already depicted in {Figs. 4-11}.  

The question of the transition from $\gamma$-softness to rigid triaxial deformation should also be addressed. Is it signifying a shape/phase transition, and of which order? 
$\gamma$-softness is expected to occur from nearly spherical to moderately deformed nuclei, characterized by the O(5) symmetry, which guarantees seniority \cite{Rakavy1957,Bes1959} to be a good quantum number, thus producing odd-even staggering of $\gamma$-soft type in the $\gamma$ bands \cite{McCutchan2007} (see also Table I and Fig. 3 of \cite{Bonatsos2004b}). Indeed, the E(5) critical point symmetry \cite{Iachello2000} characterizing the second-order shape/phase transition from spherical to 
$\gamma$-soft nuclei does have an O(5) subalgebra, guaranteeing the presence of seniority as a good quantum number. The Y(5) shape/phase transition \cite{Iachello2003} in the angle variable $\gamma$ from axially to triaxially deformed nuclei is of second order, while the X(5) shape/phase transition \cite{Iachello2001} from spherical to prolate ($\gamma =0$) axially deformed nuclei is of first order. Introducing non-zero  equilibrium $\gamma$ values in X(5), the T(5) critical point symmetry \cite{Zhang2015a} is obtained. However, X(5) and T(5) represent special approximate solutions of the Bohr Hamiltonian with a yet unknown algebraic structure.    

The apparent disagreement for nuclei beyond $Z=82$, $N=126$ of the theoretical work within various models considered in this review with the predictions of  both the proxy-SU(3) symmetry and the FRDM model should be further investigated, with an eye on its microscopic origin. Beyond mean-field calculations performed recently \cite{Egido2020} using the Gogny interaction
and taking triaxiality into account for the first time, suggest a predominance of triaxial shapes in transitional superheavy nuclei, using the $^{288-296}_{114}$Fl$_{174-182}$ isotopes as a testground. Extension of the proxy-SU(3) calculations, like the ones reported in Ref. \cite{Bonatsos2024}, to the regions of the actinides and the superheavy nuclei are also called for. 

On the other hand, the present study has been limited to nuclei above $Z=22$. Triaxiality is indeed seen in lighter nuclei, with clustering \cite{Freer2018} playing a major role in this region. A detailed study of triaxiality in light nuclei might be an interesting project. 

Finally, the present study has been focused on triaxiality in the ground state band and the $\gamma$-band of even-even nuclei. Higher-lying triaxial bands can occur in odd-odd nuclei by coupling two particles to a triaxial rotor, as first done in the framework of the tilted axis cranking theory \cite{Frauendorf1997}, based on experimental evidence for the odd-odd nucleus $^{134}_{59}$Pr$_{75}$ presented in Ref. \cite{Petrache1996}. 
The role of the rotating mean field of triaxial nuclei in breaking the chiral symmetry has been understood in Ref. \cite{Dimitrov2000}, with the relevant theoretical developments reviewed in Ref. \cite{Frauendorf2001} and the relevant experimental findings for chiral doublet bands in odd-odd and odd nuclei collected in Ref. \cite{Xiong2019}. Evidence for chiral bands in even-even nuclei has been recently observed in the nucleus $^{136}_{60}$Nd$_{76}$ \cite{Petrache2018}, with theoretical approaches with the Particle Rotor Model \cite{Chen2018} and the Triaxial Projected Shell Model \cite{Wang2019} supporting the existence of five possible chiral doublets in it. For chiral bands a separate study would be required.   

\section{Conclusions and Outlook} \label{conclusion}

The main conclusions of the present study are summarized here.

a) Triaxiality in even-even nuclei appears to be present all over the nuclear chart, albeit in a non-uniform way.

b) Values of the collective variable $\gamma$ close to maximal triaxiality ($30^{\rm o}$) appear within specific stripes on the nuclear chart (22-26, 34-48, 74-80, 116-124, 172-182),  predicted by the proxy-SU(3) symmetry in a parameter-free way, based on the Pauli principle and the short-range nature of the nucleon-nucleon interaction. 

c) Calculations within the Finite-Range Droplet Model (FRDM), which is based on completely different assumptions and uses free parameters fitted in order to reproduce basic nuclear data, 
show that the reduction of the potential energy of the ground state due to triaxiality shows non-negligible values (above 0.01 MeV) roughly within the above mentioned stripes, 
thus correlating $\gamma$ values close to maximal triaxiality ($30^{\rm o}$) to deep minima of the potential energy. 

d) The existing few results for $\gamma$ values coming from computationally demanding Monte Carlo Shell Model calculations offer support to the above conclusions. 

e) Calculations performed within a variety of theoretical approaches support the above findings up to $Z=82$, $N=126$. Further work is needed in heavier nuclei, in which discrepancies 
from the above picture might start to appear.   

f) On the microscopic front, values of $\gamma$ close to $30^{\rm o}$ (maximal triaxiality) can be reached in nuclei in which the protons or the neutrons fall within the above-mentioned nucleon intervals favoring triaxiality. Values of $\gamma$ close to $60^{\rm o}$ (oblate shapes) can be reached in nuclei in which both the protons and the neutrons fall within the above-mentioned nucleon intervals favoring triaxiality. 

Points demanding futher investigation include the distinction between rigid triaxiality and $\gamma$-softness, as well as the nature of the transition from one kind of deformation to the other. 
Relatively scarce information exists on triaxiality in the actinides and in the superheavy elements, calling for extension of the above-mentioned studies in these regions.

%%%%%%%%%%%%%%%%%%%%%%%%%%%%%%%%%%%%%%%%%%
\vspace{6pt}

\appendix
\section[\appendixname~\thesection]{Theoretical Terms and Models}
%\subsection[\appendixname~\thesubsection]{}
%\section*{Appendix: Theoretical terms and models}

In this Appendix, a list of  theoretical terms and models used in the text of this review are listed (in alphabetical order), along with the section in which their definition and/or brief description appears.

Algebraic Collective Model (ACM) \hfill \ref{ACM}, \ref{IBM1}

$\beta$ deformation \hfill \ref{RTRM}

BCS approximation \hfill \ref{meanfield}

Coherent State Model (CSM) \hfill \ref{CSM}

collective model of Bohr and Mottelson  \hfill \ref{BohrMottelson}

conformable fractional derivative \hfill \ref{modifiedB}

cranking \hfill \ref{SM}

critical point symmetry (CPS) \hfill \ref{SPT}

cubic terms \hfill \ref{IBM1}

deformation dependent mass (DDM) \hfill \ref{modifiedB}

deformed shell model \hfill \ref{SM}

density functional theory (DFT) \hfill \ref{meanfield}

E(5) CPS \hfill \ref{SPT}

E(5)-$\beta^6$ model \hfill \ref{stagg} 

Elliott SU(3) model \hfill \ref{SU3}

energy-dependent potential \hfill \ref{modifiedB}

extended Thomas-Fermi plus Strutinsky 

$\qquad$ integral method \hfill \ref{Ge}

finite-range droplet  model (FRDM) \hfill \ref{SM}

finite-range liquid-drop model (FRLDM) \hfill \ref{SM}

fractional calculus \hfill \ref{modifiedB}

$\gamma$ deformation \hfill \ref{RTRM}

$\gamma$-unstable model of Wilets and Jean \hfill \ref{RTRM}

generalized collective model (GCM) \hfill \ref{ACM}

generalized triaxial rotor model (GTRM)  \hfill \ref{RTRM}

Gogny interaction \hfill \ref{meanfield}

Hartree--Fock--Bogoliubov (HFB) method \hfill \ref{meanfield}

Hartree--Fock (HF) method \hfill \ref{meanfield}

highest weight irreducible representation 

$\qquad$ (hw irrep) \hfill \ref{proxy}

Interacting Boson Model-1 (IBM-1) \hfill \ref{IBM1}

Interacting Boson Model-2 (IBM-2) \hfill \ref{IBM2}

Interacting Vector Boson Model (IVBM) \hfill \ref{IVBM}

irreducible representation (irrep) \hfill \ref{SU3}

mean field \hfill \ref{meanfield}

minimal length (ML) \hfill \ref{modifiedB}

Monte Carlo Shell Model (MCSM) \hfill \ref{SM}

next highest weight irreducible representation 

$\qquad$ (nhw irrep) \hfill \ref{proxy}

Nilsson model \hfill \ref{SM}

Nilsson--Strutinsky model \hfill \ref{meanfield}

O(5) symmetry \hfill \ref{stagg} 

O(6) symmetry \hfill \ref{IBM1}

odd-even staggering \hfill \ref{stagg}

pairing interaction \hfill \ref{SM} \ref{meanfield}

Pauli principle \hfill \ref{proxy}

projected shell model (PSM) \hfill \ref{SM}

prolate over oblate dominance  \hfill \ref{proxy}

prolate to oblate transition  \hfill \ref{proxy}

proxy-SU(3) symmetry \hfill \ref{proxy}

pseudo-SU(3) symmetry \hfill \ref{pseudo}

quadrupole shape invariants \hfill \ref{IBM1}, \ref{shapepar}

quantum Monte Carlo diagonalization 

$\qquad$ (QMCD) method \hfill \ref{SM}

quantum phase transition (QPT) \hfill \ref{SPT}

relativistic mean field (RMF) \hfill \ref{meanfield}

rigid triaxial rotor model (RTRM)  \hfill \ref{RTRM}

rotation-vibration model (RVM) \hfill \ref{RTRM}

shape/phase transition (SPT)  \hfill \ref{SPT}

Skyrme interaction \hfill \ref{meanfield}

spherical shell model \hfill \ref{SM}

Strutinsky's method \hfill \ref{meanfield}

SU(3) symmetry \hfill \ref{SU3}, \ref{IBM1}

SU(3)$^*$ symmetry \hfill \ref{IBM2}

T(4) SPT \hfill \ref{SPT}

T(5) SPT \hfill \ref{SPT}

triaxial projected shell model (TPSM) \hfill \ref{SM}

triaxial rotation-vibration model (TRVM) \hfill \ref{RTRM}

U(5) symmetry \hfill \ref{IBM1}

U(6) symmetry \hfill \ref{IBM1}

Vector Boson Model (VBM) \hfill \ref{IVBM}

X(3) SPT \hfill \ref{SPT}

X(4) SPT \hfill \ref{SPT}

X(5) CPS \hfill \ref{SPT}

Y(5) SPT \hfill \ref{SPT}

Z(4) SPT \hfill \ref{SPT}

Z(5) SPT \hfill \ref{SPT}

\section{The ${\bf Z=24}$-26 Region} \label{Z2426}

Protons in this region fall within the $Z=22$-26 interval, in which triaxiality is favored according to the proxy-SU(3) predictions (\cite{Bonatsos2025}, see also sec. \ref{regi}). 

\subsection{The Cr (Z=24) Isotopes}

Triaxiality in $^{48}$Cr$_{24}$ has been studied in the cranked Nilsson-Strutinsky framework \cite{Juodagalvis2000}, suggesting the appearance of triaxiality 
in its ground state band at angular momentum $L=8$. Quadrupole shape invariants calculated within the configuration-interaction shell model \cite{Poves2020} suggest substantial deformation for the Cr isotopes with $N=24$, 40, albeit with large fluctuations attached to them (see Table I of \cite{Poves2020}).    

From the above it seems that some signs of triaxiality appear in the Cr isotopes at $N=24$, 40, i.e., they are in agreement with the expectations of the proxy-SU(3) symmetry \cite{Bonatsos2025}, according to which triaxiality is favored in the regions $N=22$-26 and $N=34$-48, as well as in the region $Z=22$-26, to which Cr belongs. 

\subsection{The Fe (Z=26) Isotopes}

$^{56}$Fe$_{30}$ has been among the nuclei for which substantial triaxiality has been suggested since the introduction of the Triaxial Rotor Model \cite{Davydov1960} (see Table 1 of \cite{Davydov1960}. Hartree-Bogoliubov calculations \cite{Girod1978} suggest absence of triaxiality in $^{58}$Fe$_{32}$ (see Table I of \cite{Girod1978}). 

Both nuclei belong to the region $N=28$-32, in which no triaxiality is favored within the proxy-SU(3) symmetry \cite{Bonatsos2025}, but Fe belongs to the $Z=22$-26 region,  in which the proxy-SU(3) symmetry \cite{Bonatsos2025} favors triaxiality. No conclusion can be drawn from this poor set of evidence.  

\section{The ${\bf Z=28}$-32 Region} 

Protons in this region lie in the interval 28-32, extending between the intervals 22-26 and 34-48, in which triaxiality is favored according to the proxy-SU(3) predictions 
 (\cite{Bonatsos2025}, see also sec. \ref{regi}).

% As we shall see below, most nuclei exhibiting triaxiality in this region possess neutrons in the interval 34-48, urging us to call them nuclei presenting neutron-induced triaxiality.

\subsection{The Ni (Z=28) Isotopes}

 Quadrupole shape invariants calculated within the configuration-interaction shell model \cite{Poves2020} suggest substantial deformation for $^{68}$Ni$_{40}$, albeit with large fluctuations attached to it (see Table I of \cite{Poves2020}),  in agreement with the expectations of the proxy-SU(3) symmetry \cite{Bonatsos2025}, according to which triaxiality is favored in the region $N=34$-48.    

\subsection{The Zn (Z=30) Isotopes}

Recent experimental work suggests triaxial deformation in $^{66}$Zn$_{36}$ \cite{Rocchini2021} and $^{72}$Zn$_{42}$ \cite{Hellgartner2023}, while model-independent considerations based on quadrupole shape invariants support triaxiality in $^{70}$Zn$_{40}$ \cite{Poves2020}. Early Hartree-Bogoliubov calculations \cite{Girod1978} support triaxiality in $^{62}$Zn$_{32}$, while high-speed HF calculations with the extended Thomas-Fermi plus Strutinsky integral method \cite{Dutta2000} contradict this suggestion. 

From the above it seems that some signs of triaxiality appear in the Zn isotopes at $N=36$, 40, 42, i.e., within the region $N=32$-48,  in full agreement with the expectations of the proxy-SU(3) symmetry \cite{Bonatsos2025}, according to which triaxiality is favored in the regions $N=34$-48 and $N=74$-80.

It should be mentioned that the reduction in the potential energy of the ground state due to triaxiality within the macroscopic-microscopic approach of FRDM \cite{Moeller2008}, shows non-negligible (above 0.01 MeV) values for $N=36$, 64, 74,  the first value being in agreement with the above-mentioned experimental evidence and with the proxy-SU(3) expectations, and the last value being in agreement with the expectations of the proxy-SU(3) symmetry.

\subsection{The Ge (Z=32) Isotopes} \label{Ge}

Experimental evidence for triaxiality in the Ge isotopes has been seen for $N=40$ \cite{Ayangeakaa2016}, $N=42$ \cite{Sun2014}, and $N=44$ \cite{Toh2013,Ayangeakaa2019}, 
the last case considered as an example of rigid triaxial deformation also within model-independent studies based on quadrupole invariants \cite{Ayangeakaa2019,Poves2020}. 
Empirical evidence for triaxiality for $N=46$ has also been seen by considering the odd-even staggering criterion \cite{Liao1995}.

Mean-field calculations including the high-speed HF method known as the extended Thomas-Fermi plus Strutinsky integral method \cite{Dutta2000}, the Hartree-Bogoliubov 
\cite{Girod1978} and the triaxial HFB with the Gogny D1 interaction \cite{Girod1983} methods have been focused on triaxiality in $^{74}$Ge$_{42}$, while $^{76}$Ge$_{44}$ has been considered \cite{Sugawara2018} within the Generalized Triaxial Rotor Model. Recent Triaxial Projected Shell Model calculations \cite{Jehangir2021,Rouoof2024} have been focused on $^{76}$Ge$_{44}$ (see Table 2 of \cite{Jehangir2021} and Table 3 of \cite{Rouoof2024}), but calculations have also been performed for $N=36$ \cite{Sheikh2016} and $N=38$-48 (see Table I of Ref. \cite{Bhat2014}). The region of $N=32$-44 has also been covered by self-consistent cranked shell model calculations \cite{Shen2011}, while $N=32$ has also been considered in the spherical shell model, as well as in the cranked Nilsson-Strutinsky model \cite{Srivastava2017}. 
The $N=42$, 44 isotopes have been considered as examples of the Z(4)-D \cite{Yigitoglu2018} special solution of the Bohr Hamiltonian.

From the above we conclude that signs of triaxiality appear in the Ge isotopes with $N=32$-48,  in full agreement with the expectations of the proxy-SU(3) symmetry \cite{Bonatsos2025}, according to which triaxiality is favored in the regions $N=34$-48 and $N=74$-80.

It should be mentioned that the reduction in the potential energy of the ground state due to triaxiality within the macroscopic-microscopic approach of FRDM \cite{Moeller2008}, shows non-negligible (above 0.01 MeV) values for $N=32$, 38, 42, 60, 64, 66, 74, 76,  the first values ($N=32$, 38, 42) being in agreement with the above-mentioned studies and with the proxy-SU(3) expectations, and the last values ($N=74$, 76) being in agreement with the expectations of the proxy-SU(3) symmetry. 

\section{The ${\bf Z=34}$-48 Region}

Protons in this region fall within the $Z=34$-48 interval, in which triaxiality is favored according to the proxy-SU(3) predictions (\cite{Bonatsos2025}, see also sec. \ref{regi}).

%  As we shall see, most of the nuclei 
% exhibiting triaxiality in this region have neutrons in the interval 50-72, for which no triaxiality is expected by the proxy-SU(3) symmetry. We could therefore call them nuclei 
% presenting proton-induced triaxiality.

\subsection{The Se (Z=34) Isotopes}

$^{76}$Se$_{40}$ has been among the nuclei considered as possible candidates when the Davydov model was introduced \cite{Davydov1958}. Experimental signs of triaxiality have been seen in $^{74}$Se$_{40}$ \cite{Marchini2023} and $^{76}$Se$_{40}$ \cite{Henderson2019}, while the latter has also been discussed using model-independent shape indicators  based on quadrupole shape invariants \cite{Poves2020}. Triaxiality in $^{74}$Se$_{40}$ and $^{76}$Se$_{40}$ has been studied within the Triaxial Projected Shell Model (TPSM) in Refs. \cite{Jehangir2021,Rouoof2024} (see Table 2 of \cite{Jehangir2021} and Table 3 of \cite{Rouoof2024}), while in Ref. \cite{Bhat2014} the TPSM study of triaxiality has been extended to the Se isotopes with $N=42$-48 (see Table I of \cite{Bhat2014} and in Ref. \cite{Shen2011} the Se isotopes with $N=34$-48 have been considered in the framework of the self-consistent cranked shell model (see Fig. 3 of \cite{Shen2011}). 

The above mentioned work suggests triaxiality in the Se isotopes with $N=42$-48, while  both cranked and configuration-constrained shell model calculations \cite{Xu2002} in the ground state of $^{100}$Se$_{66}$ indicate absence of triaxiality (see Table II of Ref. \cite{Xu2002}). 

From the above we conclude that signs of triaxiality appear in the Se isotopes with $N=42$-48 but not at $N=66$, in full agreement with the expectations of the proxy-SU(3) symmetry \cite{Bonatsos2025}, according to which triaxiality is favored in the regions $N=34$-48 and $N=74$-80.  

It should be mentioned however that, in contrast to the above, the reduction in the potential energy of the ground state due to triaxiality within the macroscopic-microscopic approach of FRDM \cite{Moeller2008}, shows non-negligible (above 0.01 MeV) values for $N=64$, 66, 74, 76, the last values ($N=74$, 76) being in agreement with the expectations of the proxy-SU(3) symmetry.  

\subsection{The Kr (Z=36) Isotopes}

Early self-consistent mean field calculations \cite{Bonche1985} in the Kr isotopes for $N=38$, 40, 54, 62, 64  indicate no triaxial minima (see Figs. 3, 4, 12 in Ref.
\cite{Bonche1985}). The same holds for both cranked and configuration-constrained shell model calculations \cite{Xu2002} in the ground state of $^{102}$Kr$_{66}$ (see Table II of Ref.\cite{Xu2002}). However, odd-even staggering systematics \cite{Liao1995} for the Kr isotopes with $N=42$, 44 show some degree of triaxiality (see Table II of Ref. \cite{Liao1995}). The same holds for more recent HFB calculations with the finite-range Gogny D1S interaction for the Kr isotopes with $N=36$-40 \cite{Girod2009}, as well as with relativistic mean-field plus BCS calculations, employing a five-dimensional collective Hamiltonian \cite{Xiang2016} for $^{76}$Kr$_{40}$ (see Fig. 1 of Ref. \cite{Xiang2016}). 

From the above we conclude that signs of triaxiality appear in the Kr isotopes with $N=36$-40, in agreement with the expectations of the proxy-SU(3) symmetry \cite{Bonatsos2025}, according to which triaxiality is favored in the regions $N=34$-48 and $N=74$-80.  

It should be mentioned however that, in contrast to the above, the reduction in the potential energy of the ground state due to triaxiality within the macroscopic-microscopic approach of FRDM \cite{Moeller2008}, shows non-negligible (above 0.01 MeV) values for $N=52$, 64, 74-78, the last group ($N=74$-78) being in agreement with the expectations of the proxy-SU(3) symmetry.  

\subsection{The Sr (Z=38) Isotopes}

Experimental studies for $^{98}$Sr$_{60}$ indicate a more stable behavior in its yrast band in comparison to its neighboring Mo isotopes, in which triaxiality is seen \cite{Smith1996}. Relativistic mean-field calculations using the D1S-Gogny interaction for the Sr isotopes with $N=58$-62 corroborate the sharp transition from prolate to oblate shapes at $N=60$ \cite{RodriguezGuzman2010b}, in contrast to the Mo isotopes, which do exhibit a region of triaxiality.  Relativistic mean-field plus BCS calculations, employing a five-dimensional collective Hamiltonian \cite{Xiang2016}, show that in $^{98}$Sr$_{60}$ a triaxial barrier still exists between the prolate and the oblate minimum in the potential energy surface, preventing triaxiality to appear as in the Mo case.  Both cranked and configuration-constrained shell model calculations \cite{Xu2002} corroborate the absence of triaxial deformation in the ground state of $^{104}$Sr$_{66}$ (see Table II of \cite{Xu2002}). 

Early self-consistent mean field calculations \cite{Bonche1985} in the Sr isotopes for $N=36$-64 indicate no triaxial minima.
Mean-field calculations employing the Skyrme interaction for the Sr isotopes with $N=40$-48 \cite{Heenen1993} find no signs of triaxiality in the ground state band, 
but only in the first excited band of $^{82}$Sr$_{46}$ (see Fig. 3 of \cite{Heenen1993}). 
 
The reduction in the potential energy of the ground state due to triaxiality within the macroscopic-microscopic approach of FRDM \cite{Moeller2008},
shows non-negligible (above 0.01 MeV) values only for $N=32$.

In conclusion, almost no signs of triaxiality are seen in the Sr isotopes.  

\subsection{The Zr (Z=40) Isotopes}

Experimental studies in the Zr isotopes with $N=60$-64 indicate a more stable behavior in their yrast bands in comparison to their neighboring Mo isotopes, in which triaxiality is seen \cite{Smith1996}. In corroboration of these observations, triaxial shapes in the Zr isotopes with $N=60$, 62 are seen experimentally only in excited bands \cite{Urban2019}. Relativistic mean-field calculations using the D1S-Gogny interaction corroborate the sharp transition from prolate to oblate shapes at $N=60$ \cite{RodriguezGuzman2010b}, in contrast to the Mo isotopes, which do exhibit a region of triaxiality. The disappearance of the triaxial barrier in $^{100}$Zr$_{60}$ is also seen in relativistic mean-field plus BCS calculations, employing a five-dimensional collective Hamiltonian \cite{Xiang2016}. Both cranked and configuration-constrained shell model calculations \cite{Xu2002} corroborate the absence of triaxial deformation in the ground state of $^{106}$Zr$_{66}$ (see Table II of Ref. \cite{Xu2002}). 

Early self-consistent mean field calculations \cite{Bonche1985} in the Zr isotopes for $N=36$-64 indicate triaxial minima for $N=44$ (see Fig. 12 of Ref. \cite{Bonche1985}) and 54 (see Fig. 3 of Ref. \cite{Bonche1985}). However, more recent calculations using the Strutinsky shell correction method in the $N=52$-58 Zr isotopes \cite{Nayak2022} 
indicate a triaxial band in $^{94}$Zr$_{54}$, not far from the ground state. As a consequence, only  $^{84}$Zr$_{44}$ remains as a candidate for triaxiality. 

Extended Thomas-Fermi plus Strutinsky calculations (which are a high-speed approximation to the HF method) for the Zr isotopes with $N=70$-78 \cite{Dutta2000} indicate triaxial shapes in the ground states of $^{112}$Zr$_{72}$ and  $^{114}$Zr$_{74}$ (see Table III of Ref. \cite{Dutta2000}). 

The reduction in the potential energy of the ground state due to triaxiality within the macroscopic-microscopic approach of FRDM \cite{Moeller2008},
shows non-negligible (above 0.01 MeV) values only for $N=32$ and 74, the latter being in agreement with the findings of Ref. \cite{Dutta2000}.

In conclusion, rather clear signs of triaxiality in the Zr isotopes are seen for $N=44$ and $N=74$. These are in agreement with the expectations of the proxy-SU(3) symmetry \cite{Bonatsos2025}, according to which triaxiality is favored in the regions $N=34$-48 and $N=74$-80, to which these values belong.   

\subsection{The Mo (Z=42) Isotopes} 

Experimental studies of triaxiality in the Mo isotopes have been focused on the region with $N=58$-66 \cite{Smith1996,Snyder2013,Ralet2017,Ha2020}. Theoretical studies with the Triaxial Projected Shell Model have been focused on $N=62$-66 \cite{Zhang2015b,Jehangir2021,Rouoof2024}. The same holds for mean-field calculations employing the Gogny \cite{RodriguezGuzman2010b}  and Skyrme \cite{Zhang2015b} interactions, while cranked shell model calculations have been focused on $^{108}$Mo$_{66}$ \cite{Xu2002} and calculations employing the Strutinsky method on $^{106}$Mo$_{64}$ \cite{Dutta2000}. Recent relativistic mean-field calculations \cite{Abusara2017a,ElBassem2024} cover a wider region, extending over $N=50$-68, earlier considered within constrained HF calculations \cite{Bonche1985}, while the lighter isotopes with $N=44$-58 have been considered by calculations employing the Strutinsky method \cite{Luo1993,Nayak2022}. In summary, triaxiality is seen mostly in the Mo isotopes with $N=60$-66, although signs of triaxiality are seen within a wider neutron region.     

In the macroscopic-microscopic approach of FRDM \cite{Moeller2008}, the reduction in the potential energy of the ground state due to triaxiality has been calculated. The reduction shows a maximum value of 0.30 MeV at $N=66$, while substantial reductions have been obtained in most of the Mo isotopes with $N=54$-74 (see Table 1 of Ref. \cite{Moeller2008}).

In conclusion, substantial triaxiality is seen in the Mo isotopes in an extended neutron region ($N=54$-74), in agreement to the expectations of the proxy-SU(3) symmetry \cite{Bonatsos2025}, according to which triaxiality is favored in the region $Z=34$-48, to which Mo belongs.  

\subsection{The Ru (Z=44) Isotopes} 

The Ru isotopes with $N=56$, 58, 60 were among the triaxial nuclei considered when the Davydov model was introduced \cite{Davydov1958}. $^{104}$Ru$_{60}$ has been used as the textbook example of the SU(3)$^*$ symmetry within IBM-2 \cite{Dieperink1982,Dieperink1983,Dieperink1984,Dieperink1985}. It has also been described in the framework of IBM-1, exhibiting the need for inclusion of higher order (cubic) terms for accommodating triaxiality within IBM-1 \cite{Heyde1984}. Its triaxiality has also been considered within the relativistic mean-field plus BCS \cite{Xiang2016} and triaxial projected shell model \cite{Nazir2023} frameworks, as well as examined through recent lifetime measurements \cite{Esmaylzadeh2022} in which $^{106}$Ru$_{62}$ has also been considered. $^{106}$Ru$_{62}$ has been studied in the Davydov model framework in Ref. \cite{Varshni1970}.
Triaxiality revealed by the odd-even staggering in the $\gamma$ bands of $^{104-108}$Ru$_{60-64}$ has been studied within the Vector Boson Model framework in Ref. \cite{Lalkovski2005}. 
Lighter Ru isotopes with $N=54$-60 have been studied in the macroscopic-microscopic formalism using the Strutinsky method in Ref. \cite{Nayak2022}, 
as well as within special solutions of the Bohr Hamiltonian involving the sextic potential (Z(4)-sextic \cite{Budaca2016b}) and the hyperbolic P\"{o}schl-Teller potential 
\cite{Naderi2017}, while odd-even staggering in $^{98}$Ru$_{54}$ has been considered in Ref. \cite{Liao1995}. In summary, triaxiality is seen in $^{104}$Ru$_{60}$ and in a few isotopes around it. 

Among heavier Ru isotopes, $^{112}$Ru$_{68}$ is one of the few nuclei in which the odd-even staggering quantity of Eq. (\ref{SL}) clearly exhibits rigid triaxiality
\cite{McCutchan2007}. Experimental signs of triaxiality in the Ru isotopes with $N=64$-70 have been seen in \cite{Shannon1994} ($^{108-114}$Ru$_{64-70}$), \cite{Snyder2013} ($^{108-112}$Ru$_{64-68}$), \cite{Doherty2017} ($^{110}$Ru$_{66}$), while for the next heavier isotopes ($^{116,118}$Ru$_{72,74}$) triaxiality has been observed in Ref. \cite{Soderstrom2013}. Triaxiality in $^{108-112}$Ru$_{64-68}$ has been phenomenologically interpreted in terms of IBM-1 including cubic terms \cite{Sorgunlu2008}, IBM-2 by distinguishing protons from neutrons \cite{Duarte1998}, as well as within the Generalized Triaxial Rotor Model \cite{Sugawara2019,Allmond2017}, while triaxiality in $^{110-114}$Ru$_{66-70}$ has been considered in the framework of the Z(4)-D \cite{Yigitoglu2018} special solution of the Bohr Hamiltonian. In the realm of microscopic calculations, triaxiality in $^{108-114}$Ru$_{64-70}$ has been considered within the Triaxial Projected Shell Model \cite{Zhang2015b,Jehangir2021,Rouoof2024}, the cranked shell model \cite{Xu2002}, and the cranked HFB approach \cite{Schuck2019}, while density functional theory has been employed in \cite{Zhang2015b} and \cite{Abusara2017a}, the latter calculation covering the Ru isotopes with $N=52$-68. In summary, triaxiality is seen in the Ru isotopes with $N=64$-70.

In the macroscopic-microscopic approach of FRDM \cite{Moeller2006,Moeller2008}, the reduction in the potential energy of the ground state due to triaxiality has been calculated. The reduction shows a maximum value of 0.63 MeV at $N=64$, while substantial reductions have been obtained in most of the Ru isotopes with $N=58$-76 (see Table 1 of Ref. \cite{Moeller2008}).

In conclusion, substantial triaxiality is seen in the Ru isotopes in an extended neutron region ($N=58$-74), in agreement to the expectations of the proxy-SU(3) symmetry \cite{Bonatsos2025}, according to which triaxiality is favored in the region $Z=34$-48, to which Ru belongs.  

\subsection{The Pd (Z=46) Isotopes} 

The Pd isotopes with $N=60$, 62 have been among the nuclei considered when the Davydov model was introduced \cite{Davydov1958}. They have been further studied by quadrupole shape invariants through the introduction of a model-independent cubic shape parameter \cite{Werner2005}, as well as in the IBM-2 framework in terms of the SU(3)$^*$ symmetry \cite{Dieperink1984}. Triaxiality revealed by the odd-even staggering in the $\gamma$ bands of $^{108-112}$Pd$_{62-66}$ has been studied within the Vector Boson Model framework in Ref. \cite{Lalkovski2005}. 
Triaxiality in lighter Pd isotopes with $N=56$, 58 has been considered with the Strutinsky approach \cite{Nayak2022}, as well as within the Triaxial Projected Shell Model \cite{Naz2018}, while the isotopes with $N=58$-62 have also been considered within the Z(4)-sextic \cite{Budaca2016b} special solution of the Bohr Hamiltonian.  

Heavier Pd isotopes with $N=66$-70 have been studied in IBM-1 using cubic terms \cite{Sorgunlu2008}, and by the conformable fractional Bohr Hamiltonian using the Kratzer potential \cite{Hammad2021c}, as well as the Morse, Tietz-Hua and multi-parameter exponential-type potentials \cite{Hammad2023}. The $N=62$-70 isotopes have been considered within the Z(4)-DDM-D \cite{Buganu2017} special solution of the Bohr Hamiltonian, while $^{114}$Pd$_{68}$ has been studied within the Z(4)-DDM and Z(4)-ML \cite{AitElKorchi2020} special solutions of the Bohr Hamiltonian. Cranked shell model calculations have also been performed for $N=66$ \cite{Xu2002}. 

These two regions have been bridged by recent relativistic Hartree-Bogoliubov calculations using the DD-ME2 energy density functional for N$=50$-68 \cite{Naz2018}.    

The reduction in the potential energy of the ground state due to triaxiality within the macroscopic-microscopic approach of FRDM \cite{Moeller2008},
shows a maximum value of 0.32 MeV at $N=64$, while substantial reductions have been obtained in most of the Pd isotopes with $N=62$-76 (see Table 1 of Ref. \cite{Moeller2008}).

In conclusion, substantial triaxiality is seen in the Pd isotopes in an extended neutron region ($N=56$-74), in agreement to the expectations of the proxy-SU(3) symmetry \cite{Bonatsos2025}, according to which triaxiality is favored in the region $Z=34$-48, to which Pd belongs.

\subsection{The Cd (Z=48) Isotopes}

The Cd isotopes with $N=62$-68 have been considered already when the Davydov model was introduced \cite{Davydov1958,Davydov1960}. They have also been considered through quadrupole shape invariants, leading to model-independent shape parameters \cite{Werner2005}. The isotopes with $N=60$, 62 have also been considered within the Z(4)-sextic \cite{Budaca2016b} special solution of the Bohr Hamiltonian.  

Lighter Cd isotopes with $N=56$, 58 have been considered by the Strutinsky shell-correction method \cite{Nayak2022}, while $^{106}$Cd$_{58}$ has been recently studied experimentally in Ref. \cite{Gray2022}.

These two regions have recently been bridged by Triaxial Projected Shell Model calculations for $N=56$-74 \cite{Rajput2022}, resulting in $\gamma$ values ranging from 
32$^{\rm o}$ to 40$^{\rm o}$ (see Table 1 of Ref. \cite{Rajput2022}).    

No significant (larger than 0.01 MeV) reduction in the potential energy of the ground state due to triaxiality has been seen in the macroscopic-microscopic approach of FRDM \cite{Moeller2008} for the Cd isotopes. 

In conclusion, some weak evidence for triaxiality exists in the Cd isotopes in the $N=56$-74 region, in agreement to the expectations of the proxy-SU(3) symmetry \cite{Bonatsos2025}, according to which triaxiality is favored in the region $Z=34$-48, of  which Cd marks the upper border.

\section{The ${\bf Z=52}$-66 Region}

Protons in this region lie in the interval 50-72, extending between the intervals 34-48 and 74-80, in which triaxiality is favored according to the proxy-SU(3) predictions (\cite{Bonatsos2025}, see also sec. \ref{regi}). The nuclei up to the $Z=66$ midshell of the 50-82 major shell are considered in this section, while the nuclei beyond the midshell are considered in the next section. 

% As we shall see below, most nuclei exhibiting triaxiality in this region possess neutrons in the intervals 74-80 and 116-124, urging us to call them nuclei presenting neutron-induced 
% triaxiality.

\subsection{The Te (Z=52) Isotopes}

$^{122,126}$Te$_{70,74}$ have been among the nuclei suggested as candidates for substantial triaxiality when the Triaxial Rotor Model was introduced \cite{Davydov1958,Davydov1959} (see Tables 3, 5 of \cite{Davydov1958} and Table 5 of \cite{Davydov1959}). However, Woods-Saxon-Strutinsky calculations \cite{Kern1987}
suggest the absence of triaxiality in the Te isotopes in the region $N=60$-72, leaving only  $^{126}$Te$_{74}$ as an undisputed suggestion.  

The reduction in the potential energy of the ground state due to triaxiality within the macroscopic-microscopic approach of FRDM \cite{Moeller2008},
shows non-negligible (above 0.01 MeV) values for Te isotopes with $N=114$, 120, for which no experimental information exists \cite{ensdf}.

In conclusion, the only suggestion for substantial triaxiality in the Te isotopes has been made for $N=74$, in agreement with the expectations of the proxy-SU(3) symmetry \cite{Bonatsos2025}, according to which  triaxiality is favored in the interval 74-80.
 
\subsection{The Xe (Z=54) Isotopes}

$^{128}$Xe$_{74}$ is among the nuclei suggested as possible candidates for triaxiality when the Triaxial Rotor Model was introduced \cite{Davydov1958}, a suggestion corroborated later by detailed  experimental studies \cite{Orce2006}, as well as by relativistic mean-field (RMF) calculations employing the PC-PK1 energy density functional
\cite{Nomura2021a,Nomura2021b} (see Fig. 1 of \cite{Nomura2021a} and Table I of \cite{Nomura2021b}). These RMF calculations \cite{Nomura2021a,Nomura2021b} predict 
triaxiality in $^{128}$Xe$_{74}$  but not in $^{130}$Xe$_{76}$ (see Fig. 1 of \cite{Nomura2021a} and Table I of \cite{Nomura2021b}), in contrast to relativistic HFB calculations using the DD-ME2 effective interaction \cite{Naz2018}, which predict  triaxiality in both of these isotopes (see Table 1 of \cite{Naz2018}),  but no triaxiality in the heavier isotopes with $N=78$-86 \cite{Naz2018} (see Table 1 of \cite{Naz2018}), again in contrast to Triaxial Projected Shell Model calculations \cite{Naz2018} predicting substantial triaxiality in the Xe isotopes with $N=78$, 80 (see Table 3 of \cite{Naz2018}).  

Triaxiality within the wider region of Xe isotopes with $N=70$-80 has been considered within several theoretical frameworks involving the Bohr-Mottelson collective model. Studies within the Triaxial Rotor Model have suggested triaxiality for the Xe isotopes in the region $N=68$-78 \cite{Yan1993} (see Table I of \cite{Yan1993}), while weaker signs of triaxiality have been seen in the region $N=64$-78 using the Bohr Hamiltonian with a cubic term \cite{Rohozinski1974} (see Table 1 of \cite{Rohozinski1974}). The Xe isotopes with $N=74$-78 have been considered within special solutions of the Bohr Hamiltonian involving the Davidson potential \cite{Yigitoglu2011} (see Table III of \cite{Yigitoglu2011}) and the inverse square potential \cite{Ajulo2022}, while the wider region with $N=72$-82 has been studied using the Hulth\'{e}n potential \cite{Chabab2015}. The Xe isotopes with $N=74$-78 have also been considered within the Z(4) special solution of the Bohr Hamiltonian \cite{Bonatsos2005} (see Fig. 3 of \cite{Bonatsos2005}), which is a $\gamma$-rigid version of the Z(5) critical point symmetry \cite{Bonatsos2004}, involving an infinite square well potential, as well as with several variations of Z(4),  involving the Davidson \cite{Yigitoglu2018} and sextic \cite{Buganu2015,Buganu2015b,Budaca2016b} potentials, as well as the Davidson \cite{Buganu2017} and Kratzer \cite{AitElKorchi2022} with deformation-dependent mass (DDM). In addition, a modified version of Z(4), in which the value of $\gamma$ is allowed to deviate from $30^{\rm o}$ has been introduced \cite{Ajulo2024}.  The Xe isotopes with $N=72$-76 have also been considered within the conformable fractional Bohr Hamiltonian including a Kratzer potential \cite{Hammad2021c}, as well as the Morse, Tietz-Hua, and various exponential-type potentials \cite{Hammad2023}. Furthermore, triaxiality in $^{126}$Xe$_{72}$ has been considered within the Triaxial Rotation Vibration Model \cite{Meyer1997,Meyer1998}. 

The Xe isotopes with $N=70$-78 have been considered as a textbook example of manifestation of the O(6) dynamical symmetry of the Interacting Boson Model \cite{Casten1985a,Casten1985b}, in which triaxiality is taken into account by including cubic terms in IBM-1 \cite{Casten1985a,Casten1985b,Sorgunlu2008}, or terms inspired by the SU(3)$^*$  dynamical symmetry in IBM-2 \cite{Sevrin1987b}. The odd-even staggering in the $\gamma$-bands suggests a $\gamma$-soft behavior \cite{Sorgunlu2008}
(see Fig. 8 of \cite{Sorgunlu2008}). 

Substantial triaxiality has been seen in the Xe isotopes with $N=70$, 72 through model-independent measures of triaxiality based on quadrupole shape invariants \cite{Werner2001} (see Table 7 of \cite{Werner2001}), as well as with $N=72$-76 through Woods-Saxon-Strutinsky calculations \cite{Kern1987}, which find no triaxiality in the ligher Xe isotopes with $N=60$-70.  $^{124}$Xe$_{70}$ has been used as the test-ground for modified versions of the X(3) \cite{Bonatsos2006} special solution of the Bohr Hamiltonian, allowing for deformation-dependent mass (X(3)-DDM, \cite{AitElKorchi2020}), as well as for minimal length (ML) formalism (X(3)-ML, 
\cite{AitElKorchi2020}), while the Xe isotopes with $N=64$-74 have been used as a test-ground for a special solution of the Bohr Hamiltonian involving the Kratzer potential with the coupling constant of the hyperbolic term depending explicitly on energy \cite{Budaca2020}. 

The reduction in the potential energy of the ground state due to triaxiality within the macroscopic-microscopic approach of FRDM \cite{Moeller2008},
shows non-negligible (above 0.01 MeV) values for Xe isotopes with $N=64$, 74, 76, 120. No experimental information exists \cite{ensdf} for $^{174}$Xe$_{120}$.

In conclusion, evidence for substantial triaxiality is seen in the Xe isotopes in the region ($N=72$-78), in agreement with the FRDM calculations \cite{Moeller2008} suggesting triaxiality for the Xe isotopes with $N=74$, 76, as well as with the expectations of the proxy-SU(3) symmetry \cite{Bonatsos2025}, according to which  triaxiality is favored in the interval 74-80.
 In contrast, no evidence for triaxiality is seen in the Xe isotopes in the neutron region beyond $N=80$, in agreement with the FRDM calculations \cite{Moeller2008}, as well as with the expectations of the proxy-SU(3) symmetry \cite{Bonatsos2025}, according to which  triaxiality is favored neither in the region $N=82$-114, nor in $Z=50$-72, to which Xe belongs.

\subsection{The Ba (Z=56) Isotopes}

Evidence for substantial triaxiality in the Ba isotopes has been provided for $^{130}$Ba$_{74}$ within the Triaxial Rotation Vibration Model \cite{Meyer1997,Meyer1998}, as well as within Woods-Saxon-Strutinsky calculations \cite{Kern1987} (see Table I of \cite{Kern1987}). For  $^{132}$Ba$_{76}$  substantial triaxiality has been proposed by extended Thomas-Fermi plus Strutinsky integral calculations \cite{Dutta2000} (see Table II of \cite{Dutta2000}), as well as by relativistic Hartree-Bogoliubov calculations using the DD-ME2 effective interaction \cite{Naz2018} (see Table 1 of \cite{Naz2018}), while no triaxiality is suggested by early Hartree-Bogoliubov calculations \cite{Girod1978} (see Table I of \cite{Girod1978}). Triaxial Projected Shell Model calculations indicate substantial triaxiality in $^{134}$Ba$_{78}$, which has also been considered as the textbook example \cite{Casten2000} of the E(5) critical point symmetry \cite{Iachello2000}, the special solution of the Bohr Hamiltonian corresponding to the second-order shape/phase transition (SPT) from spherical to $\gamma$-unstable shapes. This SPT in $^{134}$Ba$_{78}$ has also been discussed within the IBM-1 framework including cubic and sextic terms \cite{Jolos2004a,Jolos2004b} (see Tables 1, 2 of \cite{Jolos2004a} and Tables 7, 8 of \cite{Jolos2004b}), while $^{134}$Ba$_{78}$ has also been used as an example of the SU(3)$^*$ symmetry within IBM-2 \cite{Dieperink1984}. 

The Ba isotopes in the region $N=72$-78 have been considered as textbook examples of manifestation of the O(6) dynamical symmetry of IBM \cite{Casten1985a,Casten1985b}, with triaxiality accommodated within the IBM-1 framework through the inclusion of cubic terms \cite{Casten1985a,Casten1985b,Sevrin1987b,Sorgunlu2008} (see Fig. 9 of \cite{Sorgunlu2008}), while in IBM-2 triaxiality has been described by using terms related to the SU(3)$^*$ symmetry \cite{Sevrin1987b}.
 A mapping of the triaxial rotor onto the IBM has also been successfully used for  $^{128}$Ba$_{72}$ \cite{Zhang2014}. 

The Bohr Hamiltonian including cubic terms \cite{Rohozinski1974} has also been used for considering triaxiality of the Ba isotopes in the region $N=68$-78  (see Table 1 of \cite{Rohozinski1974}), along with the Triaxial Rotor Model \cite{Varshni1970,Yan1993} (see Table 1 of \cite{Varshni1970} and Table I of \cite{Yan1993}).
The Triaxial Rotor Model has also been used for the heavier Ba isotopes with $N=86$-90 \cite{Bindra2018} (see Table 1 of \cite{Bindra2018}). 

The reduction in the potential energy of the ground state due to triaxiality within the macroscopic-microscopic approach of FRDM \cite{Moeller2008},
shows non-negligible (above 0.01 MeV) values for Ba isotopes with $N=64$, 76, 78, 120, 122. No experimental information exists \cite{ensdf} for $^{176,178}$Ba$_{120,122}$.

In conclusion, evidence for substantial triaxiality is seen in the Ba isotopes in the region ($N=72$-78), in agreement with the FRDM calculations \cite{Moeller2008} suggesting triaxiality for the Ba isotopes with $N=76$, 78, as well as with the expectations of the proxy-SU(3) symmetry \cite{Bonatsos2025}, according to which  triaxiality is favored in the interval 74-80.
 In contrast, very little evidence for triaxiality is seen in the Ba isotopes in the neutron region $N=86$-90, for which the FRDM calculations \cite{Moeller2008} suggest that the triaxial minima should be relatively shallow. This finding is in rough agreement with the expectations of the proxy-SU(3) symmetry \cite{Bonatsos2025}, according to which  triaxiality is favored neither in the region $N=82$-114, nor in $Z=50$-72, to which Ba belongs.

\subsection{The Ce (Z=58) Isotopes}

Triaxiality in $^{134}$Ce$_{76}$ has been the focus of several investigations, including mean-field calculations by extended Thomas-Fermi plus Strutinksy integral \cite{Dutta2000} (see Table II of \cite{Dutta2000}), Hartree-Bogoliubov \cite{Girod1978} (see Table I of \cite{Girod1978}), 3d-cranked HFB \cite{Oi2003}, and HFB with Gogny D1 interaction \cite{Girod1983} (see Fig. 11 of \cite{Girod1983}) methods, as well as by quadrupole shape invariants within IBM-1 \cite{Werner2001} (see Table 7 of \cite{Werner2001}).  Additional studies, focused on triaxiality in Ce isotopes with $N=70$-80 have been performed within the Triaxial Rotor Model for $N=70$-78 \cite{Varshni1970,Yan1993} (see Table 1 of \cite{Varshni1970} and Table I of \cite{Yan1993}), the Triaxial Projected Shell Model \cite{Sheikh2009,Sheikh2016} (see Table 1 of \cite{Sheikh2009} and Table 1 of \cite{Sheikh2016}), as well as with the Bohr Hamiltonian including a cubic term \cite{Rohozinski1974} (see Table 1 of \cite{Rohozinski1974}), finding substantial evidence for triaxiality in this region ($N=70$-80). A special triaxial solution of the Bohr Hamiltonian with a Davidson potential and deformation-dependent mass (Z(4)-DDM-D, \cite{Buganu2017}) has also been used for the Ce isotopes with $N=78$-80. In contrast, Woods-Saxon-Strutinsky calculations \cite{Kern1987} predict substantial triaxiality in the Ce isotopes $N=74$-80, but not in $N=60$-72. 

Beyond $N=80$, $^{148}$Ce$_{90}$ has been suggested  as a candidate for the T(5) critical point symmetry \cite{Zhang2015a}, related to the shape/phase transition from spherical to triaxially deformed shapes (see Table 2 of \cite{Zhang2015a}). Furthermore, triaxiality in the Ce isotopes with $N=86$-90 has been predicted within the Triaxial Rotor Model \cite{Bindra2018} (see Table 1 of \cite{Bindra2018}).  

The reduction in the potential energy of the ground state due to triaxiality within the macroscopic-microscopic approach of FRDM \cite{Moeller2008},
shows non-negligible (above 0.01 MeV) values for Ce isotopes with $N=76$-80, 122. No experimental information exists \cite{ensdf} for $^{180}$Ce$_{122}$.

In conclusion, evidence for substantial triaxiality is seen in the Ce isotopes in an extended neutron region ($N=70$-80), in agreement with the FRDM calculations \cite{Moeller2008} suggesting triaxiality for the Ce isotopes with $N=76$-80, as well as with the expectations of the proxy-SU(3) symmetry \cite{Bonatsos2025}, according to which  triaxiality is favored in the interval 74-80.
 In contrast, very little evidence for triaxiality is seen in the Ce isotopes in the neutron region $N=84$-88, for which the FRDM calculations \cite{Moeller2008} suggest that the triaxial minima should be relatively shallow. This finding is in rough agreement with the expectations of the proxy-SU(3) symmetry \cite{Bonatsos2025}, according to which  triaxiality is favored neither in the region $N=82$-114, nor in $Z=50$-72, to which Ce belongs.

\subsection{The Nd (Z=60) Isotopes}

$^{150}$Nd$_{90}$ has been among the nuclei suggested as candidates for moderate triaxiality since the early days the Triaxial Rotor Model \cite{Varshni1970,Allmond2017} (see Table 1 of \cite{Varshni1970} and Table 1 of \cite{Allmond2017}), as well as within cranked HFB calculations \cite{Schuck2019} (see Table I of \cite{Schuck2019}. It is also one of the best examples \cite{Kruecken2002} of manifestation of the X(5) critical point symmetry \cite{Iachello2001}, related to the shape/phase transition from spherical to prolate deformed shapes. Triaxiality in Nd isotopes around $^{150}$Nd$_{90}$ has been suggested by Triaxial Rotor Model calculations \cite{Bindra2018} in $N=84$-90 (see Table 1 of \cite{Bindra2018}), as well as within Triaxial Projected Shell Model calculations \cite{Naz2018} in $N=90$, 94 (see Table 3 of \cite{Naz2018}). 
In contrast, relativistic HB calculations for Nd isotopes using the DD-ME2 parametrization \cite{Naz2018} do not find any triaxiality in the $N=82$-96 region (see Table 2 of \cite{Naz2018}). 

On the other hand, substantial triaxiality in the Nd isotopes with $N=72$-78 has been suggested \cite{Sheikh2009,Sheikh2016} by Triaxial Projected Shell Model calculations 
(see Table 1 of \cite{Sheikh2009} and Table 1 of \cite{Sheikh2016}), as well as by Woods-Saxon-Strutinsky calculations \cite{Kern1987} for $N=74$, 76 (see Table I of \cite{Kern1987}) and calculations with a Bohr Hamiltonian including a cubic term \cite{Rohozinski1974} (see Table 1 of \cite{Rohozinski1974}). Experimental evidence for many triaxial bands in $^{138}$Nd$_{78}$ has been provided in Ref. \cite{Petrache2015}.    

The reduction in the potential energy of the ground state due to triaxiality within the macroscopic-microscopic approach of FRDM \cite{Moeller2008},
shows non-negligible (above 0.01 MeV) values for Nd isotopes with $N=76$, 78, 122. No experimental information exists \cite{ensdf} for $^{182}$Nd$_{122}$.  

In conclusion, evidence for triaxiality is seen in the Nd isotopes in an extended neutron region ($N=84$-94), but the FRDM calculations \cite{Moeller2008} suggest that the triaxial minima should be relatively shallow. This finding is in rough agreement with the expectations of the proxy-SU(3) symmetry \cite{Bonatsos2025}, according to which  triaxiality is favored neither in the region $N=82$-114, nor in $Z=50$-72, to which Nd belongs. In contrast, the FRDM calculations \cite{Moeller2008} suggest substantial triaxiality for the Nd isotopes with $N=76$, 78, in agreement with existing experimental evidence and theoretical results, as well as with the expectations of the proxy-SU(3) symmetry \cite{Bonatsos2025}, according to which  triaxiality is favored in the interval 74-80.

\subsection{The Sm (Z=62) Isotopes} \label{Sm}

$^{152}$Sm$_{90}$ has been among the nuclei suggested as candidates for triaxiality when the Triaxial Rotor Model was introduced \cite{Davydov1958,Davydov1960} (see Table 3 of \cite{Davydov1958} and Table 1 of \cite{Davydov1960}). Model-independent shape parameters based on quadrupole shape invariants \cite{Kumar1972,Werner2005} (see Table I of \cite{Kumar1972} and Table II of \cite{Werner2005}), also support the existence of moderate triaxiality in this nucleus, which is the textbook example \cite{Casten2001} of manifestation of the X(5) critical point symmetry \cite{Iachello2001}, related to the shape/phase transition from spherical to prolate deformed shapes. 
Further investigations within the Triaxial Rotor Model \cite{Varshni1970,Bindra2018} have considered triaxiality in the region $N=86$-98 (see Table 1 of \cite{Varshni1970} and Table 1 of \cite{Bindra2018}).

On the microscopic front, triaxiality has been suggested by Triaxial Projected Shell Model calculations performed for the Sm isotopes in the $N=88$, 90 region \cite{Naz2018} (see Table 3 of \cite{Naz2018}), as well as by adiabatic time-dependent HF calculations \cite{Gupta1982} for $N=88$-92 (see Table II of \cite{Gupta1982}). 
In contrast, relativistic HB calculations for Sm isotopes using the DD-ME2 parametrization \cite{Naz2018} do not find any triaxialty in the $N=82$-96 region (see Table 2 of \cite{Naz2018}). 

Regarding $^{154}$Sm$_{92}$, Monte Carlo shell model calculations \cite{Otsuka2023} predict a small value of $\gamma$, $3.7^{\rm o}$ (see Figs. 18, 24 of \cite{Otsuka2023}).
Recently \cite{Kleemann2025} empirical values of the deformation parameters have been extracted through Smekal-Raman scattering (a new probe of nuclear ground-state deformation). The values $\beta=0.2925(25)$ and $\gamma=5.0(15)^{\rm o}$ have been obtained \cite{Kleemann2025}, in amazing agreement with the predictions of the proxy-SU(3) symmetry ($\beta=0.296$, 
$\gamma= 4.31^{\rm o}$) reported in Ref. \cite{Bonatsos2024} (see Table 3 of \cite{Bonatsos2024}).   

On the other hand, experimental evidence for substantial triaxiality in the Sm isotopes with $N=74$, 76 has been seen in Ref. \cite{Kern1987}, also supported by Woods-Saxon-Strutinsky calculations for $N=76$-80 (see Table I of \cite{Kern1987}). Evidence for triaxiality in $^{138}$Sm$_{76}$ has also been seen in Hartree-Fock+BCS calculations \cite{Redon1986} (see Fig. 1 of \cite{Redon1986}), as well as through the extended Thomas-Fermi plus Strutinsky integral method \cite{Dutta2000} (see Table II of \cite{Dutta2000}). 

The reduction in the potential energy of the ground state due to triaxiality within the macroscopic-microscopic approach of FRDM \cite{Moeller2008},
shows non-negligible (above 0.01 MeV) values for Sm isotopes with $N=74$-78, 122. No experimental information exists \cite{ensdf} for $^{184}$Sm$_{122}$.  

In conclusion, evidence for triaxiality is seen in the Sm isotopes in an extended neutron region ($N=86$-94), but the FRDM calculations \cite{Moeller2008} suggest that the triaxial minima should be relatively shallow. This finding is in rough agreement with the expectations of the proxy-SU(3) symmetry \cite{Bonatsos2025}, according to which  triaxiality is favored neither in the region $N=82$-114, nor in $Z=50$-72, to which Sm belongs. In contrast, the FRDM calculations \cite{Moeller2008} suggest substantial triaxiality for the Sm isotopes with $N=74$-78, in agreement with existing experimental evidence and theoretical results, as well as with the expectations of the proxy-SU(3) symmetry \cite{Bonatsos2025}, according to which  triaxiality is favored in the interval 74-80.

\subsection{The Gd (Z=64) Isotopes}

$^{154-158}$Gd$_{90-94}$ have been among the nuclei suggested as candidates for triaxiality when the Triaxial Rotor Model was introduced \cite{Davydov1958,Davydov1959,Davydov1960} (see Table 3 of \cite{Davydov1958}, Table 5 of \cite{Davydov1959}, and Table 1 of \cite{Davydov1960}).
Further investigations within the Triaxial Rotor Model \cite{Varshni1970,Bindra2018,Allmond2017} have considered triaxiality in the region $N=88$-96 (see Table 1 of \cite{Varshni1970}, Table 1 of \cite{Bindra2018}, and Table 1 of \cite{Allmond2017}).
Odd-even staggering systematics \cite{Liao1995} in the $\gamma$-band of $^{160}$Gd$_{96}$ (see Table II of \cite{Liao1995}), as well as model-independent shape parameters based on quadrupole shape invariants \cite{Werner2005} in $^{154-160}$Gd$_{90-96}$ (see Table II of \cite{Werner2005}), also support the existence of moderate triaxiality in these isotopes, in accordance to IBM-1 calculations including cubic terms \cite{Gill1992} performed for the Gd isotopes  with $N=92$, 94 (see Table 1 of \cite{Gill1992}). It should be noticed that 
$^{154}$Gd$_{90}$ has been suggested \cite{Dewald2004,Tonev2004} as a good example of the X(5) critical point symmetry \cite{Iachello2001}.   

On the microscopic front, triaxiality has been suggested by Triaxial Projected Shell Model calculations performed for the Gd isotopes in the $N=90$, 92 region \cite{Jehangir2021,Rouoof2024,Naz2018} (see Table 2 of \cite{Jehangir2021}, Table 3 of \cite{Rouoof2024}, Table 3 of \cite{Naz2018}), as well as for $^{156}$Gd$_{92}$ by cranked HFB calculations \cite{Schuck2019} (see Table I of \cite{Schuck2019}). Triaxiality has also been considered in the Gd isotopes with $N=94$-98 in relation to Monte Carlo Shell Model calculations \cite{Otsuka2023} (see Fig. 16 of \cite{Otsuka2023}). In contrast, relativistic HB calculations using the DD-ME2 parametrization \cite{Naz2018} do not find any triaxiality in the $N=82$-94 region (see Table 2 of \cite{Naz2018}). 

Triaxiality in the neutron-deficient $^{140}$Gd$_{76}$ isotope has been seen through the odd-even staggering in the $\gamma$-band of this nucleus \cite{Paul1989}. 
Triaxiality for the Gd isotopes with $N=76$-80 is supported by Woods-Saxon-Strutinsky calculations \cite{Kern1987} (see Table I of \cite{Kern1987}).  

The reduction in the potential energy of the ground state due to triaxiality within the macroscopic-microscopic approach of FRDM \cite{Moeller2008},
shows non-negligible (above 0.01 MeV) values for Gd isotopes with $N=66$, 76, 78.  No experimental information exists \cite{ensdf} for $^{130}$Gd$_{66}$.  

In conclusion, evidence for triaxiality is seen in the Gd isotopes in an extended neutron region ($N=88$-98), but the FRDM calculations \cite{Moeller2008} suggest that the triaxial minima should be relatively shallow. This finding is in rough agreement with the expectations of the proxy-SU(3) symmetry \cite{Bonatsos2025}, according to which  triaxiality is favored neither in the region $N=82$-114, nor in $Z=50$-72, to which Gd belongs. In contrast, the FRDM calculations \cite{Moeller2008} suggest substantial triaxiality for the Gd isotopes with $N=76$, 78, in agreement with existing experimental evidence and theoretical results, as well as with the expectations of the proxy-SU(3) symmetry \cite{Bonatsos2025}, according to which  triaxiality is favored in the interval 74-80.

\subsection{The Dy (Z=66) Isotopes}

$^{160}$Dy$_{94}$ has been among the nuclei suggested as candidates for triaxiality when the Triaxial Rotor Model was introduced \cite{Davydov1958,Davydov1960} (see Table 3 of \cite{Davydov1958} and Table 1 of \cite{Davydov1960}). Further investigations within the Triaxial Rotor Model \cite{Varshni1970,Bindra2018} have considered triaxiality in the region $N=88$-98 (see Table 1 of \cite{Varshni1970} and Table 2 of \cite{Bindra2018}). $^{164}$Dy$_{98}$ has been suggested as a possible candidate for the Y(5) critical point symmetry \cite{Iachello2003}, regarding the shape/phase transition from axially to triaxially deformed shapes (see Table III of \cite{Iachello2003}). 
Odd-even staggering systematics \cite{Liao1995} in the $\gamma$-band of $^{164}$Dy$_{98}$ (see Table II of \cite{Liao1995}), as well as model-independent shape parameters based on quadrupole shape invariants \cite{Werner2005} (see Table II of \cite{Werner2005}), also support the existence of moderate triaxiality in this nucleus, in accordance to IBM-1 calculations including cubic terms \cite{Gill1992} performed for the Dy isotopes  with $N=94$-98 (see Table 1 of \cite{Gill1992}). It should be noticed that 
$^{156}$Dy$_{90}$ has been suggested \cite{Caprio2002,Dewald2004} as a good example of the X(5) critical point symmetry \cite{Iachello2001}.   

On the microscopic front, triaxiality has been suggested by Triaxial Projected Shell Model calculations performed for the Dy isotopes in the $N=88$-98 region \cite{Sheikh2001,Sun2002,Sheikh2016,Naz2018,Jehangir2021,Rouoof2024} (see Table 1 of \cite{Sun2002}, Table 2 of \cite{Sheikh2016}, Table 3 of \cite{Naz2018}, Table 2 of \cite{Jehangir2021}, Table 3 of \cite{Rouoof2024}). Triaxiality has also been considered in the Dy isotopes with $N=92$-100, 104 in relation to Monte Carlo Shell Model calculations \cite{Otsuka2023} (see Fig. 16 of \cite{Otsuka2023}). In contrast, 
relativistic HB calculations using the DD-ME2 parametrization \cite{Naz2018} do not find any triaxilaity in the $N=82$-94 region (see Table 2 of \cite{Naz2018}). 
 
The reduction in the potential energy of the ground state due to triaxiality within the macroscopic-microscopic approach of FRDM \cite{Moeller2008},
shows non-negligible (above 0.01 MeV) values for Dy isotopes with $N=76$, 78, for which little experimental information exists \cite{ensdf}. Woods-Saxon-Strutinsky calculations \cite{Kern1987} for $N=76$ corroborate this prediction (see Table I of \cite{Kern1987}).  

In conclusion, evidence for triaxiality is seen in the Dy isotopes in an extended neutron region ($N=88$-104), but the FRDM calculations \cite{Moeller2008} suggest that the triaxial minima should be relatively shallow. This finding is in rough agreement with the expectations of the proxy-SU(3) symmetry \cite{Bonatsos2025}, according to which  triaxiality is favored neither in the region $N=82$-114, nor in $Z=50$-72, to which Dy belongs. In contrast, the FRDM calculations \cite{Moeller2008} suggest substantial triaxiality for the Dy isotopes with $N=76$, 78, in agreement with existing theoretical calculations and 
the expectations of the proxy-SU(3) symmetry \cite{Bonatsos2025}, according to which  triaxiality is favored in the interval 74-80.

\section{The ${\bf Z=68}$-72 Region} \label{Z6872}

Protons in this region lie in the interval 50-72, extending between the intervals 34-48 and 74-80, in which triaxiality is favored according to the proxy-SU(3) predictions 
 (\cite{Bonatsos2025}, see also sec. \ref{regi}). The nuclei above the $Z=66$ midshell of the 50-82 major shell are considered in this section, while the nuclei up to the midshell have been considered in the previous section. 

% The $Z=68$-72 region appears to be a desert, as far as substantial triaxiality is concerned.

\subsection{The Er (Z=68) Isotopes}

Much attention has been attracted by the Er isotopes with $N=96$-102. $^{166,168}$Er$_{98,100}$ have been suggested as candidates for triaxiality within the Triaxial Rotor Model since its introduction \cite{Davydov1960,Wood2004,Allmond2017} (see Table 1 of \cite{Davydov1960} and Table I of \cite{Allmond2017}). They have also been suggested as candidates for the critical point symmetry Y(5) \cite{Iachello2003}, corresponding to the shape/phase transition from axially to triaxially deformed shapes (see Table III of \cite{Iachello2003}). Triaxiality in $^{166}$Er$_{98}$ has attracted special attention within recent Monte Carlo Shell Model calculations \cite{Otsuka2019,Tsunoda2021}, aiming at pointing out that a certain degree of triaxiality is present in nuclei considered as prolate deformed, in agreement with earlier experiments \cite{Fahlander1992}, as well as with a successful description of this nucleus within the Algebraic Collective Model including a cubic term \cite{Thiamova2015}. Further evidence for triaxiality in the Er isotopes with $N=96$-102 has been provided within the Triaxial Projected Shell Model \cite{Jehangir2021,Rouoof2024} (see Table 2 of \cite{Jehangir2021} and 
Table 3 of \cite{Rouoof2024}), cranked HFB calculations \cite{Schuck2019}  (see Table I of \cite{Schuck2019}), calculations with the extended Thomas-Fermi plus Strutinsky integral method \cite{Dutta2000} (see Table II of \cite{Dutta2000}), mean-field calculations involving the Gogny D1S interaction \cite{Tagami2016}, and IBM-1 calculations involving a triaxial term \cite{Gill1992} (see Table 1 of \cite{Gill1992}), also supported by systematics of the odd-even staggering in the $\gamma$-band \cite{Liao1995} (see Table II of \cite{Liao1995}), which single out $^{170}$Er$_{102}$ as a clear example of rigid triaxiality \cite{McCutchan2007} (see Fig. 4 of \cite{McCutchan2007}). In addition, IBM-1 calculations with parameters determined through a Skyrme energy density functional \cite{Vasileiou2025} (see Table 1 of \cite{Vasileiou2025} for the $N=92$-112 Er isotopes) have been performed.
Moderate triaxiality is seen in several cases. 

Calculations covering triaxiality in Er isotopes below the $N=96$-102 region have been performed within the Triaxial Rotor Model for $N=88$-102 \cite{Varshni1970,Bindra2018}
(see Table 1 of \cite{Varshni1970} and Table 2 of \cite{Bindra2018}), the Triaxial Projected Shell Model for $N=88$-102 \cite{Sheikh1999,Sun2000,Sheikh2001,Sheikh2008}
(see Table I of \cite{Sun2000}), as well as for $N=88$-104 within HFB calculations with the generator coordinate method using a pairing-plus-quadrupole residual interaction 
\cite{Chen2017} (see Table I of \cite{Chen2017}). Furthermore, triaxiality in the Er isotopes with  $N=92$-100 has been considered in relation to Monte Carlo Shell Model calculations \cite{Otsuka2023} (see Fig. 16 of \cite{Otsuka2023}). Finally, triaxiality in $^{158}$Er$_{90}$  has been studied within a Skyrme-Hartree-Fock framework 
\cite{Shi2012}, while the same nucleus has been considered as an example for the T(4) special solution of the Bohr Hamiltonian \cite{Zhang2017}, which provides a connection of the X(4)\cite{Budaca2016}                                                                                                                                                                                                                                                                                                                                                                                                                                                                                                                                                                                                                                                                                                                                                                                                                                                                                                                                                                                                                                                                                                                                                                                                                                                                                                                                                                                                                                                                                                                                                                                                                                                                          and Z(4) \cite{Bonatsos2005} critical point symmetries. 

In conclusion, evidence for triaxiality is seen in the Er isotopes in an extended neutron region ($N=88$-104), but the FRDM calculations \cite{Moeller2008} suggest that the triaxial minima should be relatively shallow. This finding is in rough agreement with the expectations of the proxy-SU(3) symmetry \cite{Bonatsos2025}, according to which  triaxiality is favored neither in the region $N=82$-114, nor in $Z=50$-72, to which Er belongs.

\subsection{The Yb (Z=70) Isotopes}

Substantial values of the collective variable $\gamma$ have been obtained for Yb isotopes within the Triaxial Rotor Model for $N=92$-106 in Ref. \cite{Varshni1970} (see Table 1 of \cite{Varshni1970}), as well as for $N=90$-104 in Ref. \cite{Bindra2018} (see Table 2 of \cite{Bindra2018}). Signs of triaxiality in 
$^{164-168,178}$Yb$_{94-98,108}$ have been reported in relation to Monte Carlo Shell Model calculations \cite{Otsuka2023} (see Fig. 16 of \cite{Otsuka2023}).
Special attention has been given to $^{172}$Yb$_{102}$ within the Triaxial Rotor Model \cite{Wood2004,Allmond2017} (see Table II of \cite{Wood2004} and Table I of \cite{Allmond2017}), as well as within cranked HFB calculations \cite{Schuck2019}). Studies of lighter isotopes have been performed within the Triaxial Projected Shell Model for $^{158}$Yb$_{88}$ \cite{Sheikh1999}, the IBM-1 with a cubic term for  $^{168}$Yb$_{98}$ \cite{Gill1992} (see Table 1 of \cite{Gill1992}), as well as with the special solution T(5) of the Bohr Hamiltonian \cite{Zhang2015a}, related to the shape/phase transition from spherical to triaxially deformed nuclei (see Table 2 of \cite{Zhang2015a}). Heavier Yb isotopes with $N=110$-122 have been considered by self-consistent HFB calculations employing the Gogny D1S and Skyrme Sly4 interactions \cite{Robledo2009}, 
pointing out triaxiality in $^{186}$Yb$_{116}$ (see Table 1 of \cite{Robledo2009}). 

In conclusion, evidence for triaxiality is seen in the Yb isotopes in an extended neutron region ($N=88$-116), but the FRDM calculations \cite{Moeller2008} suggest that the triaxial minima should be relatively shallow. This finding is in rough agreement with the expectations of the proxy-SU(3) symmetry \cite{Bonatsos2025}, according to which  triaxiality is favored neither in the region $N=82$-114, nor in $Z=50$-72, to which Yb belongs.

\subsection{The Hf (Z=72) Isotopes}

Substantial values of the collective variable $\gamma$ have been obtained for Hf isotopes in extended neutron regions. Numerical results have been reported within the Triaxial Rotor Model for $N=94$-108 in Ref. \cite{Varshni1970} (see Table 1 of \cite{Varshni1970}), $N=92$-110 in Ref. \cite{Esser1997} (see Table II of \cite{Esser1997}). and $N=100$-104 in Ref. \cite{Bindra2018} (see Table 2 of \cite{Bindra2018}), as well as from studies within IBM-1 with parameters determined through mean-field calculations involving a Skyrme energy density functional for $N=90$-112 \cite{Vasileiou2024} (see Table II of \cite{Vasileiou2024}). 

Special attention has been given to $^{180}$Hf$_{108}$ within the Triaxial Projected Shell Model \cite{Sheikh2016,Jehangir2021,Rouoof2024} (see  Table 2 of \cite{Sheikh2016}, Table 2 of \cite{Jehangir2021}, and Table 3 in \cite{Rouoof2024}), the Bohr Hamiltonian with a sextic potential in the collective variable $\beta$ and a Mathieu wave function in $\gamma$ \cite{Buganu2012,Buganu2012b,Raduta2013}, as well as within the Coherent State Model \cite{Buganu2012,Buganu2012b,Raduta2013}. Staggering systematics \cite{Liao1995} suggest triaxial deformation in $^{170,180}$Hf$_{98,108}$ (see Table II of \cite{Liao1995}), while IBM-1 calculations with a cubic term \cite{Gill1992} point out $^{178}$Hf$_{106}$ as a possible candidate for triaxiality (see Table 1 of \cite{Gill1992}). Signs of triaxiality in $^{168,178,182}$Hf$_{96,106,110}$ have been reported in relation to Monte Carlo Shell Model calculations \cite{Otsuka2023} (see Fig. 16 of \cite{Otsuka2023}), while in heavier Hf isotopes triaxiality has been suggested for $^{188,190}$Hf$_{116,118}$
within mean-field HFB calculations involving the Gogny D1S and Skyrme Sly4 interactions \cite{Robledo2009} (see Table 1 of \cite{Robledo2009}).  

The reduction in the potential energy of the ground state due to triaxiality within the macroscopic-microscopic approach of FRDM \cite{Moeller2008},
does not show any non-negligible (above 0.01 MeV) values for Hf isotopes. 

In conclusion, evidence for triaxiality is seen in the Hf isotopes in an extended neutron region ($N=90$-118), but the FRDM calculations \cite{Moeller2008} suggest that the triaxial minima should be relatively shallow. This finding is in rough agreement with the expectations of the proxy-SU(3) symmetry \cite{Bonatsos2025}, according to which  triaxiality is favored neither in the region $N=82$-114, nor in $Z=50$-72, to which Hf belongs.  

\section{The ${\bf Z=74}$-80 Region} \label{Z7480}

The protons of this region belong to the 74-80 interval, in which triaxiality is favored according to the proxy-SU(3) symmetry (\cite{Bonatsos2025}, see also sec. \ref{regi}). 

% As we shall see, most of the nuclei exhibiting triaxiality in this region have neutrons in the interval 82-114, for which no triaxiality is expected by the proxy-SU(3) symmetry. We 
% could therefore call them nuclei presenting proton-induced triaxiality.

\subsection{The W (Z=74) Isotopes}

The W isotopes with $N=108$-112 have been suggested as possible candidates for triaxiality since the introduction of the Davydov model \cite{Davydov1958,Davydov1959,Davydov1960} (see Table 2 of \cite{Davydov1958}, Table 5 of \cite{Davydov1959}, Table 1 of \cite{Davydov1960}).
Further studies within the Triaxial Rotor Model \cite{Wood2004,Allmond2017} have corroborated this suggestion (see Table II of \cite{Wood2004} and Table 1 of \cite{Allmond2017}), in agreement with odd-even staggering systematics \cite{Liao1995} (see Table II of \cite{Liao1995}). Phenomenological studies involving the Bohr Hamiltonian with a sextic potential in the collective variable $\beta$ and a Mathieu wave function in $\gamma$ \cite{Buganu2012,Raduta2013}, the Coherent State Model \cite{Buganu2012,Raduta2013}, 
and the IBM-1 with cubic terms \cite{Gill1992} (see Table 1 of \cite{Gill1992}) corroborate this suggestion. The same holds for mean-field calculations using the Hartree-Bogoliubov approach \cite{Girod1978}, the extended Thomas--Fermi plus Strutinsky integral method \cite{Dutta2000}, and the cranked HFB approach \cite{Schuck2019} (see Table I of \cite{Girod1978}, Table II of \cite{Dutta2000}, and Table I of \cite{Schuck2019}). 

In addition, several studies have been performed, including the $N=108$-112 W isotopes and extended to higher and/or lower neutron numbers. These include calculations within the Triaxial Rotor Model for $N=98$-112 \cite{Varshni1970}, $N=104$-112 \cite{Esser1997}, and $N=106$-112 \cite{Bindra2018} (see Table 1 of \cite{Varshni1970}, Table II of \cite{Esser1997}, and Table 3 in \cite{Bindra2018}, the Monte Carlo Shell Model \cite{Otsuka2023} (see Fig. 16 of \cite{Otsuka2023} for $N=102$-114), as  well as IBM-1 calculations with parameters determined through a Skyrme energy density functional \cite{Vasileiou2024} (see Table II of \cite{Vasileiou2024} for the $N=94$-112 W isotopes).
Substantial triaxiality is seen in all cases. 

Studies for W isotopes above the $N=108$-112 region include HFB calculations with the Gogny D1S and Skyrme SLy4 interactions \cite{Robledo2009}, in which triaxiality is seen for $N=116$-120 (see Table 1 of \cite{Robledo2009}), HF calculations with a Skyrme energy density functional complemented with the tensor force \cite{Fracasso2012} revealing triaxiality for $N=116$ (see Table VIII of \cite{Fracasso2012}), as well as covariant density functional theory calculations involving a five-dimensional collective Hamiltonian \cite{Yang2021a} pointing out triaxiality for $N=120$. For W isotopes below the $N=108$-112 region, IBM-1 calculations including a cubic term \cite{Pan2024} indicate triaxiality for $N=92$, while Triaxial Projected Shell Model calculations \cite{Sheikh1999} find triaxiality for $N=102$. 

The reduction in the potential energy of the ground state due to triaxiality within the macroscopic-microscopic approach of FRDM \cite{Moeller2008},
shows a non-negligible (above 0.01 MeV) value only for $N=84$. 

In conclusion, substantial triaxiality is seen in the W isotopes in an extended neutron region ($N=92$-120), in agreement to the expectations of the proxy-SU(3) symmetry \cite{Bonatsos2025}, according to which triaxiality is favored in the region $Z=74$-80, to which W belongs. However, the FRDM calculations \cite{Moeller2008} suggest that the triaxial minima should be relatively shallow.

\subsection{The Os (Z=76) Isotopes}

The Os isotopes with $N=110$-116 have been suggested as possible candidates for triaxiality since the introduction of the Davydov model \cite{Davydov1958,Davydov1959,Davydov1960} (see Tables 3 and 4 of \cite{Davydov1958}, Table 5 of \cite{Davydov1959}, Table 1 of \cite{Davydov1960}).
Further studies within the Triaxial Rotor Model \cite{Wood2004,Allmond2008,Allmond2017,Sugawara2018} have corroborated this suggestion (see, for example, Table II of \cite{Allmond2008} and Table 1 of \cite{Allmond2017}), in agreement with additional studies employing the Rotation Vibration Model \cite{Faessler1964b}, as well as 
exactly separable solutions of the Bohr Hamiltonian employing the Kratzer, Davidson, and infinite square-well potentials \cite{Fortunato2006}, or the sextic potential with a centrifugal term in the $\beta$ variable and a potential leading to the Mathieu equation in the $\gamma$ variable \cite{Raduta2011,Raduta2013,Raduta2014}, the latter  results being also compared to the findings of the Coherent State Model \cite{Raduta2011,Raduta2013}. The analogue of the Z(4) solution of the Bohr Hamiltonian involving the Davidson potential, called the Z(4)-D solution \cite{Yigitoglu2018} has also been used for $^{192}$Os$_{116}$. Odd-even staggering systematics \cite{Liao1995,McCutchan2007} suggest $^{192}$Os$_{116}$ as a clear example of a rigid triaxial nucleus, 
while model-independent measures of triaxiality based on quadrupole shape invariants \cite{Werner2005} support triaxiality in the Os isotopes with $N=112$-116. Experimental studies 
also indicate almost complete triaxiality in $^{186}$Os $_{110}$ \cite{Wheldon1999}. 

Triaxiality in the Os isotopes with $N=110$-116 has also been suggested by mean-field methods including the Hartree-BCS theory \cite{Sahu1979} (see Table I of \cite{Sahu1979} for 
$^{188}$Os$_{112}$), the HF+BCS theory \cite{Redon1986} (see Fig. 3 of \cite{Redon1986} for $^{192}$Os$_{116}$), the extended Thomas-Fermi plus Strutinsky integral method \cite{Dutta2000} (see Table II of \cite{Dutta2000} for  $^{188,192}$Os$_{112,116}$), the HFB theory with angular momentum projection \cite{Hayashi1984} (see Fig. 1(b) of \cite{Hayashi1984} for $^{188}$Os$_{112}$), and the cranked HFB theory \cite{Schuck2019} (see Table I of \cite{Schuck2019} for  $^{186-192}$Os$_{110-118}$), as well as by relativistic mean-field calculations \cite{Nomura2012a,Nomura2021b} (see Fig. 1 of
\cite{Nomura2012a} for $^{190}$Os$_{114}$ and Table I of \cite{Nomura2021b} for $^{188-192}$Os$_{112-118}$).

In the realm of algebraic collective models, triaxiality in the Os isotopes with $N=110$-116 has been considered within the IBM realization of the Triaxial Rotor Model 
\cite{Zhang2024} (see Fig. 6 of \cite{Zhang2024} for  $^{190,192}$Os$_{114,116}$), as well as within IBM-2 \cite{Walet1987} and the two-fluid Interacting Vector Boson  Model \cite{Ganev2011}, in both of which $^{192}$Os$_{116}$ has been considered as the textbook example of the SU(3)$^*$ symmetry. Triaxiality in the Os isotopes with $N=112$-116 has also been considered within the Algebraic Collective Model \cite{Thiamova2010,Thiamova2014}.

Studies covering triaxiality in the $N=110$-116 region but also extended to heavier Os isotopes appear to be limited in the $N=118$-122 region, including mean-field calculations with the Gogny D1S and Skyrme Sly4 interactions for $N=110$-122 \cite{Robledo2009} (see Table 1 of \cite{Robledo2009}), covariant density functional theory calculations employing a five-dimensional collective Hamiltonian \cite{Yang2021a}, in which rigid triaxial deformation is found for $N=116$-120, as well as Rigid Triaxial Rotor Model calculations \cite{Esser1997} for $N=98$-118 (see Table II of \cite{Esser1997}), while $^{196}$Os$_{120}$ has been investigated experimentally in Ref. \cite{John2014}. 

Triaxiality in the $N=110$-116 region but also extended to lighter Os isotopes has been considered within the Triaxial Rotor Model \cite{Varshni1970,Esser1997,Bindra2018} (see Table 1 of \cite{Varshni1970} for $N=102$-116, Table II of \cite{Esser1997} for $N=98$-118, and Table 3 of \cite{Bindra2018} for $N=106$-116), the Triaxial Projected Shell Model \cite{Sheikh1999,Rouoof2024} (see Table 3 of \cite{Rouoof2024} for $N=106$-116), the Monte Carlo Shell Model \cite{Otsuka2023} (see Fig. 16 of \cite{Otsuka2023} for $N=106$-112), as well as within IBM-1 including a cubic term \cite{Gill1992,Sorgunlu2008} (see Table 1 of \cite{Gill1992} for $N=108$, 110 and Figs. 11, 13 of \cite{Sorgunlu2008} for $N=104$-110). 

Isolated cases of triaxiality in light Os isotopes have been considered by mapping a triaxial rotor model on the IBM-1 with cubic terms in Ref. \cite{Zhang2022} for $N=92$, 
in Ref. \cite{Pan2024} for $N=94$-98, and in Ref. \cite{Zhang2024} for $N=92$, 94. In addition, mean-field calculations with a Skyrme interaction have been performed in Ref. \cite{Fracasso2012} for $N=102$, while relativistic mean-field plus BCS calculations have been done in Ref. \cite{Xiang2016} for $N=106$. It may be noticed that $^{178}$Os$_{102}$, and 
$^{176}$Os$_{100}$ to a lesser extent, have been considered \cite{Dewald2005} as candidates for the experimental manifestation of the X(5) critical point symmetry \cite{Iachello2001}. 

The reduction in the potential energy of the ground state due to triaxiality within the macroscopic-microscopic approach of FRDM \cite{Moeller2008},
does not show any non-negligible (above 0.01 MeV) values for Os isotopes. 

In conclusion, substantial triaxiality is seen in the Os isotopes in an extended neutron region ($N=92$-120), in agreement to the expectations of the proxy-SU(3) symmetry \cite{Bonatsos2025}, according to which triaxiality is favored in the region $Z=74$-80, to which Os belongs. However, the FRDM calculations \cite{Moeller2008} suggest that the triaxial minima should be relatively shallow. 

\subsection{The Pt (Z=78) Isotopes}

Much attention has been focused on the Pt isotopes with $N=114$-118. $^{192,196}$Pt$_{114,118}$ have been considered as candidates for triaxiality already in the first article in which the Davydov model was introduced \cite{Davydov1958} (see Table 3 of \cite{Davydov1958}), while $^{194,196}$Pt$_{116,118}$ have been studied in further work on the Triaxial Rotor Model \cite{Allmond2009,Allmond2010}. $^{192,194}$Pt$_{114,116}$ have been suggested as candidates for triaxiality through studies of the odd-even staggering in their $\gamma$ bands  \cite{Liao1995,McCutchan2007}, the former being singled out as a clear example of rigid triaxiality \cite{McCutchan2007} (see Fig. 4 of \cite{McCutchan2007}). $^{194,196}$Pt$_{116,118}$ have been pointed out as candidates for triaxiality through model-independent quantities based on quadrupole shape invariants \cite{Werner2005} (see Table II of \cite{Werner2005}). The possibility of equivalence between $\gamma$-instability and rigid triaxiality has been discussed in terms of   $^{196}$Pt$_{118}$ in Ref. \cite{Otsuka1987}. $^{192-196}$Pt$_{114-118}$ have been considered as possible candidates for the Z(5) critical point symmetry of the shape/phase transition from prolate to oblate shapes \cite{Bonatsos2004}, as well as in relation to the T(5) shape/phase transition from spherical to triaxial shapes \cite{Zhang2015a}. They have also been considered within special solutions of the Bohr Hamiltonian for triaxial nuclei involving the Morse potential (Z(5)-M, \cite{Inci2014}), the Davidson potential (Z(5)-D, \cite{Lee2013}), the Hulth\'{e}n potential (Z(5)-H, \cite{Chabab2015}), and the inverse square potential (Z(5)-IS, \cite{Ajulo2022}), as well as within special solutions of the conformable fractional Bohr Hamiltonian involving the Kratzer potential \cite{Hammad2021c,Hammad2023}, various exponential-type potentials \cite{Hammad2023} and a four inverse power terms potential \cite{Ahmadou2022}. 
The Pt isotopes with $N=114$-118 have also been considered within several variations of the Z(4) special solution of the Bohr Hamiltonian \cite{Bonatsos2005},  which is a $\gamma$-rigid version of the Z(5) critical point symmetry \cite{Bonatsos2004}, involving an infinite square well potential. Variations of Z(4), involving the Davidson potential (Z(4)-D, \cite{Yigitoglu2018}) and the sextic potential (Z(4)-sextic, \cite{Buganu2015,Buganu2015b,Buganu2015c,Budaca2016b}), as well as the Davidson \cite{Buganu2017} and Kratzer \cite{AitElKorchi2021,AitElKorchi2022} potentials with deformation-dependent mass (DDM), called Z(4)-DDM-D and Z(4)-DDM-K respectively, have been considered. In addition, a modified version of Z(4), in which the value of $\gamma$ is allowed to deviate from $30^{\rm o}$ has been introduced \cite{Ajulo2024}. 

In addition, several studies have been performed, including the $N=114$-118 Pt isotopes and extended to higher and/or lower neutron numbers. These include mean-field calculations with the Gogny D1S interaction for $N=110$-122 \cite{Robledo2009}, finding considerable triaxiality for $N=110$-118 (see Table 1 of Ref. \cite{Robledo2009}), 
as well as with the Gogny D1S, D1N, and D1M interactions for $N=88$-126 \cite{Rodriguez2010}, finding triaxiality for $N=94$ and $N=106$-116 (see Figs. 2-4 of \cite{Rodriguez2010}). Relativistic HB calculations for $N=108$-126 \cite{Niksic2010} show substantial triaxiality in $N=112$-120 (see Table I of \cite{Niksic2010}),
while RMF+BCS calculations involving the PC-PK1 and DD-PC1 relativistic energy density functionals \cite{Nomura2021b} show substantial triaxiality in $N=114$-118
(see Table I of \cite{Nomura2021b}).  Calculations within the Triaxial Rotor Model show substantial triaxiality in the region $N=102$-120 \cite{Esser1997,Bindra2018} 
(see Table II of \cite{Esser1997} and Table 3 of \cite{Bindra2018}). Triaxial Projected Shell Model calculations \cite{Bhat2012,Rouoof2024} predict substantial triaxiality in the Pt isotopes with $N=102$-112 (see Table I of \cite{Bhat2012}) and $N=114$, 116 (see Table 3 of \cite{Rouoof2024}). 

Furthermore, there are studies focusing on nuclei below $N=114$. Moving gradually from $N=114$ to lower neutron numbers we see that Hartree-BCS calculations employing the pairing plus quadrupole-quadrupole interaction \cite{Sahu1979}
predict substantial triaxiality in $N=110$ (see Table I of \cite{Sahu1979}), but lattice Hartree-Fock+BCS calculations \cite{Redon1986} predict in $N=108$ a valley connecting smoothly the oblate and prolate minima instead of a triaxial minimum (see Fig. 2 of \cite{Redon1986}), in contrast to IBM-1 calculations involving cubic terms 
 \cite{Sorgunlu2008}, which do predict triaxiality in $N=108$ (see Fig. 12 of \cite{Sorgunlu2008}). Triaxial Rotor Model calculations \cite{Varshni1970} suggest substantial triaxiality in $N=104$, 106 (see Table 1 of \cite{Varshni1970}), while IBM-1 calculations including cubic terms \cite{Zhang2022,Pan2024} predict triaxiality in $N=94$-98. 

The reduction in the potential energy of the ground state due to triaxiality within the macroscopic-microscopic approach of FRDM \cite{Moeller2008},
shows non-negligible (above 0.01 MeV) values for $N=90$-96 and $N=110$-116.

In conclusion, substantial triaxiality is seen in the Pt isotopes in an extended neutron region ($N=94$-120), in agreement to the expectations of the proxy-SU(3) symmetry \cite{Bonatsos2025}, according to which triaxiality is favored in the region $Z=74$-80, to which Pt belongs. 

\subsection{The Hg (Z=80) Isotopes}

$^{198}$Hg$_{118}$ has been among the nuclei suggested as candidates for triaxiality \cite{Davydov1958,Davydov1960} when the Davydov model was introduced (see Table 3 of \cite{Davydov1958} and Table 1 of \cite{Davydov1960}). Calculations using the Gneuss--Greiner potential in the Bohr--Mottelson Hamiltonian suggested \cite{Alimohammadi2019}
a rather rigid triaxial shape around $\gamma=30^{\rm o}$ for this nucleus. Strong triaxiality has been suggested \cite{Esser1997} for the Hg isotopes with $N=108$-122 using the Rigid Triaxial Rotor Model (see Table 2 of \cite{Esser1997}). Strong triaxiality in $^{180}$Hg$_{100}$ has been suggested \cite{Sheikh2016} in the framework of the Triaxial Projected Shell Model (see Table 2 of \cite{Sheikh2016}). 

The reduction in the potential energy of the ground state due to triaxiality within the macroscopic-microscopic approach of FRDM \cite{Moeller2008},
shows non-negligible (above 0.01 MeV) values for $N=96$, 98.

In conclusion, the existing little amount of relevant work suggests that substantial triaxiality appears to be expected in the Hg isotopes in an extended neutron region ($N=96$-122), in agreement to the expectations of the proxy-SU(3) symmetry \cite{Bonatsos2025}, according to which triaxiality is favored in the region $Z=74$-80, to which Hg belongs. 

\section{The ${\bf Z=88}$-98 Region} \label{Z8898}

Protons in this region lie in the interval 82-114, extending between the intervals 74-80 and 116-124, in which triaxiality is favored according to the proxy-SU(3) predictions 
(\cite{Bonatsos2025}, see also sec. \ref{regi}).

\subsection{The Ra (Z=88) Isotopes}

The Ra isotopes with $N=132$, 136-140 have been among the nuclei suggested as candidates for substantial triaxiality since the early days of the Triaxial Rotor Model \cite{Varshni1970} (see Table 1 of \cite{Varshni1970}). In addition, $^{222}$Ra$_{134}$ has also been singled out as a candidate for triaxiality by calculations using the extended Thomas-Fermi plus Strutinsky integral method \cite{Dutta2000} (see Table II of \cite{Dutta2000}). 

In summary, evidence for triaxiality in the Ra isotopes has been suggested in the region with $N=132$-140.  

In contrast, the reduction in the potential energy of the ground state due to triaxiality within the macroscopic-microscopic approach of FRDM \cite{Moeller2008},
shows non-negligible (above 0.01 MeV) values for the Ra isotopes with $N=110$-114 and 154-160, for which no experimental information exists \cite{ensdf}. 
 
In conclusion, substantial triaxiality has been suggested in the Ra isotopes in the neutron region $N=132$-140, in contrast to the expectations of the proxy-SU(3) symmetry \cite{Bonatsos2025}, according to which triaxiality is neither favored in the region with $N=126$-170, nor in the region $Z=82$-114, to which Ra belongs. 

\subsection{The Th (Z=90) Isotopes}

$^{232}$Th$_{142}$ has been among the nuclei suggested as candidates for substantial triaxiality when the Triaxial Rotor Model was introduced \cite{Davydov1959} (see Table 5 of \cite{Davydov1959}). It has also been singled out as one of the best examples of rigid triaxiality, based on the odd-even staggering within its $\gamma$-band \cite{McCutchan2007} (see Fig. 4 of \cite{McCutchan2007}, as well as Fig. 10 of \cite{Minkov2000}). Recent calculations within the Triaxial Projected Shell Model \cite{Jehangir2021,Rouoof2024} also suggest substantial triaxiality in $^{232}$Th$_{142}$ (see Table 2 of \cite{Jehangir2021} and Table 3 of \cite{Rouoof2024}). 

Triaxial Rotor Model calculations \cite{Varshni1970} suggest substantial triaxiality in the Th isotopes with $N=136$-142 (see Table 1 of \cite{Varshni1970}). Triaxiality in  $^{228}$Th$_{138}$ has been seen in the odd-even staggering within its $\gamma$-band \cite{Minkov2000} (see Fig. 9 in \cite{Minkov2000}).  
It has also been suggested by Triaxial Rotation-Vibration Model \cite{Meyer1998} and Coherent State Model \cite{Meyer1998} calculations, 
while triaxiality in $^{228,230}$Th$_{138,140}$ has been proposed by a Bohr Hamiltonian with a sextic potential plus a centrifugal term in the $\beta$ variable and a potential leading to Mathieu functions in the $\gamma$ variable \cite{Raduta2011,Raduta2013}, with its results compared to those of the Coherent State Model \cite{Raduta2011,Raduta2013}.

In summary, evidence for triaxiality in the Th isotopes has been suggested in the region with $N=136$-142.  

In contrast, the reduction in the potential energy of the ground state due to triaxiality within the macroscopic-microscopic approach of FRDM \cite{Moeller2008},
shows non-negligible (above 0.01 MeV) values for the Th isotopes with $N=108$-116, 156-160, for which no experimental information exists \cite{ensdf}. 
 
In conclusion, substantial triaxiality has been suggested in the Th isotopes in the neutron region $N=136$-142, in contrast to the expectations of the proxy-SU(3) symmetry \cite{Bonatsos2025}, according to which triaxiality is neither favored in the region with $N=126$-170, nor in the region $Z=82$-114, to which Th belongs. 

\subsection{The U (Z=92) Isotopes}

$^{238}$U$_{146}$ has been suggested as a candidate for substantial triaxiality in the framework of the Triaxial Projected Shell Model \cite{Sheikh2016,Jehangir2021,Rouoof2024} (see Table 2 of \cite{Sheikh2016}, Table 2 of \cite{Jehangir2021}, Table 3 of \cite{Rouoof2024}). This suggestion has also been made within mean-field calculations employing the Skyrme energy density functional complemented with the tensor force \cite{Fracasso2012}. 

Triaxiality in  $^{236}$U$_{144}$ has been suggested by Hartree-Fock+BCS calculations employing the SkM$^*$ Skyrme effective interaction \cite{Benrabia2017} (see Fig. 1 
of \cite{Benrabia2017}), as well as by calculations using the extended Thomas-Fermi plus Strutinsky integral method \cite{Dutta2000} (see Table II of \cite{Dutta2000}), which also predict mild triaxiality in  $^{262}$U$_{170}$ (see Table III of \cite{Dutta2000}).

Calculations with the Triaxial Rotor Model suggest substantial triaxiality for the U isotopes in the region $N=138$-146 \cite{Varshni1970} (see Table 1 of \cite{Varshni1970}). 

In summary, evidence for triaxiality in the U isotopes has been suggested in the region with $N=138$-146.  

In contrast, the reduction in the potential energy of the ground state due to triaxiality within the macroscopic-microscopic approach of FRDM \cite{Moeller2008},
shows non-negligible (above 0.01 MeV) values for the U isotopes with $N=112$-116, 156, 158, for which no experimental information exists \cite{ensdf}. 
 
In conclusion, substantial triaxiality has been suggested in the U isotopes in the neutron region $N=138$-146, in contrast to the expectations of the proxy-SU(3) symmetry \cite{Bonatsos2025}, according to which triaxiality is neither favored in the region with $N=126$-170, nor in the region $Z=82$-114, to which U belongs. 

\subsection{The Pu (Z=94) Isotopes}

The Pu isotopes with $N=142$-146 have been among the nuclei suggested as candidates for substantial triaxiality since the early days of the Triaxial Rotor Model \cite{Davydov1958,Varshni1970} (see Table 3 of \cite{Davydov1958} and Table 1 of \cite{Varshni1970}). $^{240}$Pu$_{146}$ has also been singled out as candidate for triaxiality by calculations using the extended Thomas-Fermi plus Strutinsky integral method \cite{Dutta2000} (see Table II of \cite{Dutta2000}), as well as by 
Hartree-Fock+BCS calculations using the SkM$^*$ Skyrme effective interaction \cite{Benrabia2017} (see Fig. 2 of \cite{Benrabia2017}).  

In summary, evidence for triaxiality in the Pu isotopes has been suggested in the region with $N=142$-146.  

In contrast, the reduction in the potential energy of the ground state due to triaxiality within the macroscopic-microscopic approach of FRDM \cite{Moeller2008},
shows non-negligible (above 0.01 MeV) values for the Pu isotopes with $N=116$, 158, for which no experimental information exists \cite{ensdf}. 
 
In conclusion, substantial triaxiality has been suggested in the Pu isotopes in the neutron region $N=142$-146, in contrast to the expectations of the proxy-SU(3) symmetry \cite{Bonatsos2025}, according to which triaxiality is neither favored in the region with $N=126$-170, nor in the region $Z=82$-114, to which Pu belongs. 

\subsection{The Cm (Z=96) Isotopes}

The Cm isotopes with $N=146$-152 have been among the nuclei suggested as candidates for substantial triaxiality since the early days of the Triaxial Rotor Model \cite{Varshni1970} (see Table 1 of \cite{Varshni1970}). $^{244}$Cm$_{148}$ has also been singled out as candidate for triaxiality by calculations using the extended Thomas-Fermi plus Strutinsky integral method \cite{Dutta2000} (see Table II of \cite{Dutta2000}), while $^{248}$Cm$_{152}$ has been singled out by 
Hartree-Fock+BCS calculations using the SkM$^*$ Skyrme effective interaction \cite{Benrabia2017} (see Fig. 3 of \cite{Benrabia2017}).  

In summary, evidence for triaxiality in the Cm isotopes has been suggested in the region with $N=146$-152.  

In contrast, the reduction in the potential energy of the ground state due to triaxiality within the macroscopic-microscopic approach of FRDM \cite{Moeller2008},
shows non-negligible (above 0.01 MeV) values for the Cm isotope with $N=130$, for which no experimental information exists \cite{ensdf}. 
 
In conclusion, substantial triaxiality has been suggested in the Cm isotopes in the neutron region $N=146$-152, in contrast to the expectations of the proxy-SU(3) symmetry \cite{Bonatsos2025}, according to which triaxiality is neither favored in the region with $N=126$-170, nor in the region $Z=82$-114, to which Cm belongs. 

\subsection{The Cf (Z=98) Isotopes}

$^{250}$Cf$_{152}$ has been among the nuclei suggested as candidates for substantial triaxiality since the early days of the Triaxial Rotor Model \cite{Varshni1970} (see Table 1 of \cite{Varshni1970}). $^{246}$Cf$_{154}$ has also been singled out as candidate for triaxiality by calculations using the extended Thomas-Fermi plus Strutinsky integral method \cite{Dutta2000} (see Table II of \cite{Dutta2000}), as well as by  
Hartree-Fock+BCS calculations using the SkM$^*$ Skyrme effective interaction \cite{Benrabia2017} (see Fig. 4 of \cite{Benrabia2017}).  

In summary, evidence for triaxiality in the Cf isotopes has been suggested in the region with $N=152$-154.  

In contrast, the reduction in the potential energy of the ground state due to triaxiality within the macroscopic-microscopic approach of FRDM \cite{Moeller2008},
shows non-negligible (above 0.01 MeV) values for Cf isotopes in the region  $N=130$-138, for which no experimental information exists \cite{ensdf}. 
 
In conclusion, substantial triaxiality has been suggested in the Cf isotopes in the neutron region $N=152$-154, in contrast to the expectations of the proxy-SU(3) symmetry \cite{Bonatsos2025}, according to which triaxiality is neither favored in the region with $N=126$-170, nor in the region $Z=82$-114, to which Cf belongs.

\end{document}